\documentclass{article}
\usepackage{KLMM/klmm}
\usepackage{microtype}
\usepackage{tikz}
\usepackage{placeins}
\usepackage{array,xurl}
\hypersetup{
  pdftitle={Three-Block Bounds for the Asymptotic Rank of Weighted Graph Tensors},
  pdfauthor={Zhengyang Wang, Yichuan Cao, Hongbo Li, Xiao-Shan Gao},
  bookmarksdepth=3
}

\pdfmapfile{=zi4.map}

\numberwithin{equation}{section}
\newtheorem{conjecture}[theorem]{Conjecture}

\newcommand{\rk}{\operatorname{R}}
\newcommand{\ark}{\widetilde{\operatorname{R}}}
\newcommand{\diam}{\operatorname{diam}}
\newcommand{\pos}[1]{\bigl[#1\bigr]_{+}}
\newcommand{\MC}{\operatorname{MaxCut}}
\newcommand{\BP}{B_{\mathrm{PZ}}}
\newcommand{\tr}{\operatorname{tr}}
\newcommand{\trans}{^{\mathsf{T}}}
\newcommand{\onevec}{\mathbf{1}}
\newcommand{\Prb}{\operatorname{Pr}}
\newcommand{\Exp}{\mathbb{E}}

\title{Three-Block Bounds for the Asymptotic Rank of Weighted Graph Tensors}
\author{\fontsize{11}{16}\selectfont
    Zhengyang Wang\institution{State Key Laboratory of Mathematical Sciences, Academy of Mathematics and Systems Science, Chinese Academy of Sciences; University of Chinese Academy of Sciences}\email{wangzhengyang22@mails.ucas.ac.cn}
    \and Yichuan Cao\instref{1}\email{caoyichuan@amss.ac.cn}
    \and Hongbo Li\instref{1}\email{hli@mmrc.iss.ac.cn}
    \and Xiao-Shan Gao\instref{1}\email{xgao@mmrc.iss.ac.cn}
}

\begin{document}

\maketitle

\begin{abstract}
A weighted graph $(G,w)$ defines a multilinear contraction tensor over a fixed field $F$ with one factor per vertex and edge dimensions growing as $2^{\lfloor t w_e\rfloor}$. Grouping vertices into at most three blocks restricts it to rectangular matrix multiplication (exponent $U$), giving a lower bound $L_3(G,w)$ on the asymptotic rank exponent $\Gamma_F(G,w)$ that refines the maximum-cut bound. We investigate when $\Gamma_F(G,w)=L_3(G,w)$ and how tightly $L_3$ controls $\Gamma_F$ on near-bipartite graphs.
Over every field $F$ and all nonnegative real weights $w$, we prove $\Gamma_F(G,w)=L_3(G,w)$ whenever at most five vertices are incident to positive-weight edges, or $G$ is a cycle; on an odd cycle $C_m$, an optimal partition isolates one vertex and alternates two blocks along the remaining path. For arbitrary $(G,w)$, we characterize $L_3(G,w)$ as the maximum weighted edge sum over sums of at most two tree metrics obeying Zhou's quadratic mass inequality, and show this fails for three trees. When the positive support becomes bipartite after deleting at most two edges, we prove $L_3\le\Gamma_F\le (100/93)L_3$ ($19/18$ if it is $K_4$-free), establish exactness on the cone generated by triangle edge indicators, and classify $\Gamma_F$ for positive integer weights made bipartite by deleting a unit-weight edge; general exactness on this class remains open. Finally, we record the exact $m^2/4$ law for unit-weight complete graphs $K_m$, determine the first-order asymptotic correction under a dominant edge on any graph, and contrast separated tensor rank with division-free arithmetic circuits reusing intermediate products, showing that every weighted matrix cycle $C_m$ has circuit exponent $\max_i U(\min_j w_j,w_i,w_{i+1})$ ($9/4$ at unit weights).
\end{abstract}

\noindent\textbf{Keywords:} Graph tensor, Asymptotic tensor rank, Matrix multiplication exponent, Asymptotic spectrum, Ptolemaic metric, Tree metric, Arithmetic circuit

\section{Introduction}
\label{sec:intro}

\subsection{Motivation and background}

Matrix multiplication is the simplest nontrivial instance of tensor network contraction: multiplying two order-two tensors $A\in F^{a\times b}$ and $B\in F^{b\times c}$ contracts a shared index of dimension $b$, and pairing the output with a third matrix $C\in F^{c\times a}$ produces the trilinear trace form $\tr(ABC)$. Its coefficient tensor $M_{a,b,c}$ is a three-factor tensor whose shared indices form a triangle $K_3$, and the minimal number of scalar multiplications in a bilinear algorithm for $AB$ coincides with the exact tensor rank of $M_{a,b,c}$. Replacing the triangle by an arbitrary finite graph $G=(V,E)$ equipped with edge weights $w_e\ge 0$ leads directly to \emph{weighted graph tensors}: each vertex $v\in V$ holds an arbitrary input tensor of order $\deg_G(v)$, each edge $e\in E$ carries an independently summed index of dimension $N_e(t)=2^{\lfloor t w_e\rfloor}$, and the resulting closed contraction has a coefficient tensor $T_F(G,w;t)$ of order $|V|$ over a fixed field $F$. Determining the asymptotic rank exponent
\[
\Gamma_F(G,w)=\lim_{t\to\infty}\frac{\log_2 \rk_F\bigl(T_F(G,w;t)\bigr)}{t}
\]
generalizes the matrix multiplication exponent to arbitrary pairwise tensor contractions.

For unweighted graph tensors over $\mathbb{C}$, Christandl and Zuiddam~\cite[Problem~1.2]{surgery} asked for the asymptotic rank exponent of general graph tensors, and Christandl, Vrana, and Zuiddam~\cite[Question~1.57]{graph} asked whether the maximum cut of $G$---obtained by grouping the vertex factors into two blocks and taking matrix rank---always determines this exponent. Recent developments have fundamentally changed the picture around both questions. First, OpenAI established that the square matrix multiplication exponent over every field is $\omega=9/4$~\cite[Theorem~1.1]{square} (see also Pass~\cite[Theorem~1.1]{pass} for a constructive upper bound for integer matrices) and determined the exact rectangular matrix multiplication exponent $\omega_F(x,y,z)=U(x,y,z)$ for all nonnegative dimension exponents $x,y,z$~\cite{rectangle}, matching the earlier upper curves proved by Uzna\'nski~\cite[Theorem~1.1]{uzn} in characteristic zero and by Amanov, van den Berg, and Zuiddam~\cite[Theorem~1.2]{abz} over arbitrary fields. Already on the triangle $K_3$, the value $\Gamma_F(K_3)=9/4$ strictly exceeds the maximum cut $2$, showing that two-block flattenings do not suffice in general. Second, building on Strassen's theory of the asymptotic spectrum of tensors~\cite{strassen}, Zhou~\cite{zhou} expressed $\Gamma_F(G,w)$ as the support function of spectral edge-length tables subject to a quadratic mass inequality, while OpenAI~\cite{geometry} proved that every spectral pseudometric embeds isometrically into a complete geodesic $\mathrm{CAT}(0)$ space and therefore obeys the four-point Ptolemy inequality of Foertsch, Lytchak, and Schroeder~\cite[Theorem~1.3]{ptolemy}.

Partitioning the vertices of $(G,w)$ into at most three blocks---fixing internal edge indices and combining crossing indices between each pair of blocks---restricts $T_F(G,w;t)$ to a rectangular matrix multiplication tensor. Maximizing $U(x,y,z)$ over all maps $c\colon V\to\{1,2,3\}$ produces the \emph{three-block lower bound} $L_3(G,w)$, which contains all two-block cuts as degenerate cases and already captures the excess on the triangle. Let $\BP(G,w)$ denote the geometric upper bound obtained by maximizing $\sum_{i<j} w_{ij}d_{ij}$ over all pseudometrics $d$ on $V$ satisfying both Ptolemy's inequality and Zhou's quadratic mass inequality. Every weighted graph then obeys the sandwich
\begin{equation}\label{eq:intro-sandwich}
\MC(G,w)\le L_3(G,w)\le \Gamma_F(G,w)\le \BP(G,w).
\end{equation}
This brings us to the central question of the present paper: \emph{when does the three-block rectangular lower bound $L_3(G,w)$ coincide with the asymptotic rank exponent $\Gamma_F(G,w)$, and how closely does it control $\Gamma_F(G,w)$ on graphs close to bipartite?}

\subsection{Overview of main results}

Our results organize naturally into four themes, progressing from general metric characterizations and small-support exactness to arbitrary cycles, near-bipartite graphs, and broader asymptotic and circuit-complexity comparisons.

\paragraph{1. Tree-metric characterization and exactness on at most five active vertices.}
On an arbitrary weighted graph $(G,w)$, we first characterize $L_3(G,w)$ purely in terms of metric geometry (Theorem~\ref{thm:trees}): among all pseudometrics $d$ that are sums of at most two tree metrics and obey Zhou's quadratic mass inequality, the maximum of $\sum_{i<j} w_{ij}d_{ij}$ equals $L_3(G,w)$, attained over every field by a three-leaf tripod spectral table. In particular, this yields exactness $\Gamma_F=\BP=L_3$ on every graph with at most four active vertices and on every graph admitting a vertex cover of size at most two (Corollary~\ref{cor:four}). The threshold of two trees is sharp: a five-point cycle split metric decomposed into three trees attains $25/6$ on the diagonals of $C_5$ (recovering Zhou's conic relaxation value~\cite[Proposition~5.13]{zhou}), strictly exceeding the true exponent $49/12$ and violating Ptolemy's inequality.

Combining the Ptolemy geometry with a second-order variational analysis at fixed stationary mass, we extend exactness to every graph having at most five vertices incident to positive-weight edges, with no structural assumption on the graph or edge weights (Theorem~\ref{thm:five}):
\[
\Gamma_F(G,w)=\BP(G,w)=L_3(G,w).
\]
In particular, all ten edge weights on $K_5$ may vary independently. Furthermore, whenever a weighted graph partitions into independent sets admitting correlated random representatives whose pair marginals match the normalized inter-block weights, all three quantities $\Gamma_F$, $\BP$, and $L_3$ coincide with those of the weighted quotient (Proposition~\ref{prop:rep}), lifting small-graph equalities to graphs of unbounded size.

\paragraph{2. Odd Ptolemy polygons and arbitrary weighted cycles.}
To handle cycles of arbitrary length and general odd-cycle packings, we prove an unconditional geometric inequality for odd polygons in any Ptolemy pseudometric space (Theorem~\ref{thm:polygon}): for $2k+1$ cyclically ordered points of diameter at most $D$, the sum $S$ of all pairwise distances and the perimeter $Y$ satisfy $S-kY+k(k-1)D\ge 0$, together with an exterior-diameter row bound (Lemma~\ref{lem:exterior}). Using these inequalities, we show that for every odd cycle $C_m$ and any Ptolemy--Zhou pseudometric of global diameter $D$, a single distribution over three-block spectral tables simultaneously dominates every cycle edge and attains expected diameter at least $D$ (Theorem~\ref{thm:cyclelaw}). Consequently, over every fixed field and for all nonnegative real weights $w_0,w_1,\dots,w_{m-1}$ with total $W$,
\[
\BP(C_m,w)=\Gamma_F(C_m,w)=L_3(C_m,w)=\max_{0\le i<m} U\bigl(W-w_{i-1}-w_i,\,w_{i-1},\,w_i\bigr),
\]
with subscripts modulo $m$, while every even cycle has exponent $W$. Every optimal three-block partition on an odd cycle isolates one vertex and alternates the other two blocks along the complementary path.

At unit weights, this yields $\Gamma_F(C_m)=m-1+\frac{1}{4(m-2)}$ for every odd $m\ge 3$ (Corollary~\ref{cor:main}), recovering $\Gamma_F(C_3)=9/4$~\cite{square}, sharpening Zhou's upper bound $m-1+\frac{1}{m+1}$ for $m\ge 5$~\cite[Theorem~6.6]{zhou}, and classifying all extremal spectral pseudometrics together with the exact one-sided directional derivatives of $\Gamma_F$ at the all-ones vector (Proposition~\ref{prop:rigid}). As an elementary algebraic consequence, whenever $m\ge 7$, $a=\min_i w_i>0$, and $w_{i-1}+w_i\ge 9a/4$ for all $i$, the exponent collapses to the maximum cut $W-a$ (Corollary~\ref{cor:chamber}). The simultaneous diameter domination also yields explicit bounds and exactness criteria for odd cycles with bipartite attachments and layered graphs over odd cycles (Corollaries~\ref{cor:attachments} and~\ref{cor:cyclelayers}).

\paragraph{3. Near-bipartite graphs: packings, uniform comparison, and exact subclasses.}
Consider next the class of graphs whose positive-weight support becomes bipartite after deleting at most two edges---a natural hierarchy encompassing all odd cycles (one deletion), books, theta graphs, and $K_4$ (two deletions). Combining fractional odd-cycle packings controlled by a single global diameter (Theorem~\ref{thm:pack}) with Hu's two-commodity flow theorem~\cite{hu} (Proposition~\ref{prop:hu}), we establish the following results across this class:
\begin{itemize}
\item \emph{Uniform comparison:} For every nonnegative weighting and every field $F$,
\[
L_3(G,w)\le \Gamma_F(G,w)\le \frac{100}{93}\,L_3(G,w),
\]
with the upper factor tightening to $19/18$ whenever the positive support contains no $K_4$ (Theorem~\ref{thm:globalratio}).
\item \emph{Exactness on the triangle cone:} If $w$ is any nonnegative linear combination of edge indicators of triangles on a graph in the two-deletion class, then $\Gamma_F(G,w)=L_3(G,w)$, with no bound on the number of vertices or triangles (Theorem~\ref{thm:trianglecone}).
\item \emph{Single-deletion classification and cut-deficit bounds:} For a connected nonbipartite support with positive integer weights made bipartite by deleting one edge of weight $1$, $\Gamma_F=L_3$ equals $W-1+\frac{1}{4(W-2)}$ if and only if some vertex of weighted degree $2$ meets every odd cycle, and equals $W-1$ otherwise (Theorem~\ref{thm:unit}). At cut deficit $W-M=2$ in integer weights, $\Gamma_F\le W-2+\frac{1}{W-4}$ (Theorem~\ref{thm:two}).
\item \emph{Explicit families and two-distance tables:} We obtain exact $O(k^2)$-computable exponents for arbitrarily weighted $k$-page books, explicit formulas and open exactness chambers for six-vertex theta graphs and odd subdivisions of $K_4$ (Section~\ref{sec:examples}), and exact domination for every Ptolemy--Zhou pseudometric on a two-deletion graph taking at most two distinct positive values across all vertex pairs (Theorem~\ref{thm:twoprices}).
\end{itemize}
Whether $\Gamma_F(G,w)=L_3(G,w)$ holds for arbitrary weights throughout the entire two-deletion class remains open (Conjecture~\ref{conj:twodel}).

\paragraph{4. Complete graphs, structural boundaries, dominant edges, and arithmetic circuits.}
As an immediate warm-up of the three-block restriction, we record the exact law $\Gamma_F(K_m)=m^2/4$ and finite-rank bound $r^4\ge n^{m^2}$ for uniform complete graphs (Proposition~\ref{prop:clique}); the upper bound and even-order equality are due to Khoruzhii~\cite{khor} and Zhou~\cite{zhou}, while the matching odd-order lower bound was obtained independently here and by Pokutta~\cite{pokutta}.

Beyond five vertices, we compare $\Gamma_F$ and $\BP$ with the Goemans--Williamson semidefinite cut bound $S(w)$~\cite{gw}, showing that on the two-deletion class $\BP\le S(w)$ and that equality $\BP=S(w)$ forces both to coincide with $L_3(w)$ (Section~\ref{sec:sdp}). We also identify a geometric boundary of the Ptolemy--Zhou relaxation: on six points in an $\ell_2$ product of metric trees, a graph requiring three edge deletions achieves $\BP-L_3=27/1846>0$, whereas under uniform stationary mass every six-point Ptolemy pseudometric lies in the convex hull of $10$ balanced cuts and $15$ equilateral three-block tables (Appendix~\ref{app:six}).

Finally, we address two general questions valid on arbitrary graphs. First, when the weight $q$ of a single edge $ab$ (or a bipartite set of heavy edges) tends to infinity on any fixed graph, we determine the exact first-order asymptotic excess above the anchored maximum cut $q+M$:
\[
\lim_{q\to\infty} q\bigl(\Gamma_F(q)-q-M\bigr)=\frac{H^2}{4},
\]
uniformly in $F$, where $H$ is half the largest residual weighted degree of vertices that can switch sides freely and independently in optimal anchored cuts; exact equality $\Gamma_F(q)=q+M$ holds for all large $q$ if and only if $H=0$ (Theorem~\ref{thm:dominant}). Second, to clarify the computational meaning of separated tensor rank relative to general tensor contraction algorithms, we determine the exact operation exponent $\tau_F(G,w)$ of division-free arithmetic circuits that may store and reuse intermediate products (Section~\ref{sec:circuits}). For every weighted matrix cycle $C_m$, $\tau_F(C_m,w)=\max_i U(\min_j w_j,w_i,w_{i+1})$, yielding $\tau_F(C_m)=9/4$ at unit weights regardless of $m$ (Proposition~\ref{prop:circuit})---in sharp contrast to $\Gamma_F(C_5)=49/12$. Together with a block--cut decomposition at articulation vertices (Proposition~\ref{prop:block}), this determines $\tau_F$ on all trees and cactus graphs, while for the unit tetrahedron $K_4$ we establish $27/8\le \tau_F(K_4)\le 4-23/120000$ via Fischer's swap identity~\cite{fischer} (Proposition~\ref{prop:k4circuit}).

\paragraph{Organization of the paper.}
Section~\ref{sec:prelim} builds the bridge from matrix multiplication to general graph tensors, establishes the three-block lower bound $L_3$ (with the $K_m$ application), and introduces Strassen's spectral representation and the Ptolemy--Zhou relaxation $\BP$. Section~\ref{sec:trees} proves the two-tree metric characterization and the representative-lifting principle. Section~\ref{sec:five} proves exactness on at most five active vertices. Section~\ref{sec:cycles} establishes the odd Ptolemy polygon and exterior-diameter inequalities, deduces exactness for all weighted cycles, and derives uniform rigidity, cut chambers, and extensions to attached and layered graphs. Section~\ref{sec:deletions} treats fractional odd-cycle packings, single-deletion classifications, the $100/93$ uniform bound, triangle-indicator combinations, explicit weighted families, and two-distance tables. Section~\ref{sec:beyond} presents the semidefinite cut comparison, inertia and anchor bounds, and the dominant-edge asymptotic expansion. Section~\ref{sec:circuits} analyzes division-free arithmetic circuits. Section~\ref{sec:open} formulates the two-deletion conjecture, minimal-counterexample reductions, and the six-point relaxation obstruction. Appendices~\ref{app:local}--\ref{app:lean} supply self-contained proofs of a local integer-mass cycle certificate, the six-position two-book inequality used in Theorem~\ref{thm:trianglecone}, the six-point uniform-row representation, and the Lean~4 formalization details for uniform odd cycles.

\subsection{Lean formalization}
\label{sec:lean}

The accompanying Lean~4 formalization covers the uniform odd-cycle theorem ($C_m$, Corollary~\ref{cor:main}), together with its internal tensor-rank, spectral, and geometric derivations over an arbitrary field $F$ and every odd length $m\ge 3$. The source code, build instructions, and theorem correspondence are hosted at \url{https://github.com/MechMath/tensor-rank}, which serves as the repository for the ongoing Lean~4 formalization of the paper and currently contains the complete verified proof for uniform odd cycles $C_m$. The development uses Lean~4.35.0 and \texttt{mathlib}~\cite{mathlib2020}, building on the upstream Lean formalizations of OpenAI's $\mathrm{CAT}(0)$ spectral geometry and rectangular matrix multiplication exponent~\cite{geometry,rectangle} and Khoruzhii's tensor-rank library~\cite{khor}.

Specifically, for uniform odd cycles $C_m$, the formalization defines the exact $m$-factor graph tensor $T_F(C_m,n)$, its exact tensor rank $r_F(m,n)$, and its dyadic asymptotic rank exponent $\Gamma_F(C_m)$ directly in the original $m$-factor coordinate model, and proves without additional mathematical hypotheses: (i)~the three-block $(m-2,1,1)$ grouping reduction and finite-rank lower bound $r_F(m,n)^{4(m-2)}\ge n^{(2m-3)^2}$ for every integer $n\ge 2$; (ii)~the spectral-to-metric bridge establishing Ptolemy's inequality and Zhou's quadratic mass inequality for maximizing spectral pseudometrics; (iii)~the sharp odd-cycle perimeter inequality $4(m-2)\sum_{i=0}^{m-1} d(i,i+1)\le (2m-3)^2$; and (iv)~the exact exponent equality $\Gamma_F(C_m)=(2m-3)^2/[4(m-2)]$, the dyadic rate limit, and the explicit integer decomposition witnesses of Corollary~\ref{cor:main}(i)--(iii). Only the uniform odd-cycle result $C_m$ is claimed as formalized in Lean; all other results of this paper---including arbitrarily weighted cycles, complete graphs, two-tree metrics, five active vertices, near-bipartite bounds, dominant-edge asymptotics, and arithmetic circuits---are established by the written proofs presented here.

All proofs in the uniform odd-cycle Lean development are free of project-specific axioms and \texttt{sorry} placeholders; the audited dependency closure of the main declarations, checked both by Lean and by the independent kernel checker \texttt{leanchecker}, uses only Lean's standard axioms \texttt{propext}, \texttt{Classical.choice}, and \texttt{Quot.sound}. Appendix~\ref{app:lean} records the principal modules, declaration correspondence, axiom boundary, and reproducibility instructions.

\section{Preliminaries: Graph Tensors, Three-Block Bounds, and Spectral Geometry}
\label{sec:prelim}

To make the exposition self-contained across combinatorics, algebraic complexity, and tensor geometry, we first derive the graph-tensor model directly from matrix multiplication, then establish the three-block lower bound $L_3$ and the spectral upper-bound framework.

\subsection{From matrix multiplication to general tensor network contractions}
\label{sec:dictionary}

Fix a commutative field $F$. Let $A\in F^{a\times b}$ and $B\in F^{b\times c}$ be matrices, regarded as order-two tensors. Matrix multiplication contracts the shared index $j\in [b]=\{1,\dots,b\}$ while leaving $i\in [a]$ and $k\in [c]$ free:
\[
(AB)_{ik}=\sum_{j=1}^{b} A_{ij}B_{jk}.
\]
To treat the inputs and output symmetrically, pair the product $AB\in F^{a\times c}$ with an arbitrary test matrix $C\in F^{c\times a}$ via the trace form
\begin{equation}\label{eq:pretrace}
\langle AB,C\rangle=\tr(ABC)=\sum_{i=1}^{a}\sum_{j=1}^{b}\sum_{k=1}^{c} A_{ij}B_{jk}C_{ki}.
\end{equation}
Evaluating~\eqref{eq:pretrace} on the matrix unit $C=E_{ki}$ recovers the entry $(AB)_{ik}$ over any field $F$, in every characteristic; hence closing the output in this way preserves the entire bilinear map $(A,B)\mapsto AB$. Writing $a_{ij}^*$, $b_{jk}^*$, and $c_{ki}^*$ for the coordinate functionals on $F^{a\times b}$, $F^{b\times c}$, and $F^{c\times a}$, the coefficient tensor of~\eqref{eq:pretrace} is the matrix multiplication tensor
\[
M_{a,b,c}=\sum_{i=1}^{a}\sum_{j=1}^{b}\sum_{k=1}^{c} a_{ij}^*\otimes b_{jk}^*\otimes c_{ki}^*.
\]
Note the distinction between input order and coefficient-tensor order: each input $A,B,C$ is an order-two tensor (a matrix), whereas $M_{a,b,c}$ has order three because it occupies the tensor product of the three dual input spaces.

The \emph{exact tensor rank} (or exact CP rank) $\rk_F(M_{a,b,c})$ is the least integer $r\ge 0$ for which~\eqref{eq:pretrace} admits a separated expansion
\begin{equation}\label{eq:prebilinear}
\tr(ABC)=\sum_{s=1}^{r}\alpha_s(A)\,\beta_s(B)\,\gamma_s(C)
\end{equation}
with linear functionals $\alpha_s\in (F^{a\times b})^*$, $\beta_s\in (F^{b\times c})^*$, and $\gamma_s\in (F^{c\times a})^*$. By nondegeneracy of the trace pairing, each $\gamma_s$ corresponds to a unique matrix $D_s\in F^{a\times c}$ with $\gamma_s(C)=\tr(D_sC)$. Testing~\eqref{eq:prebilinear} on all matrix units shows that it is equivalent to
\[
AB=\sum_{s=1}^{r}\alpha_s(A)\,\beta_s(B)\,D_s.
\]
Thus a rank-$r$ decomposition of $M_{a,b,c}$ is precisely a bilinear algorithm that computes every entry of $AB$ as a linear combination of the same $r$ scalar products of input linear forms (see Landsberg~\cite[Section~1.2]{landsberg}).

\begin{figure}[htbp]
\centering
\begin{tikzpicture}[x=1cm,y=1cm,site/.style={circle,draw,fill=black!4,inner sep=2pt,minimum size=5.7mm},cedge/.style={line width=.65pt}]
 \node[site] (a) at (-0.9,0) {$A$}; \node[site] (b) at (0.9,0) {$B$}; \node[site] (c) at (0,1.20) {$C$};
 \draw[cedge] (a)--node[below] {\scriptsize $j\in[b]$} (b);
 \draw[cedge] (b)--node[right] {\scriptsize $k\in[c]$} (c);
 \draw[cedge] (c)--node[left] {\scriptsize $i\in[a]$} (a);
 \node at (0,-.73) {\small (a) Three matrix inputs ($K_3$)};
 \begin{scope}[xshift=5.8cm]
 \node[site] (p) at (-.85,0) {$1$}; \node[site] (q) at (.85,0) {$2$}; \node[site] (r) at (0,1.45) {$3$}; \node[site] (s) at (0,.47) {$4$};
 \draw[cedge] (p)--(q)--(r)--(p); \draw[cedge] (s)--(p) (s)--(q) (s)--(r);
 \node at (0,-.73) {\small (b) Four order-three inputs ($K_4$)};
 \end{scope}
\end{tikzpicture}
\caption{Graph representation of closed tensor contractions. Each vertex carries an input tensor whose order equals its degree, and each edge carries an independently summed index. (a)~The triangle $K_3$ represents the matrix multiplication tensor $M_{a,b,c}$. (b)~The complete graph $K_4$ represents the contraction of four order-three tensors along six shared indices, yielding a four-factor coefficient tensor.}
\label{fig:dictionary}
\end{figure}
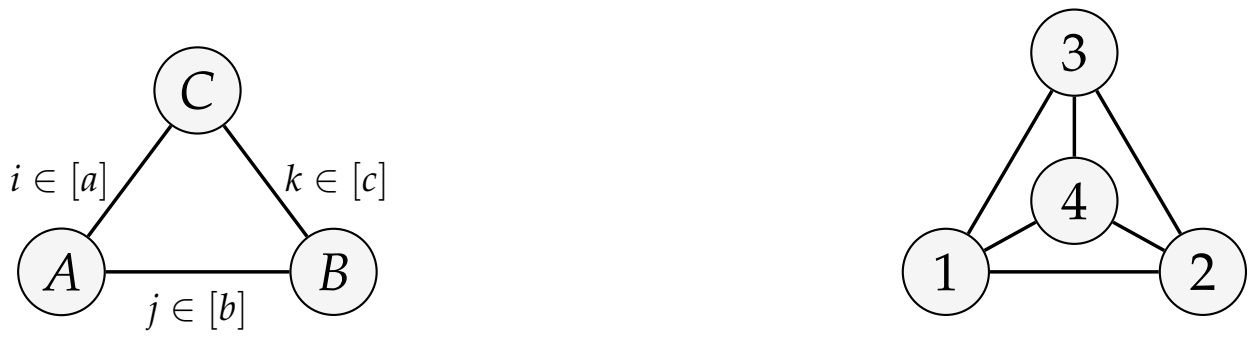

In~\eqref{eq:pretrace}, the pairs $(A,B)$, $(B,C)$, and $(C,A)$ share the summed indices $j\in [b]$, $k\in [c]$, and $i\in [a]$, respectively. Placing a vertex at each input space and joining the two occurrences of each contracted index by an edge yields a triangle $K_3$ with edge dimensions $b,c,a$. This correspondence extends immediately to arbitrary pairwise tensor contractions, following the graph-tensor convention of~\cite{surgery,graph}; see Figure~\ref{fig:dictionary}.

Formally, let $G=(V,E)$ be a finite loopless graph and assign to each edge $e\in E$ a positive integer dimension $N_e$. Each vertex $v\in V$ carries the local tensor space $H_v=\bigotimes_{e\ni v} F^{N_e}$ (in a fixed incident-edge order), so an input array $X_v\in H_v$ has order $\deg_G(v)$. The closed contraction of $(X_v)_{v\in V}$ along $G$ is the multilinear form
\begin{equation}\label{eq:prenetwork}
\mathcal{C}_G\bigl((X_v)_{v\in V}\bigr)
=\sum_{(i_e)\in\prod_{e\in E}[N_e]}\ \prod_{v\in V} X_v\bigl((i_e)_{e\ni v}\bigr).
\end{equation}
Its coefficient tensor in $\bigotimes_{v\in V} H_v^*$, expressed in the dual standard basis, is the \emph{graph tensor}
\begin{equation}\label{eq:graphtensor}
T_F\bigl(G,(N_e)\bigr)=\sum_{(i_e)\in\prod_{e\in E}[N_e]}\ \bigotimes_{v\in V} e^{(v)}_{(i_e:\,e\ni v)},
\end{equation}
where $e^{(v)}_{(i_e:\,e\ni v)}$ is the standard basis vector of $H_v$ indexed by the incident edge coordinates. Since equality of multilinear forms on basis inputs determines every coefficient, the exact tensor rank of $T_F(G,(N_e))$ over $F$ satisfies
\[
\rk_F\bigl(T_F(G,(N_e))\bigr)
=\min\left\{r\in\mathbb{Z}_{\ge 0}:\ \mathcal{C}_G\bigl((X_v)_{v\in V}\bigr)
=\sum_{s=1}^{r}\prod_{v\in V}\ell_{s,v}(X_v)\ \text{for all } X_v\in H_v,\ \ell_{s,v}\in H_v^*\right\}.
\]
In this dictionary, $\deg_G(v)$ records the order of the individual input array $X_v$, whereas $|V|$ counts the number of factor spaces (the order) of the coefficient tensor $T_F(G,(N_e))$. Suppressing a dimension-$1$ edge or adjoining an isolated vertex (a scalar factor) leaves $\rk_F$ unchanged, and parallel edges between the same two vertices combine into a single edge of product dimension; we may therefore restrict attention to finite simple graphs without loss of generality. Open contractions with free output indices fit into the same framework by pairing the output with one test tensor carrying all free indices (for instance, leaving one vertex of $K_m$ open contracts $m-1$ tensors of order $m-1$ to one tensor of order $m-1$). Likewise, a repeated index shared by three or more arrays can be encoded via a fixed equality tensor, which acts as a restriction of a larger pairwise network and may have strictly smaller rank.

To study asymptotic growth with inhomogeneous edge dimensions, assign each edge $e\in E$ a nonnegative real weight $w_e\ge 0$ (putting $w_{ij}=0$ on nonedges) and set $N_e(t)=2^{\lfloor tw_e\rfloor}$ for $t>0$. Write $T_F(G,w;t)=T_F(G,(N_e(t)))$. For integer weights, the vertexwise tensor product $T_F(G,w;t_1)\boxtimes T_F(G,w;t_2)$ is isomorphic to $T_F(G,w;t_1+t_2)$ by concatenating binary index strings along each edge. Submultiplicativity of tensor rank and Fekete's lemma therefore guarantee the existence of the \emph{asymptotic rank exponent}
\begin{equation}\label{eq:limit}
\Gamma_F(G,w)=\lim_{t\to\infty}\frac{\log_2 \rk_F(T_F(G,w;t))}{t}
=\inf_{t\ge 1}\frac{\log_2 \rk_F(T_F(G,w;t))}{t},
\end{equation}
where the infimum identity holds for integer weights. For arbitrary real weights $w\ge 0$, rounding $\lfloor tw_e\rfloor$ changes each edge dimension by at most a factor of $2$, which alters $\log_2 \rk_F$ by at most $|E|$ (by local coordinate restriction in one direction and tensoring with single binary edges in the other). Rational approximation and the same estimate show that the limit in~\eqref{eq:limit} exists for all $w\in\mathbb{R}_{\ge 0}^E$ and defines a positively homogeneous, $1$-Lipschitz function in the $\ell_1$ norm on weights. When all edges carry unit weight we write $\Gamma_F(G)$ and $T_F(G,n)$ for uniform edge dimension $n$, and denote the asymptotic rank by $\ark_F(T)=\lim_{k\to\infty}\rk_F(T^{\boxtimes k})^{1/k}$.

Note carefully the distinction between $\Gamma_F(G,w)$ and the minimum arithmetic cost of evaluating the scalar polynomial $\mathcal{C}_G$. The rank $\rk_F$ counts completely separated rank-one products in which each linear form $\ell_{s,v}$ reads only $X_v$, forbidding the storage and reuse of intermediate contractions across gates. We examine division-free arithmetic circuits with reuse explicitly in Section~\ref{sec:circuits}.

\subsection{Three-block restrictions and the complete-graph warm-up}
\label{sec:threeblock}

For any three nonnegative reals $x,y,z$ sorted in nonincreasing order $x\ge y\ge z\ge 0$, define
\begin{equation}\label{eq:rectangle}
U(x,y,z)=\begin{cases}
x+y+\dfrac{\pos{(x+y)z-xy}^2}{4xyz}, & z>0,\\[6pt]
x+y, & z=0,
\end{cases}
\end{equation}
where $\pos{u}=\max\{u,0\}$, and extend $U$ symmetrically to all triples in $\mathbb{R}_{\ge 0}^3$. By the exact rectangular matrix multiplication exponent theorem of OpenAI~\cite{rectangle} (matching the upper curves proved earlier by Uzna\'nski~\cite[Theorem~1.1]{uzn} in characteristic zero and by Amanov, van den Berg, and Zuiddam~\cite[Theorem~1.2]{abz} over all fields, and recovered from a conic program by Zhou~\cite[Proposition~6.2]{zhou}), $U(x,y,z)$ is the exact asymptotic rank exponent $\omega_F(x,y,z)$ of the rectangular matrix multiplication tensor with dimension exponents $(x,y,z)$ over every field $F$; moreover, for integer exponents and any base $n\ge 2$, any exact rank-$r$ decomposition of $M_{n^x,n^y,n^z}$ over $F$ satisfies $\log_n r\ge U(x,y,z)$~\cite{rectangle}. Whenever $(x+y)z>xy$ (the \emph{strictly curved branch}), we may also write
\[
U(x,y,z)=\frac{(xy+xz+yz)^2}{4xyz}.
\]

By a \emph{three-block partition} of $G=(V,E)$ we mean any map $c\colon V\to\{1,2,3\}$; empty blocks and monochromatic edges (adjacent vertices placed in the same block) are permitted. We reserve the term \emph{proper $3$-coloring} strictly for a three-block partition with no monochromatic positive-weight edges. Given a three-block partition $c$, let
\[
W_{ab}(c)=\sum_{\substack{uv\in E\\ \{c(u),c(v)\}=\{a,b\}}} w_{uv} \qquad (1\le a<b\le 3)
\]
denote the three aggregate weights crossing between distinct blocks, and define the \emph{three-block bound}
\begin{equation}\label{eq:L3def}
L_3(G,w)=\max_{c\colon V\to\{1,2,3\}} U\bigl(W_{12}(c),W_{13}(c),W_{23}(c)\bigr).
\end{equation}
When one block is empty, $U(W_{12},0,0)=W_{12}$ is the weight of a two-block cut; hence $L_3(G,w)\ge M(G,w):=\MC(G,w)$.

\begin{lemma}[Three-block restriction]\label{lem:group}
For every finite simple graph $G=(V,E)$, every weight vector $w\in\mathbb{R}_{\ge 0}^E$, and every field $F$,
\[
\MC(G,w)\le L_3(G,w)\le \Gamma_F(G,w).
\]
Furthermore, for integer weights $w$ and every integer $n\ge 2$, any exact rank-$r$ decomposition of $T_F(G,(n^{w_e}))$ over $F$ satisfies $\log_n r\ge L_3(G,w)$.
\end{lemma}

\begin{proof}
Fix a three-block partition $c\colon V\to\{1,2,3\}$. For integer weights $w$ and any $n\ge 2$, fix every monochromatic edge coordinate to a single basis label, group the tensor factor spaces belonging to each block into one composite factor space, and concatenate the crossing indices between each pair of distinct blocks. Under these local projections and factor groupings, each rank-one product $\bigotimes_{v\in V} u_v$ maps to a single rank-one tensor (or zero), while $T_F(G,(n^{w_e}))$ specializes to the rectangular matrix multiplication tensor of dimensions $(n^{W_{12}(c)},n^{W_{13}(c)},n^{W_{23}(c)})$. By the finite-rank rectangular lower bound~\cite{rectangle}, every rank-$r$ decomposition satisfies $\log_n r\ge U(W_{12}(c),W_{13}(c),W_{23}(c))$. Maximizing over $c$ proves the finite-rank inequality, and setting $n=2^t$ yields $L_3(G,w)\le \Gamma_F(G,w)$ for integer weights. Positive homogeneity and $\ell_1$-continuity extend the inequality to all nonnegative real weights.
\end{proof}

As an immediate illustration of Lemma~\ref{lem:group}, we record the exact exponent of unit-weight complete graphs $K_m$. While the maximum cut of $K_m$ has size $\lfloor m^2/4\rfloor$, Khoruzhii~\cite[Theorem~3]{khor} and Zhou~\cite[Theorem~4.10]{zhou} independently proved the upper bound $\Gamma_F(K_m)\le m^2/4$ and equality for even $m$. The three-block restriction closes the odd-order gap directly, answering the threshold question of Christandl, Vrana, and Zuiddam~\cite[Question~1.56]{graph} (as also noted in~\cite[Theorem~4.10]{zhou}); the matching odd-order lower bound below was obtained independently of Pokutta~\cite[Theorems~1--3]{pokutta}.\footnote{Our derivation of the odd-order finite-rank bound and exact complete-graph exponent dates to 25 November 2026; Pokutta's independent note~\cite{pokutta} was posted on 27 November 2026.}

\begin{proposition}[Uniform complete graphs]\label{prop:clique}
Let $m\ge 2$ and $n\ge 2$ be integers. Over every field $F$, every exact rank-$r$ decomposition of $T_F(K_m,n)$ satisfies $r^4\ge n^{m^2}$, and consequently
\[
\Gamma_F(K_m)=\frac{m^2}{4}.
\]
\end{proposition}

\begin{proof}
For even $m$, a balanced bipartition already yields a matrix flattening of rank $n^{m^2/4}$. For $m\ge 3$, partition the $m$ vertices into three nonempty blocks of sizes $u\ge v\ge w$ with $u+v+w=m$. Their crossing edge counts are $uv\ge uw\ge vw$, so Lemma~\ref{lem:group} and~\eqref{eq:rectangle} yield
\[
\log_n r\ge U(uv,uw,vw)=u(v+w)+\frac{\pos{v+w-u}^2}{4}.
\]
Whenever $u\le v+w$, the right-hand side equals $(u+v+w)^2/4=m^2/4$. Choosing block sizes differing by at most $1$ (or taking $(k,k,1)$ when $m=2k+1$) satisfies $u\le v+w$ for every $m\ge 3$, proving $r^4\ge n^{m^2}$ and $\Gamma_F(K_m)\ge m^2/4$. Combining this with the upper bound $\Gamma_F(K_m)\le m^2/4$ of Khoruzhii~\cite{khor} and Zhou~\cite{zhou} yields equality.
\end{proof}

For example, $K_5$ has maximum cut $6$, whereas the partition $(2,2,1)$ yields crossing counts $(4,2,2)$ and $\Gamma_F(K_5)=U(4,2,2)=25/4$. Note that the grouped tensor in Proposition~\ref{prop:clique} is the complete tripartite specialization noted by Zhou~\cite[Section~6.4]{zhou}.

\subsection{Asymptotic spectrum and the Ptolemy--Zhou relaxation}
\label{sec:spectrum}

To establish matching upper bounds on $\Gamma_F(G,w)$, we pass to Strassen's asymptotic spectrum of tensors~\cite{strassen} in the graph-tensor formulation of Zhou~\cite{zhou} together with the $\mathrm{CAT}(0)$ spectral geometry of OpenAI~\cite{geometry}. For a finite vertex set $V$ and a fixed field $F$, let $\phi$ be a normalized spectral point on $V$-factor tensors over $F$ (a monotone semiring homomorphism normalized to value $2$ on the $V$-factor binary diagonal tensor). For each unordered pair of distinct vertices $i,j\in V$, let $E_{ij}(2)$ denote the binary edge tensor consisting of a single dimension-$2$ edge between $i$ and $j$ (padded with one-dimensional spaces at $V\setminus\{i,j\}$), and define the \emph{spectral distance}
\[
d_\phi(i,j)=\log_2\phi\bigl(E_{ij}(2)\bigr), \qquad d_\phi(i,i)=0.
\]
We call $d_\phi$ a \emph{realizable} (or \emph{spectral}) \emph{pseudometric} over $F$. For any symmetric nonnegative matrix $d$ on $V$ with zero diagonal and any vector $u\in\mathbb{R}_{\ge 0}^V$, define the quadratic mass energy
\[
M_d(u)=\sum_{\{i,j\}\subset V} u_i u_j\, d(i,j)=\frac{1}{2}u\trans d\, u.
\]

\begin{proposition}[Spectral representation and metric constraints]\label{prop:spectral}
Let $G=(V,E)$ be a finite simple graph with edge weights $w\in\mathbb{R}_{\ge 0}^E$, and let $F$ be a field. Then
\begin{equation}\label{eq:spectrum}
\Gamma_F(G,w)=\max_{\phi}\sum_{\{i,j\}\in E} w_{ij}\, d_\phi(i,j),
\end{equation}
where the maximum is attained over all normalized $V$-factor spectral points $\phi$ over $F$. Every realizable spectral pseudometric $d=d_\phi$ is a symmetric pseudometric on $V$ satisfying the four-point Ptolemy inequality
\begin{equation}\label{eq:ptolemy}
d(a,b)d(c,e)\le d(a,c)d(b,e)+d(a,e)d(b,c) \qquad (a,b,c,e\in V)
\end{equation}
and Zhou's quadratic mass inequality
\begin{equation}\label{eq:mass}
4M_d(u)\le \biggl(\sum_{v\in V} u_v\biggr)^2 \qquad (u\in\mathbb{R}_{\ge 0}^V).
\end{equation}
In particular, $\diam(d):=\max_{i,j\in V} d(i,j)\le 1$.
\end{proposition}

\begin{proof}
For integer weights, $T_F(G,w;t)$ is isomorphic to the vertexwise tensor product over $\{i,j\}\in E$ of the binary edge tensors $E_{ij}(2)^{\boxtimes tw_{ij}}$. By Strassen's spectral duality~\cite{strassen} and Zhou's graph-tensor formulation~\cite[Theorem~2.20, Lemmas~2.25 and~2.28]{zhou}, multiplicativity and compactness of the asymptotic spectrum give~\eqref{eq:spectrum} with attainment; positive homogeneity and $\ell_1$-continuity extend~\eqref{eq:spectrum} to all $w\in\mathbb{R}_{\ge 0}^E$. Every $d_\phi$ is a pseudometric whose zero-distance quotient embeds isometrically into a complete geodesic $\mathrm{CAT}(0)$ spectral space~\cite[Section~2.2, Corollary~3.4, and Theorem~4.3]{geometry}. Since every $\mathrm{CAT}(0)$ space satisfies Ptolemy's inequality (including for repeated points) by Foertsch, Lytchak, and Schroeder~\cite[Theorem~1.3]{ptolemy},~\eqref{eq:ptolemy} follows.

Inequality~\eqref{eq:mass} is Zhou~\cite[Lemma~5.1]{zhou}: for integer masses $u_v$ with $N=\sum_v u_v\ge 2$, grouping the $N$ vertices of $K_N$ into fibers of sizes $u_v$ produces $u_i u_j$ parallel binary edges between fibers $i$ and $j$, so pulling back $\phi$ bounds $M_d(u)$ by $\Gamma_F(K_N)\le N^2/4$; homogeneity and density extend~\eqref{eq:mass} to all $u\in\mathbb{R}_{\ge 0}^V$. Taking $u_i=u_j=1$ and $u_k=0$ elsewhere gives $d(i,j)\le 1$.
\end{proof}

Let $\mathcal{D}_V$ denote the set of all pseudometrics $d$ on $V$ satisfying~\eqref{eq:ptolemy} and~\eqref{eq:mass}; we call elements of $\mathcal{D}_V$ \emph{Ptolemy--Zhou pseudometrics}. Equivalently,~\eqref{eq:mass} states that $2p\trans dp\le 1$ for every probability vector $p$ on $V$. Because $\mathcal{D}_V$ is defined by finitely many closed inequalities inside $[0,1]^{\binom{V}{2}}$, it is compact, and the geometric relaxation
\[
\BP(G,w)=\max_{d\in\mathcal{D}_V}\sum_{\{i,j\}\in E} w_{ij}\, d(i,j)
\]
is attained. Conversely, grouping the factors of $G$ along any three-block partition $c\colon V\to\{1,2,3\}$ defines a homomorphism of tensor semirings, so pulling back an optimal three-factor spectral point $\psi$ yields a realizable $V$-factor spectral pseudometric $d(u,v)=d_\psi(c(u),c(v))$ occupying at most three metric positions and scoring $L_3(G,w)$. Thus $L_3(G,w)$ is the support function of the compact convex hull of pullbacks of realizable three-point spectral pseudometrics, establishing~\eqref{eq:intro-sandwich} for every $(G,w)$ and every field $F$. Note that on any bipartite graph, $d_{ij}\le 1$ on every edge while the bipartition cut achieves weight $W=\sum_{e\in E}w_e$, giving Zhou's bipartite equality $\MC=L_3=\Gamma_F=\BP=W$~\cite[Proposition~6.4]{zhou}.

\section{Two-Tree Metric Characterization and Lifting Principles}
\label{sec:trees}

Before restricting to specific graph classes, we address a structural geometric question valid on an \emph{arbitrary} weighted graph $(G,w)$: which family of pseudometrics obeying Zhou's mass inequality~\eqref{eq:mass} has support function exactly equal to $L_3(G,w)$? We show that sums of at most two tree metrics provide an exact characterization, deduce exactness for four active vertices and vertex covers of size two, and establish a general lifting principle via consistent representatives.

\subsection{A common mass for two tree metrics}

For any subset $S\subseteq V$, let $c(S)=w(\delta S)$ be the total weight of edges crossing between $S$ and $V\setminus S$, and let $\delta_S(i,j)=\mathbf{1}_{\{|S\cap\{i,j\}|=1\}}$ denote the corresponding cut pseudometric. Following the split-decomposition framework of Bandelt and Dress~\cite{bandelt}, a \emph{tree pseudometric} on $V$ is a nonnegative linear combination $\sum_a \ell_a \delta_{S_a}$ of pairwise compatible splits (those induced by the edges of a tree carrying the vertex labels of $V$, allowing zero edge lengths and coincident labels).

\begin{theorem}[Two tree metrics]\label{thm:trees}
Let $(G,w)$ be an arbitrary finite weighted graph with $w\ge 0$. The maximum of $\sum_{i<j} w_{ij}d_{ij}$ over all pseudometrics $d$ expressible as a sum of at most two tree pseudometrics and satisfying Zhou's mass inequality~\eqref{eq:mass} equals $L_3(G,w)$. Over every field $F$, this maximum is attained by a realizable spectral pseudometric supported on a single three-leaf tripod.
\end{theorem}

\begin{proof}
The zero-weight case $L_3(G,w)=0$ is trivial, so assume $L=L_3(G,w)>0$. For any $K\ge L$ and $S\subseteq V$, we have $c(S)\le \MC(G,w)\le L\le K$, so we may set
\[
b_K(S)=\frac{1-\sqrt{1-c(S)/K}}{2}\in\bigl[0,\tfrac{1}{2}\bigr].
\]
Note that $b_K(S)$ is the smaller root $p$ of $4Kp(1-p)=c(S)$; hence for any $p\in[0,1]$, the inequality $c(S)\le 4Kp(1-p)$ is equivalent to $b_K(S)\le p\le 1-b_K(S)$. We proceed in three steps.

\emph{Step 1: A partition budget inequality.}
We claim that every partition $\mathcal{P}$ of $V$ into nonempty blocks satisfies
\begin{equation}\label{eq:partition}
\sum_{S\in\mathcal{P}} b_L(S)\le 1.
\end{equation}
Suppose to the contrary that some partition violates~\eqref{eq:partition}, and select one with the smallest number of blocks $k$. As $K$ increases continuously from $L$ to $\infty$, the sum $\sum_{S\in\mathcal{P}} b_K(S)$ decreases strictly to $0$, so there exists $K>L$ at which $\sum_{i=1}^k p_i=1$, where $p_i=b_K(S_i)$. Because $c(S_i)\le L<K$, every $p_i<\frac{1}{2}$, forcing $k\ge 3$. Let $w'_{ij}$ be the total weight between blocks $S_i$ and $S_j$ ($i\ne j$), and define the symmetric zero-diagonal residual matrix $R$ by $R_{ij}=w'_{ij}-4Kp_i p_j$ for $i\ne j$. Since $c(S_i)=4Kp_i(1-p_i)=4Kp_i\sum_{j\ne i}p_j$, every row of $R$ sums to zero.

If $k\ge 4$, we claim there is a pair $i\ne j$ with $p_i+p_j\le \frac{1}{2}$ and $R_{ij}\le 0$. Indeed, since $\sum p_r=1$, the family $\mathcal{H}$ of unordered pairs $\{i,j\}$ with $p_i+p_j>\frac{1}{2}$ is pairwise intersecting, and therefore lies either inside a star centered at some index $r_0$ or inside a triangle $\{r_1,r_2,r_3\}$. If $R_{ij}>0$ held for every pair outside $\mathcal{H}$, then in the star case every $i\ne r_0$ would have $R_{i,r_0}=-\sum_{j\notin\{i,r_0\}}R_{ij}<0$, contradicting $\sum_{i\ne r_0}R_{i,r_0}=0$; in the triangle case, any $i\notin\{r_1,r_2,r_3\}$ would satisfy $R_{ij}>0$ for all $j\ne i$, contradicting $\sum_{j\ne i}R_{ij}=0$. Merging such a pair $(i,j)$ forms a block $S_i\cup S_j$ of cut weight
\[
c(S_i\cup S_j)=c(S_i)+c(S_j)-2w'_{ij}\ge 4K\bigl(p_i(1-p_i)+p_j(1-p_j)-2p_ip_j\bigr)
=4K(p_i+p_j)(1-p_i-p_j).
\]
As $p_i+p_j\le\frac{1}{2}$, we obtain $b_K(S_i\cup S_j)\ge p_i+p_j$, so the merged $(k-1)$-block partition still has $b_K$-sum at least $1$ and hence $b_L$-sum strictly exceeding $1$, contradicting the minimality of $k$. Thus $k=3$. With three blocks, the three zero-row-sum equations force $R_{ij}=0$ identically, so the crossing weights are $x=4Kp_1p_2$, $y=4Kp_1p_3$, and $z=4Kp_2p_3$ with $p_1+p_2+p_3=1$ and $\max_i p_i<\frac{1}{2}$. Substituting into~\eqref{eq:rectangle} gives $U(x,y,z)=K>L$, contradicting the definition of $L=L_3(G,w)$. This proves~\eqref{eq:partition}. (In fact, $L$ is the smallest $K\ge\MC(G,w)$ satisfying all partition inequalities, since when $L>\MC(G,w)$ the optimal three-block partition lies in the strictly curved branch and its three masses proportional to $xy,xz,yz$ sum to $1$ at $K=L$.)

\emph{Step 2: Simultaneous feasibility for two laminar families.}
Write $b=b_L$. Completing any family of pairwise disjoint subsets to a partition shows via~\eqref{eq:partition} that their $b$-values sum to at most $1$. Given two laminar families $\mathcal{L}_1$ and $\mathcal{L}_2$ of nonempty subsets of $V$, we claim there exists a probability distribution $\nu$ on $V$ with $\nu(S)\ge b(S)$ for all $S\in\mathcal{L}_1\cup\mathcal{L}_2$. Consider the linear program minimizing $\sum_{v\in V}\nu_v$ subject to $\nu_v\ge 0$ and $\sum_{v\in S}\nu_v\ge b(S)$ for all $S\in\mathcal{L}_1\cup\mathcal{L}_2$. Its dual maximizes $\sum_{S} y_S b(S)$ subject to $y_S\ge 0$ and $\sum_{S\ni v} y_S\le 1$ for every $v\in V$. For any dual feasible vector $y$, assign to each $S\in\mathcal{L}_1$ a half-open interval $I_S\subseteq [0,1)$ of length $y_S$ placed immediately after the intervals of all strict ancestors of $S$ in $\mathcal{L}_1$, starting from $0$; symmetrically, assign to each $S\in\mathcal{L}_2$ an interval $J_S\subseteq [0,1)$ of length $y_S$ placed immediately before its strict ancestors, working backward from $1$. At each vertex $v\in V$, the constraint $\sum_{S\ni v} y_S\le 1$ guarantees that the left-packed chain of sets containing $v$ in $\mathcal{L}_1$ and the right-packed chain in $\mathcal{L}_2$ are disjoint up to endpoints. Hence for almost every $t\in [0,1)$, the family $\mathcal{F}_t$ of sets whose interval contains $t$ is pairwise disjoint, giving $\sum_{S\in\mathcal{F}_t} b(S)\le 1$. Integrating over $t\in[0,1)$ yields $\sum_S y_S b(S)\le 1$. By linear programming duality, the primal optimum is at most $1$; padding any leftover mass arbitrarily produces the desired probability distribution $\nu$.

\emph{Step 3: Centroid rooting and quadratic penalty minimization.}
Now let $d=d^{(1)}+d^{(2)}$ be a sum of two tree pseudometrics obeying~\eqref{eq:mass}, and let $\mathcal{A}$ consist of both sides of every split occurring in either tree. Minimize the convex quadratic functional
\[
h(\mu)=\sum_{S\in\mathcal{A}}\pos{b(S)-\mu(S)}^2
\]
over the compact simplex of probability distributions $\mu$ on $V$, and let $\mu^*$ attain the minimum value $h^*$. Root each of the two trees at a mass centroid with respect to $\mu^*$ (so every subtree directed away from the root has $\mu^*$-mass at most $\frac{1}{2}$). Any strictly deficient split $S\in\mathcal{A}$ satisfies $\mu^*(S)<b(S)\le\frac{1}{2}$ and therefore points away from the root of its tree. Thus the deficient sets in the two trees form two laminar families $\mathcal{L}_1,\mathcal{L}_2$, and Step~2 supplies a probability distribution $\nu$ satisfying $\nu(S)\ge b(S)$ on every deficient $S$. Along the segment $\mu_t=(1-t)\mu^*+t\nu$ ($0\le t\le 1$), each deficient term satisfies $\pos{b(S)-\mu_t(S)}\le (1-t)\pos{b(S)-\mu^*(S)}$ while each nondeficient term increases by at most $t$, yielding $h(\mu_t)\le (1-t)^2 h^*+|\mathcal{A}|t^2$. Taking $t>0$ small forces $h^*=0$.

Therefore $\mu^*(S)\ge b(S)$ on both sides of every split $S$ of either tree, whence
\[
c(S)\le 4L\,\mu^*(S)\bigl(1-\mu^*(S)\bigr)
\]
for every split. Multiplying by the nonnegative split length $\ell_S$ and summing over both trees yields
\[
\sum_{i<j} w_{ij}d_{ij}=\sum_{S}\ell_S c(S)\le 4L\sum_S \ell_S\mu^*(S)\bigl(1-\mu^*(S)\bigr)
=4L\,M_d(\mu^*)\le L,
\]
using~\eqref{eq:mass} in the final step. Conversely, every three-point pseudometric embeds isometrically into a three-leaf tripod, so pulling back an optimal three-factor spectral point via Lemma~\ref{lem:group} attains $L_3(G,w)$ over every field $F$.
\end{proof}

\begin{remark}[Sharpness at two trees]\label{rem:threetrees}
The restriction to at most two tree metrics in Theorem~\ref{thm:trees} is best possible. Arrange five points cyclically and assign them $5/12$ times their shortest-path distance on $C_5$. Each adjacent pair defines a split of weight $5/24$, and these five splits partition into three families of pairwise disjoint pairs (three trees). Since each split contributes the concave function $\mu(S)(1-\mu(S))$ to $M_d(\mu)$, cyclic averaging shows that $M_d$ is maximized at uniform mass, where $4M_d=1$. On the five diagonals of $C_5$ (themselves forming a five-cycle), each edge has length $5/6$, giving total score $25/6$. This matches Zhou's conic relaxation value on $C_5$~\cite[Proposition~5.13]{zhou}, exceeds the exact exponent $49/12$ by $1/12$, and violates Ptolemy's inequality on four consecutive vertices (which in units of $5/12$ would require $4\le 3$).
\end{remark}

\begin{corollary}[Four active vertices and vertex covers of size two]\label{cor:four}
For every weighted graph $(G,w)$ with $w\ge 0$ and every field $F$, one has $\Gamma_F(G,w)=\BP(G,w)=L_3(G,w)$ whenever either:
\begin{enumerate}
\item[{\rm (i)}] at most four vertices are incident to positive-weight edges; or
\item[{\rm (ii)}] every positive-weight edge meets a fixed set $\{s,t\}$ of at most two vertices.
\end{enumerate}
\end{corollary}

\begin{proof}
In case~(i), restrict to the at most four active vertices. By Bandelt and Dress~\cite{bandelt}, every four-point pseudometric $d$ is a nonnegative sum of singleton cuts and at most two nontrivial $2+2$ splits: letting $a_S$ denote the sum of the two within-side distances of each $2+2$ split $S$ and subtracting $\frac{1}{2}(\max_T a_T-a_S)\delta_S$ for each of the at most two splits with $a_S<\max_T a_T$ leaves equal opposite-edge sums, hence a star metric (nonnegative arm lengths follow from the triangle inequality via a split of zero coefficient). Pairing the star with one $2+2$ split writes $d$ as a sum of two tree metrics, and Theorem~\ref{thm:trees} applies (compare Zhou's conic four-point bound~\cite[Proposition~5.12]{zhou}).

In case~(ii), fix $d\in\mathcal{D}_V$ and set $a_v=d(s,v)$ and $b_v=d(t,v)$ for $v\in V$. By the triangle inequality, the contracted pseudometric
\[
d'(u,v)=\max\bigl\{|a_u-a_v|,\,|b_u-b_v|\bigr\}
=\frac{|(a_u+b_u)-(a_v+b_v)|}{2}+\frac{|(a_u-b_u)-(a_v-b_v)|}{2}
\]
obeys $d'\le d$ entrywise, agrees with $d$ on every edge incident to $\{s,t\}$, and is the sum of two line (hence tree) metrics. Because entrywise decrease preserves~\eqref{eq:mass}, Theorem~\ref{thm:trees} gives $\sum w_{ij}d_{ij}=\sum w_{ij}d'_{ij}\le L_3(G,w)$.
\end{proof}

\subsection{Lifting equalities by consistent representatives}
\label{sec:lifting}

The next proposition transfers exactness from small weighted quotients to arbitrarily large graphs whenever inter-block edge weights can be coupled as pairwise marginals of a single random selection of representatives.

\begin{proposition}[Consistent representatives]\label{prop:rep}
Let $(G,w)$ be a weighted graph whose positive-weight support is partitioned into independent vertex sets $C_1,\dots,C_k$, and let $\overline{w}_{ij}=\sum_{x\in C_i,\,y\in C_j} w_{xy}$ be the aggregate weight between $C_i$ and $C_j$, defining a weighted quotient graph $(\overline{G},\overline{w})$ on $\{1,\dots,k\}$. Suppose there exists a random tuple $(X_1,\dots,X_k)\in\prod_{i=1}^k C_i$ whose pairwise marginals satisfy
\[
\Prb(X_i=x,\ X_j=y)=\frac{w_{xy}}{\overline{w}_{ij}} \qquad (x\in C_i,\ y\in C_j)
\]
for every pair $(i,j)$ with $\overline{w}_{ij}>0$. Then over every field $F$,
\[
\Gamma_F(G,w)=\Gamma_F(\overline{G},\overline{w}),\qquad
\BP(G,w)=\BP(\overline{G},\overline{w}),\qquad
L_3(G,w)=L_3(\overline{G},\overline{w}).
\]
\end{proposition}

\begin{proof}
For any pseudometric $d$ on $V$, evaluating the quotient objective on the random restriction $d|_{\{X_1,\dots,X_k\}}$ gives
\[
\Exp\biggl[\sum_{i<j} \overline{w}_{ij}\, d(X_i,X_j)\biggr]
=\sum_{i<j}\sum_{x\in C_i,\,y\in C_j} w_{xy}\, d(x,y).
\]
Every realization $(X_1,\dots,X_k)$ inherits~\eqref{eq:ptolemy} and~\eqref{eq:mass}, proving $\BP(G,w)\le\BP(\overline{G},\overline{w})$. If $d=d_\phi$ is a realizable spectral pseudometric over $F$, then for any fixed $(x_1,\dots,x_k)$ the quotient tensor padded with one-dimensional spaces at $V\setminus\{x_1,\dots,x_k\}$ has the same exact rank as the quotient tensor itself; hence spectral duality and rational approximation bound $\sum_{i<j}\overline{w}_{ij} d_\phi(x_i,x_j)$ by $\Gamma_F(\overline{G},\overline{w})$, and taking expectations yields $\Gamma_F(G,w)\le\Gamma_F(\overline{G},\overline{w})$. The same averaging argument on three-block spectral pullbacks yields $L_3(G,w)\le L_3(\overline{G},\overline{w})$.

Conversely, grouping all tensor factors within each class $C_i$ produces the quotient tensor $(\overline{G},\overline{w})$, giving $\Gamma_F(\overline{G},\overline{w})\le \Gamma_F(G,w)$, while pulling back optimal quotient tables and three-block partitions gives $\BP(\overline{G},\overline{w})\le\BP(G,w)$ and $L_3(\overline{G},\overline{w})\le L_3(G,w)$.
\end{proof}

Independent uniform representatives recover the independent blowup of Zhou~\cite[Proposition~6.9]{zhou}. Correlated representatives permit dependent edge marginals---for instance, merging two nonadjacent vertices whose incident weight vectors are proportional. Combined with Theorem~\ref{thm:five} below, Proposition~\ref{prop:rep} produces arbitrarily large exact families from quotients of at most five vertices, with compatibility of the prescribed pair marginals reducing to a finite linear feasibility problem on $\prod_i C_i$.

\section{Exactness on Five Active Vertices}
\label{sec:five}

We now prove that the Ptolemy--Zhou geometric relaxation is exact on every graph having at most five vertices incident to positive-weight edges, with no assumption on the edge weights or deletion distance.

\begin{theorem}[Five active vertices]\label{thm:five}
Let $(G,w)$ be a finite weighted graph with $w\ge 0$, and let $F$ be any fixed field. If at most five vertices of $G$ are incident to positive-weight edges, then
\[
\Gamma_F(G,w)=\BP(G,w)=L_3(G,w).
\]
More generally, on a graph of arbitrary size, if the endpoints of all positive-weight edges occupy at most five distinct metric positions in a single pseudometric $d\in\mathcal{D}_V$, then $\sum_{i<j} w_{ij}d_{ij}\le L_3(G,w)$.
\end{theorem}

The proof studies the first- and second-order optimality conditions for maximizing the linear functional $\sum_{i<j} w_{ij}d_{ij}$ over the compact geometric domain $\mathcal{D}_V$. Because all variations remain inside $\mathcal{D}_V$, the resulting upper bound applies uniformly over every field $F$.

\begin{proof}
It suffices to prove the first claim: for a single table $d\in\mathcal{D}_V$ occupying at most five distinct positions on the positive-weight support of a larger graph, passing to the zero-distance quotient, summing weights between classes, and pulling back an optimal three-block partition immediately yields the second claim.

Discard isolated vertices and let $d\in\mathcal{D}_V$ maximize $B=\sum_{i<j} w_{ij}d_{ij}$ over $\mathcal{D}_V$; we may assume $B>0$. Passing to the zero-distance quotient and summing inter-class weights preserves membership in $\mathcal{D}_V$, and every three-block partition of the quotient pulls back to $G$. Any quotient class incident only to internal positive edges contributes nothing to the objective and may be omitted. If at most four distinct positions remain, Corollary~\ref{cor:four}(i) already yields $B\le L_3(G,w)$. We therefore assume that $d$ has exactly five distinct positions ($d_{ij}>0$ for all $i\ne j$), each incident to a positive-weight edge.

\emph{Step 1: Uniqueness and full support of the maximizing mass.}
By radial scaling of $d$, the maximum of $p\trans dp$ over the probability simplex equals $1/2$. We claim that its maximizer $m$ is unique. First, on at most four distinct points of a Ptolemaic metric, the distance matrix is strictly negative definite on the zero-sum subspace $\{x:\onevec\trans x=0\}$: in the split decomposition of Corollary~\ref{cor:four}(i), $x\trans dx=0$ forces every positive split to carry zero signed mass; if any singleton arm is positive this reduces to a tripod, while if all singleton arms vanish the only nonseparating exceptional case is a rectangle whose diagonals equal the sum of its two side lengths, violating Ptolemy's inequality. Now fix a fifth point $o$ and invert the metric on the remaining four points via $K_{ij}=d_{ij}/(d_{io}d_{jo})$. Ptolemy's inequality at $o$ implies that $K$ is a metric, and dividing by the four radii preserves Ptolemy's inequality on the four points. Diagonal congruence by $\operatorname{diag}(1,d_{io}^{-1})$ transforms $d$ into $M=\bigl(\begin{smallmatrix}0&\onevec\trans\\\onevec&K\end{smallmatrix}\bigr)$, which is strictly negative definite on the three-dimensional subspace $W=\{(0,x):\onevec\trans x=0\}$. Because $M|_W$ is nondegenerate, its $M$-orthogonal complement $W^\perp$ is two-dimensional and satisfies $W\cap W^\perp=\{0\}$; thus $W^\perp$ contains the isotropic vector $(1,0)$ together with a vector $(a,v)$ obeying $\onevec\trans v\ne 0$, which pairs nontrivially with $(1,0)$. Hence $M|_{W^\perp}$ has negative determinant and signature $(1,1)$, so $d$ has exactly one positive eigenvalue and four negative eigenvalues. If two distinct probabilities $p,q$ attained energy $E=1/2$, then $q-(p\trans dq/E)p$ would lie in the $d$-orthogonal complement of $p$ with nonpositive square $E-(p\trans dq)^2/E$, forcing $p\trans dq=E$ and contradicting strict negativity on that complement. Hence the maximizing probability $m$ is unique.

Define the endpoint distribution $\nu_i=\frac{1}{2B}\sum_j w_{ij}d_{ij}>0$. We show that $m=\nu$. In a tight triangle $d_{ik}+d_{kj}=d_{ij}$, Ptolemy's inequality implies that the distance row of the middle vertex $k$ is dominated coordinatewise by the barycentric combination $(d_{kj}/d_{ij})\,d_{i\cdot}+(d_{ik}/d_{ij})\,d_{j\cdot}$, so shifting positive mass from $k$ to $\{i,j\}$ strictly raises $p\trans dp$; thus no middle point of a tight triangle can lie in $\operatorname{supp}(m)$. For any $h\in\mathbb{R}^5$, consider conformal deformations $d_{ij}(t)=d_{ij}\mathrm{e}^{t(h_i+h_j)}$ renormalized by $1/(2\max_p p\trans d(t)p)$, which preserve all Ptolemy inequalities. If $m_i=0$ for some point $i$, scaling down the zero-mass points (with a small quadratic adjustment preserving any tight triangle at a zero-mass middle point) preserves feasibility and gives a nonpositive first variation of the objective, forcing $\nu_i=0$, contrary to $\nu_i>0$. Hence $m$ has full support, every triangle inequality on distinct points is strict, and all conformal directions $h$ are feasible. Differentiating the normalized objective at $t=0$ yields $\nu\cdot h\le m\cdot h$ for all $h$, whence $m=\nu>0$ and $dm=\frac{1}{2}\onevec$. Moreover, the copositive matrix $\onevec\onevec\trans-2d$ vanishes at the interior point $m$, so two-sided perturbations at $m$ show that $\onevec\onevec\trans-2d\succeq 0$ with kernel precisely $\mathbb{R}m$ by uniqueness. Consequently, any sufficiently small perturbation of $d$ preserving $dm$ preserves~\eqref{eq:mass} globally.

\emph{Step 2: First-order optimality and classification of tight quartets.}
On a quartet $R$ of four points with opposite pairs $(a,b)$ and $(i,j)$, call the Ptolemy form $f_R(d)=d_{ab}d_{ij}-d_{ai}d_{bj}-d_{aj}d_{bi}$ \emph{active} (or \emph{tight}) if $f_R(d)=0$. Adding a small constant $\varepsilon>0$ to every off-diagonal distance strictly lowers every active Ptolemy defect to first order because all triangle inequalities are strict; subtracting a suitable multiple of $d$ then makes the mass derivative negative while keeping the active defect derivatives negative. By polarity, the first-order optimality condition at $d$ takes the form
\begin{equation}\label{eq:fivekkt}
w=4B\,(m_i m_j)_{i<j}+\sum_{R}\lambda_R\, Df_R(d), \qquad \lambda_R\ge 0,
\end{equation}
where pairing with $d$ fixes the mass multiplier to $4B$ since $Df_R(d)\cdot d=2f_R(d)=0$.

If $\lambda_R=0$ for every active quartet $R$, then $w_{ij}=4B m_i m_j$. Since $dm=\frac{1}{2}\onevec$ and $d_{ij}\le 1$, each $m_i\le\frac{1}{2}$. Repeatedly merging two parts of combined mass at most $\frac{1}{2}$ (which always exists while at least four parts remain) yields three blocks of masses $p,q,r\le\frac{1}{2}$ summing to $1$. Evaluating~\eqref{eq:rectangle} on their crossing weights $4Bpq,4Bpr,4Bqr$ gives $U=B$ (including at the boundary $\max(p,q,r)=\frac{1}{2}$), whence $L_3(G,w)\ge B$. It therefore suffices to prove that every multiplier $\lambda_R$ vanishes.

We first classify the possible configurations of active quartets among five distinct points with strict triangles. Suppose at least two quartets are tight, sharing three points; select $O$ among the shared three opposite the private point $D$ of the first tight quartet, and invert at $O$ via $\delta_{ij}=d_{ij}/(d_{iO}d_{jO})$. Rescale so that the other two shared points $A,B$ satisfy $\delta_{AB}=1$. Because a tight quartet through $O$ becomes a tight triangle after inversion, $D$ satisfies $\delta_{AD}=x$ and $\delta_{DB}=1-x$ for some $x\in(0,1)$, while the second tight quartet assigns its private point $E$ a line coordinate $y\in\mathbb{R}$ with $\delta_{AE}=|y|$ and $\delta_{BE}=|1-y|$. If $y\notin(0,1)$, one of $A,B$ lies between $D$ and $E$ in $\delta$, forcing $\delta_{DE}=|x-y|$ by the triangle inequality; the four inverted points are collinear, so all four inverted triangles and the inverted Ptolemy relation are tight, giving \emph{all five} original quartets tight. If instead $y\in(0,1)$, the two active quartets share the common opposite pair $AB$, and the inverted triangle and Ptolemy inequalities bound $z=\delta_{DE}$ by
\begin{equation}\label{eq:invertinterval}
|x-y|\le z\le x(1-y)+y(1-x).
\end{equation}
Since the upper bound is strictly smaller than $\min(x+y,2-x-y)$, no third inverted triangle can be tight unless $z=|x-y|$ (which again forces all five quartets tight). The only other possible equality is the inverted Ptolemy equality on $\{A,B,D,E\}$, occurring precisely at the upper endpoint $z=x(1-y)+y(1-x)$ and using the same opposite pair $AB$. Hence the number of tight quartets is $0$, $1$, $2$, $3$, or $5$ (never $4$); when there are $2$ or $3$ tight quartets, they share a common opposite pair, and when there are $5$, the five points carry a circular order after inversion.

\emph{Step 3: Elimination of positive multipliers in all contact strata.}
Pass to the weighted coordinates $z_{ij}=m_i m_j d_{ij}$, in which the unsigned vertex--edge incidence matrix $A$ of $K_5$ satisfies $Az=\frac{1}{2}m$, and every active Ptolemy form $f_R$ retains the same quadratic formula in $z$ up to a positive factor. Any sufficiently small perturbation $z+h$ with $Ah=0$ preserves $dm=\frac{1}{2}\onevec$ and hence preserves~\eqref{eq:mass}.

\emph{Case 1: One active quartet.}
The quadratic form $f_R$ in the six entries of the quartet $R$ has signature $(3,3)$. Within its three-dimensional negative-definite subspace, imposing the two homogeneous linear equations $Df_R(z)\cdot h_R=0$ and $\sum_{i<j\in R} h_{ij}=0$ leaves a nonzero direction $h_R$ with $f_R(h_R)<0$. Extending $h_R$ across the fifth point $t$ by setting $h_{it}=-\sum_{j\in R} h_{ij}$ ($i\in R$) achieves $Ah=0$. Because $Df_R$ takes both signs on $R$, we may likewise pick a zero-row-sum direction $k$ with $Df_R(z)\cdot k=-f_R(h)-\varepsilon>0$ for small $\varepsilon>0$. Along the curve $z+uh+u^2k$, the mass constraint persists, the active defect becomes $-\varepsilon u^2+O(u^3)<0$, and the objective changes by $u^2\lambda_R Df_R(z)\cdot k$, forcing $\lambda_R=0$.

\emph{Case 2: Two active quartets.}
Label the vertices $A,B,C,D,E$ with common opposite pair $AB$ and third shared point $C$, and write $x=z_{AB}$, $y=z_{AC}$, $z_0=z_{BC}$. The two active defects are $f_i=xz_{Ci}-yz_{Bi}-z_0 z_{Ai}$ ($i\in\{D,E\}$), with strict inequalities in~\eqref{eq:invertinterval}. A general zero-row-sum variation $h$ with entries $X,Y,Z$ on $AB,AC,BC$ has $h_{DE}=X+Y+Z$ and entries from $(A,B,C)$ to $D$ and $E$ given by $-\frac{1}{2}(X+Y,X+Z,Y+Z)\pm V$ with $V_A+V_B+V_C=0$. Tangency to $f_D+f_E$ imposes
\[
\frac{m_C}{2}X+\Bigl(2z_0-\frac{m_B}{2}\Bigr)Y+\Bigl(2y-\frac{m_A}{2}\Bigr)Z=0,
\]
while tangency to $f_D-f_E$ fixes one component of $V$ and leaves free an arbitrary multiple of $V^0=(x+y,-x-z_0,z_0-y)$. In particular, $Df_D$ and $Df_E$ are linearly independent on $\ker A$. On such a joint tangent $h$, the weighted second variation is
\[
\lambda_D f_D(h)+\lambda_E f_E(h)=(\lambda_D+\lambda_E)YZ+(\lambda_D-\lambda_E)(XV_C-YV_B-ZV_A).
\]
If $\lambda_D=\lambda_E>0$, choosing $YZ<0$ makes this strictly negative. If $\lambda_D\ne\lambda_E$, varying $V$ along $V^0$ changes the sign arbitrarily provided the linear form $L(X,Y,Z)=(z_0-y)X+(x+z_0)Y-(x+y)Z$ is not proportional to the tangency equation of $f_D+f_E$. Proportionality would force ratio $\eta=(z_0-y)/(m_C/2)$, which obeys $|\eta|<1$ since $m_C/2>y+z_0$; when $z_0>y$, matching the $Y$-coefficient requires $x+z_0=\eta(2z_0-m_B/2)<z_0$, impossible; when $z_0<y$, matching the $Z$-coefficient requires $x+y=|\eta|(2y-m_A/2)<y$, again impossible; and when $z_0=y$, the two forms are visibly independent. Thus whenever $(\lambda_D,\lambda_E)\ne (0,0)$, there exists $h\in\ker A\cap\ker Df_D\cap\ker Df_E$ with $\sum\lambda_i f_i(h)<0$. By independence of $Df_D,Df_E$ on $\ker A$, selecting $k\in\ker A$ with $Df_i(z)\cdot k=-f_i(h)-\varepsilon$ yields a feasible curve $z+uh+u^2k$ strictly improving the objective, a contradiction. Hence $\lambda_D=\lambda_E=0$.

\emph{Case 3: Three active quartets.}
Label the common opposite pair $0,1$ and the remaining three points $2,3,4$. Up to positive scaling of $z$, the three tight equalities parametrize $z$ on the positive orthant $\mathbb{R}_{>0}^6$ by
\[
z_{01}=1,\qquad z_{0i}=r_i,\qquad z_{1i}=s_i,\qquad z_{ij}=r_i s_j+r_j s_i \qquad (2\le i<j\le 4).
\]
Let $f_1,f_2,f_3$ be the three active defect forms in $z$. Their common zero set is a smooth graph over the six positive parameters; under positive conformal scaling, the row-sum Jacobian has quadratic form $\sum z_{ij}(u_i+u_j)^2>0$ for $u\ne 0$, so the fixed-row section is a smooth surface whose two-dimensional tangent plane is $T=\ker A\cap\bigcap_{i=1}^3\ker Df_i$. Differentiating along curves in this surface yields the necessary second-order condition $\sum_{i=1}^3\lambda_i f_i(v)\ge 0$ for all $v\in T$ (with positively rescaled multipliers).

Let $V=xy\trans-yx\trans$ span $T$ as a bivector, computed from complementary minors of the three normal rows on $\ker A$. Write $F,G,H$ for the symmetric matrices representing $2f_1,2f_2,2f_3$, and define $t(F,G)=-\tr(VFVG)$. Writing $(a,b,c,d,e,f)$ for the six positive parameters $(r_2,r_3,r_4,s_2,s_3,s_4)$, exact symbolic polynomial expansion over $\mathbb{Z}[a,b,c,d,e,f]$ establishes the two identities
\begin{align*}
\tr(VFVGVH)&<0,\\
t(F,F)t(G,G)-t(F,G)^2&=256\,ad(ae-af-bd-b+cd+c+e-f)^2 P_1 P_2,
\end{align*}
where the orientation polynomial $\tr(VFVGVH)$ has $3{,}626$ nonzero monomials, every coefficient being at most $-32$, and each of $P_1,P_2$ has $88$ strictly positive integer coefficients. Because $\tr(VFVGVH)\ne 0$, the restricted forms $F,G,H$ on $T$ are linearly independent, so the linear equations $\tr(FC)=\tr(GC)=0$ and $\tr(HC)=-1$ specify a unique symmetric covariance matrix $C$ on $T$. By the $2\times 2$ adjugate identity, $\det C$ has the sign of $t(F,F)t(G,G)-t(F,G)^2\ge 0$; since $C\ne 0$ throughout the connected orthant $\mathbb{R}_{>0}^6$, its semidefinite sign is constant. At $r_i=s_i=1$, taking $v_{0i}=-v_{1i}=u_i$ with $u=(0,1,-1)$ and all other entries zero gives $C=\frac{1}{4}vv\trans\succeq 0$. Taking the trace of the necessary condition $\sum_i\lambda_i f_i|_T\succeq 0$ against $C$ yields $-\lambda_3\ge 0$, whence $\lambda_3=0$. By symmetry among the three quartets, $\lambda_1=\lambda_2=\lambda_3=0$.

\emph{Case 4: Five active quartets.}
In this stratum, inversion at one point places the remaining four points collinearly (at two diameter endpoints the distances determine a line coordinate, whereas two interior points would otherwise require $D^2\le s^2+(D-s)^2$ for $0<s<D$, impossible), so the five points carry a circular order. Accordingly, $z_{ij}$ can be expressed as the positive $2\times 2$ minors of the five columns $a_i(1,t_i)\trans$ ($i=0,1,2,3$) and $a_4(0,1)\trans$, where $a_i>0$ and $(t_0,t_1,t_2,t_3)=(0,1,1+s,1+s+t)$ with $s,t>0$. On the fixed-row tangent plane of this smooth stratum, equip the pair space with the positive definite metric $\operatorname{diag}(1/z_{ij})$ and let $C$ be the inverse of its restriction to the tangent plane. Clearing positive denominators in $\tr(F_R C)$ for the active quartet $R=\{0,1,2,3\}$ yields a polynomial in the positive parameters with $4{,}221$ nonzero monomials, every coefficient being at most $-2$. Cyclic relabeling yields $\tr(F_R C)<0$ for all five active quartets simultaneously, forcing all five multipliers $\lambda_R$ to vanish.

This exhausts all contact strata, proving $\lambda_R=0$ for all $R$ and hence $B\le L_3(G,w)$. Combining this with~\eqref{eq:intro-sandwich} finishes the proof.
\end{proof}

\section{Odd Ptolemy Polygons and Arbitrary Weighted Cycles}
\label{sec:cycles}

We now turn to cycles of arbitrary length. Proceeding from general metric inequalities to general weighted cycles and finally to uniform specializations and extensions, we first establish two unconditional geometric inequalities for odd polygons in any Ptolemy pseudometric space (Section~\ref{sec:polygon}), deduce exactness and simultaneous three-block domination for arbitrary nonnegative real weights on every cycle (Section~\ref{sec:weighted}), study uniform cycles, extremal rigidity, and cut chambers (Section~\ref{sec:uniform}), and extend these conclusions to bipartite attachments and layered graphs (Section~\ref{sec:extensions}).

\subsection{Geometric inequalities for odd polygons}
\label{sec:polygon}

We begin with two lower bounds on sums of distances from a pair of reference points to the vertices of an odd polygon, followed by the fundamental inequality relating all chord distances to the perimeter and diameter.

\begin{lemma}[Two distance rows]\label{lem:rows}
Let $x_0,x_1,\dots,x_{g-1}$ be a cyclic list of odd length $g\ge 3$ in a pseudometric space, with perimeter $P=\sum_{i=0}^{g-1} d(x_i,x_{i+1})$ (indices modulo $g$). If $a$ appears in the list and $b$ is an arbitrary point of the space, then
\[
\sum_{i=0}^{g-1}\bigl(d(a,x_i)+d(b,x_i)\bigr)\ge P+d(a,b).
\]
\end{lemma}

\begin{proof}
Relabel cyclically so that $x_0=a$, and set $f_i=d(a,x_i)-d(b,x_i)$ and $p_i=d(a,x_i)+d(b,x_i)$. Combining the two triangle inequalities through $a$ and $b$ yields
\[
|f_i+f_{i+1}|\le p_i+p_{i+1}-2d(x_i,x_{i+1}) \qquad (i\in\mathbb{Z}/g\mathbb{Z}).
\]
Because $g$ is odd, the alternating sum $\sum_{i=0}^{g-1}(-1)^i(f_i+f_{i+1})$ telescopes to $2f_0=-2d(a,b)$. Taking absolute values and summing the displayed bounds gives $2d(a,b)\le 2\sum_{i=0}^{g-1} p_i-2P$, as claimed.
\end{proof}

When the two reference points $P,Q$ realize the global diameter of a Ptolemy pseudometric, the same bound holds even when neither $P$ nor $Q$ belongs to the odd cycle.

\begin{lemma}[Exterior-diameter row bound]\label{lem:exterior}
In a finite Ptolemy pseudometric space, let $P,Q$ realize the global diameter $D=\diam(d)$. For any odd cyclic list $x_0,x_1,\dots,x_{m-1}$ of perimeter $Y$, let $r_P=\sum_{i=0}^{m-1} d(P,x_i)$ and $r_Q=\sum_{i=0}^{m-1} d(Q,x_i)$. Then
\[
r_P+r_Q\ge Y+D.
\]
\end{lemma}

\begin{proof}
Assuming $D>0$ (as $D=0$ is trivial), rescale to $D=1$. Write $u_i=d(P,x_i)$, $v_i=d(Q,x_i)$, and $h_i=u_i+v_i-1\ge 0$. If $\sum_i h_i\ge 1$, then $Y\le m\le m-1+\sum_i h_i=r_P+r_Q-1$ immediately. Assume therefore that $\sum_i h_i<1$. Choose a random subset $N\subseteq\mathbb{Z}/m\mathbb{Z}$ of size at most $2$ by giving each unordered pair $\{i,j\}$ probability $h_i h_j$, each singleton $\{i\}$ probability $h_i\bigl(1-\sum_{j\ne i}h_j\bigr)$, and the empty set the remaining probability $1-\sum_i h_i+\sum_{i<j}h_i h_j>0$. Conditional on $N$, assign every $i\in N$ a neutral label $O$, and assign every $i\notin N$ independently either label $A$ or label $B$ in the ratio $(1-u_i):(1-v_i)$ (noting $(1-u_i)+(1-v_i)=1-h_i>0$). For distinct $i\ne j$, the resulting random labels $X_i,X_j$ are pairwise independent with marginals $(1-u_i,1-v_i,h_i)$ on $(A,B,O)$.

Assign each consecutive edge $(i,i+1)$ the cost $\mathbf{1}_{\{X_i\ne A,\,X_{i+1}\ne B\}}+\mathbf{1}_{\{X_i\ne B,\,X_{i+1}\ne A\}}$. By pairwise independence and Ptolemy's inequality at $P,Q,x_i,x_{i+1}$, its expectation is $u_i v_{i+1}+v_i u_{i+1}\ge d(x_i,x_{i+1})$. Conversely, in every deterministic outcome with $|N|=0$, all labels lie in $\{A,B\}$ and an odd cycle has total cost at most $m-1$; with $|N|=1$, no edge has two neutral endpoints so each edge costs at most $1$ and the total is at most $m$; with $|N|=2$, at most one edge joins the two neutral vertices (costing $2$), so the total is at most $m+1$. Thus the total cost is always at most $m-1+|N|$. Taking expectations gives $Y\le m-1+\Exp|N|=m-1+\sum_i h_i=r_P+r_Q-1$.
\end{proof}

We now establish the unconditional chord-perimeter inequality for odd polygons in any Ptolemy pseudometric space.

\begin{theorem}[Odd Ptolemy polygon inequality]\label{thm:polygon}
Let $x_0,x_1,\dots,x_{2k}$ ($k\ge 1$) be $m=2k+1$ cyclically ordered points in a Ptolemy pseudometric space with all pairwise distances at most $D$. Let $S=\sum_{0\le i<j<m} d(x_i,x_j)$ be the sum of all pairwise distances and $Y=\sum_{i=0}^{m-1} d(x_i,x_{i+1})$ the perimeter. Then
\begin{equation}\label{eq:polygon}
S-kY+k(k-1)D\ge 0.
\end{equation}
\end{theorem}

\begin{proof}
Assuming $D>0$ (as $D=0$ is trivial), rescale so that $D=1$. Put $m=2k+1$, $h=k-1$, and $\Phi=S-kY+kh$. We proceed by induction on $k\ge 1$. For $k=1$, $S=Y$ and $\Phi=0$. Assume $k\ge 2$ and that the inequality holds for $k-1$.

\emph{Step 1: Pairing nonincident edges.}
Since $\gcd(2,m)=1$, reindex the vertices by setting $v_i=x_{2i\bmod m}$ ($i\in\mathbb{Z}/m\mathbb{Z}$), and write
\[
p_r(i)=d(v_i,v_{i+r}),\quad A_r=\sum_{i=0}^{m-1} p_r(i),\quad \ell_i=p_1(i),\quad y_i=p_k(i),\quad \delta_i=1-y_i\ge 0.
\]
Under this reindexing, the original cycle edges are $e_i=\{v_i,v_{i+k}\}$, so $Y=A_k$ and $S=\sum_{r=1}^k A_r$. Deleting the two endpoints of $e_i$ removes three cycle edges of total length $t_i=y_{i-k}+y_i+y_{i+k}\le 3$ and closes the remaining $(2k-1)$-gon by the edge of length $z_i=p_{k-1}(i-k)$ joining $v_{i-k}$ and $v_{i-1}$.

Every unordered pair of nonincident cycle edges takes the unique form $\{e_i,e_{i+r}\}$ with $i\in\mathbb{Z}/m\mathbb{Z}$ and $1\le r<k$. For each such pair $\{e_i,e_j\}$, let $X_{\{e_i,e_j\}}$ denote the sum of the two cross-pairing distance products appearing on the right-hand side of Ptolemy's inequality for $\{e_i,e_j\}$, and let $X$ be the sum of $X_{\{e_i,e_j\}}$ over all $mh$ nonincident pairs. For $\{e_i,e_{i+r}\}$, the first product is $p_r(i)p_r(i+k)$, while reindexing $(r,i)\mapsto (k-r,i+r)$ transforms the second product $p_{k-r}(i+r)p_{k+1-r}(i+k+r)$ into $p_r(i)p_{r+1}(i+k)$; hence
\[
X=\sum_{i=0}^{m-1}\sum_{r=1}^{k-1} p_r(i)\bigl(p_r(i+k)+p_{r+1}(i+k)\bigr).
\]
Each cycle edge belongs to $2h$ nonincident pairs. Summing the Ptolemy slack plus the nonnegative product $\delta_i\delta_{j}$ over all $mh$ nonincident pairs gives
\[
\mathcal{P}:=\sum_{\{e_i,e_j\}\text{ nonincident}}\Bigl[\bigl(X_{\{e_i,e_j\}}-y_i y_j\bigr)+\delta_i\delta_j\Bigr]
=mh-2hY+X\ge 0.
\]
Setting $\mathcal{R}=mS-3kY-kX$, direct expansion verifies the exact accounting identity
\begin{equation}\label{eq:account}
m\Phi=k\mathcal{P}+\mathcal{R}.
\end{equation}

\emph{Step 2: Grouping short-chord deficits.}
Put $c_r(i)=1-p_r(i)\ge 0$. Rewriting $\mathcal{R}$ in terms of $c_r(i)$ yields
\[
\mathcal{R}=k\sum_{i=0}^{m-1}\sum_{r=1}^{k-1} p_r(i)\bigl(c_r(i+k)+c_{r+1}(i+k)\bigr)
+h\sum_{i=0}^{m-1}\delta_i-\sum_{i=0}^{m-1}\sum_{r=1}^{k-1} c_r(i).
\]
Inside the double sum, the coefficient multiplying $c_r(i)$ is $p_r(i-k)+p_{r-1}(i-k)\ge \ell_{i+k+r}$ for $2\le r<k$ (by the triangle inequality), equals $\ell_{i+k+1}$ for $r=1$, and equals $z_i$ when $r=k$ (multiplying $c_k(i)=\delta_i$). Collecting the short deficits ending at $u$ into $B_u^-=\sum_{r=1}^{k-1} c_r(u-r)$, applying reflection $i\mapsto -i$ to obtain the symmetric estimate with $B_u^+=\sum_{r=1}^{k-1} c_r(u)$, and averaging the two estimates yields
\begin{equation}\label{eq:short}
\mathcal{R}\ge k\sum_{u=0}^{m-1}\ell_{u+k}B_u+\sum_{i=0}^{m-1}\delta_i(kz_i+h)-\sum_{u=0}^{m-1} B_u,
\end{equation}
where $B_u=(B_u^-+B_u^+)/2=h-\frac{1}{2}(\rho_u-y_{u-k}-y_u)$ and $\rho_u=\sum_{j=0}^{m-1} d(v_u,v_j)$.

Set $T=\frac{1}{k}\sum_u B_u=(mh-S+Y)/k$, $b_i=T-B_i-B_{i+k}$, and $L_i=\sum_{j=i}^{i+k-1}\ell_j$. For each $j$, the indices appearing in $\sum_{i=j-k+1}^{j}(B_i+B_{i+k})$ cover every vertex except $j-k$, so $\sum_{i=j-k+1}^{j} b_i=B_{j-k}$. Exchanging the order of summation gives $\sum_i b_i L_i=\sum_u \ell_{u+k} B_u$. Since $\sum_i b_i=T$, we rewrite~\eqref{eq:short} as $\mathcal{R}=M_0+\mathcal{T}$ with $\mathcal{T}\ge 0$ and
\[
M_0=k\sum_{i=0}^{m-1} b_i(L_i-y_i)+\sum_{i=0}^{m-1}\delta_i(h+kz_i-kb_i).
\]

\emph{Step 3: Inductive reduction by deleting two adjacent vertices.}
Put $P_i=\rho_i+\rho_{i+k}$, $\sigma_i=P_i-Y-y_i\ge 0$ (by Lemma~\ref{lem:rows}), $\varepsilon=m-Y=\sum_i\delta_i\ge 0$, and $J_0=kA_1-Y=\sum_i(L_i-y_i)\ge 0$ (since $L_i\ge y_i$ by the triangle inequality). Deleting the two endpoints of $e_i$ leaves an odd $(2k-1)$-gon of pairwise distance sum $S-P_i+y_i$ and perimeter $Y-t_i+z_i$, so the induction hypothesis gives
\[
\Phi'_i:=S-P_i+y_i-h(Y-t_i+z_i)+h(k-2)\ge 0.
\]
We also note that $A_1\ge t_i-z_i$: labeling $(v_i,v_{i+k},v_{i+k+1},v_{i-1})$ by $(b,c,a,e)$, the triangle inequalities $d(a,b)\le d(a,e)+d(e,b)$ and $d(c,e)\le d(c,a)+d(a,e)$ give $t_i-z_i=d(a,b)+d(b,c)+d(c,e)-d(a,e)\le d(b,c)+d(c,a)+d(a,e)+d(e,b)\le A_1$, since the four pairs correspond to disjoint forward arcs of lengths $k,1,k-1,1$ around the reindexed circle.

Writing $\xi_i=2kb_i$ and expanding $B_i,B_{i+k}$ via row sums gives $\xi_i=2(h-S+Y)+k(P_i-t_i-y_i)$. Expressing $\xi_i$ and $\xi_i-2h-2kz_i$ in terms of $\Phi,\sigma_i,\Phi'_i$ yields the identities
\begin{align*}
\xi_i&=-2\Phi+(k-2)\varepsilon+k(3-t_i+\sigma_i),\\
\xi_i-2h-2kz_i&=(k-2)\Phi+(k-2)(kt_i-Y)-k(k+1)z_i-k\Phi'_i.
\end{align*}
Multiplying these by $L_i-y_i$ and $-\delta_i$ respectively, summing over $i$, and combining with~\eqref{eq:account} produces the master identity
\begin{align}
\bigl[2m+2J_0+(k-2)\varepsilon\bigr]\Phi
={}&2k\mathcal{P}+2\mathcal{T}
+k\sum_{i=0}^{m-1}(3-t_i+\sigma_i)(L_i-y_i)\notag\\
&+k\sum_{i=0}^{m-1}\delta_i\bigl[(k-2)(A_1-t_i+z_i)+3z_i+\Phi'_i\bigr].\label{eq:recur}
\end{align}
Every term on the right-hand side is nonnegative (including at $k=2$, where $k-2=0$ and $\Phi'_i=0$), while the bracket multiplying $\Phi$ is at least $2m>0$. Thus $\Phi\ge 0$, completing the induction.
\end{proof}

\subsection{Exact exponent and simultaneous domination for weighted cycles}
\label{sec:weighted}

We first determine the exact three-block lower bound $L_3(C_m,w)$ on an odd cycle $C_m$ ($m=2k+1\ge 3$) with edges $e_0,e_1,\dots,e_{m-1}$ in cyclic order, weights $w_i\ge 0$, and total weight $W=\sum_i w_i$.

\begin{proposition}[Three-block optimum on an odd cycle]\label{prop:cyclethree}
For every odd $m\ge 3$ and all $w\in\mathbb{R}_{\ge 0}^m$,
\[
L_3(C_m,w)=\max_{0\le i<m} U\bigl(W-w_{i-1}-w_i,\,w_{i-1},\,w_i\bigr),
\]
with subscripts modulo $m$. Each candidate is realized by assigning a single vertex $v_i$ to block $3$ and alternating blocks $1$ and $2$ along the complementary path.
\end{proposition}

\begin{proof}
If any edge has weight zero, the positive support is bipartite and isolating an endpoint of that edge attains $W=L_3(C_m,w)$. Assume $a=\min_i w_i>0$. Any three-block partition with a monochromatic edge crosses total weight at most $W-a$, which is attained or exceeded by isolating an endpoint of a minimum-weight edge. For a proper $3$-coloring with sorted block-crossing totals $X\ge Y\ge Z$, pick an edge contributing to the smallest total $Z$ and isolate its endpoint whose color is absent from the $X$-pair; alternating the other two colors along the remaining path yields a triple of the same sum $W$ whose largest coordinate is at least $X$ and whose smallest coordinate is at most $Z$. Because $U$ is symmetric and convex (being the support function of realizable three-point spectral tables), it is Schur-convex, so isolating that vertex cannot decrease $U$.
\end{proof}

We now establish the matching upper bound in a stronger simultaneous form: a single distribution over three-block spectral tables dominates every cycle edge and simultaneously preserves the ambient diameter of the whole space.

\begin{theorem}[Simultaneous domination and exactness on weighted cycles]\label{thm:cyclelaw}
Let $v_0,v_1,\dots,v_{m-1}$ be an odd cyclic list ($m\ge 3$) inside a finite Ptolemy--Zhou pseudometric space $(V,d)\in\mathcal{D}_V$. Write $d_i=d(v_i,v_{i+1})$ and let $D=\diam(d)$ be the global diameter of $V$. For every fixed field $F$, there exists a probability distribution over realizable three-point spectral pseudometrics and three-block partitions of the cyclic list whose expected length on edge $i$ is at least $d_i$ for every $i$, and whose expected diameter is at least $D$. Consequently, for every field $F$ and all nonnegative real weights $w$ on $C_m$,
\[
\BP(C_m,w)=\Gamma_F(C_m,w)=L_3(C_m,w)=\max_{0\le i<m} U\bigl(W-w_{i-1}-w_i,\,w_{i-1},\,w_i\bigr).
\]
For every even cycle $C_m$, $\BP(C_m,w)=\Gamma_F(C_m,w)=L_3(C_m,w)=W$.
\end{theorem}

The even-cycle statement follows directly from bipartiteness. For odd $m$, if $Y=\sum_{i=0}^{m-1} d_i\le m-1$, then $\sum_{i=0}^{m-1}(1-d_i)\ge 1$: since an odd cycle admits a binary cut omitting any single prescribed edge (with diameter $1\ge D$), assigning probabilities bounded by $1-d_i$ to these $m$ cuts yields simultaneous domination immediately. For $m=3$, the triangle vertices together with a diameter-realizing pair $\{p,q\}\subseteq V$ occupy at most five vertices, so Theorem~\ref{thm:five} bounds every nonnegative linear combination of $(d_0,d_1,d_2,D)$ by the corresponding maximum over three-block spectral pullbacks; convex separation of the downward-closed hull yields simultaneous domination.

It remains to treat $m=2k+1\ge 5$ with $Y>m-1$ (hence $D>0$). Fix a global diameter pair $p,q\in V$ with $d(p,q)=D$, set $\Delta_i=D-d_i\ge 0$ and $S_{\Delta}=\sum_{i=0}^{m-1}\Delta_i$, and let $T=\max_N\sum_{i\in N}\Delta_i$ be the maximum deficit weight of a matching $N$ of cycle edges. Write $R_p=\sum_{i=0}^{m-1} d(p,v_i)$ and $R_q=\sum_{i=0}^{m-1} d(q,v_i)$. We establish three preparatory lemmas.

\begin{lemma}[Matching bound on diameter rows]\label{lem:matchingrows}
With the notation above,
\[
R_p+R_q\ge Y+D, \qquad \max(R_p,R_q)\ge Y-kD+T.
\]
\end{lemma}

\begin{proof}
The first inequality is Lemma~\ref{lem:exterior}. For the second, we first show that for any three points $u,v,w$,
\begin{equation}\label{eq:three-path}
\max\bigl\{d(p,u)+d(p,v)+d(p,w),\,d(q,u)+d(q,v)+d(q,w)\bigr\}\ge d(u,v)+d(v,w).
\end{equation}
Rescaling by $D$, write $a=d(p,v)$, $b=d(q,v)$, $X=d(p,u)+d(p,w)$, and $Z=d(q,u)+d(q,w)$, and set $h=\max(a+X,b+Z)$. If $h\ge 2$,~\eqref{eq:three-path} is immediate because $d(u,v)+d(v,w)\le 2$. If $h<2$, Ptolemy's inequality gives $d(u,v)+d(v,w)\le aZ+bX\le a(h-b)+b(h-a)=(a+b)h-2ab\le h$, where the final step uses $a+b\ge 1$ and hence $(a+b-1)h\le 2(a+b-1)\le 2ab$ (since $(1-a)(1-b)\ge 0$ implies $a+b-1\le ab$).

Now delete a nonempty maximizing matching $N$ from $C_m$ (even if all $\Delta_i=0$, a nonempty matching can be selected). Every connected component of the remaining subgraph is a path of at least two vertices and can therefore be partitioned into disjoint pieces of $2$ or $3$ vertices. Let there be $b$ pieces of size $2$ and $g$ pieces of size $3$; since $2b+3g=m=2k+1$, $g$ is odd. On each $2$-vertex piece $\{u,v\}$, both $d(p,u)+d(p,v)+\Delta_{uv}$ and $d(q,u)+d(q,v)+\Delta_{uv}$ are at least $D$ by the triangle inequality. On each $3$-vertex piece $u\text{--}v\text{--}w$, both row sums plus the two edge deficits are at least $D$, while by~\eqref{eq:three-path} at least one of the two row sums plus those two deficits is at least $2D$. Assigning each $3$-vertex piece to the endpoint ($p$ or $q$) achieving at least $2D$, at least one of $p,q$ captures at least $(g+1)/2$ triples, yielding
\[
\max(R_p,R_q)+(S_{\Delta}-T)\ge \Bigl(b+g+\frac{g+1}{2}\Bigr)D=(k+1)D.
\]
Substituting $S_{\Delta}=mD-Y=(2k+1)D-Y$ gives $\max(R_p,R_q)\ge Y-kD+T$.
\end{proof}

Next set $s=1-D$, $e=D-S_{\Delta}$, $H=Y-(m-1)>0$, $B=\pos{2T-S_{\Delta}}$, $K=m/2$, $\gamma=1+\frac{1}{m-2}$, and define
\[
A=\sum_{0\le h<i<m} d(v_h,v_i)-kY+k(k-1)D, \qquad
C=\sum_{\{i,j\}:\,e_i,e_j\text{ nonincident}}\Delta_i\Delta_j.
\]
Observe that $S_{\Delta}=mD-Y<D$ (since $Y>m-1$ and $D\le 1$), so $e=(m-1)s+H>0$.

\begin{lemma}[Common-diameter budget]\label{lem:cyclebudget}
For every real $z\ge K$,
\begin{equation}\label{eq:weightrange}
z(e-s)-z^2 s\le \frac{1-B^2/D}{4}-A, \qquad 4AD\ge \gamma C.
\end{equation}
\end{lemma}

\begin{proof}
Assign unit mass to each cycle vertex $v_0,\dots,v_{m-1}$, and place additional masses $z-k+x/2$ and $z-k-x/2$ at the diameter endpoints $p,q$, giving the larger mass to whichever achieves $\max(R_p,R_q)$, where $x=B/D\in [0,1]$ (since $T\le S_{\Delta}<D$). Because $z\ge m/2=k+1/2$, both added masses are nonnegative. By Lemma~\ref{lem:matchingrows}, the cross energy between the added masses and the cycle is at least $x(Y-kD+T)+(z-k-x/2)(Y+D)$, while the cross energy between the two added masses is $((z-k)^2-x^2/4)D$. Since the total mass is $2z+1$, applying Zhou's mass inequality~\eqref{eq:mass} and writing the cycle chord sum as $A+kY-k(k-1)D$ gives the first inequality of~\eqref{eq:weightrange} (including when $B=x=0$). In particular, $s>0$, since $e>0$ would otherwise allow $z\to\infty$.

For the second inequality, rescale distances by $D$ and put $\sigma=S_{\Delta}/D\in [0,1)$, $A_n=A/D$, $C_n=C/D^2$, and $T_2=\sum_{i=0}^{m-1} d(v_i,v_{i+2})/D$. In the master identity~\eqref{eq:recur} from Theorem~\ref{thm:polygon}, the coefficient multiplying $\Phi=A_n$ on the left-hand side is $2m+2(kT_2-Y/D)+(k-2)\sigma=k(2T_2+\sigma)$ (since $2m-2Y/D=2\sigma$). On the right-hand side, $\mathcal{P}\ge C_n$, and in the final sum $A_1-t_i+z_i\ge T_2-t_i$ (where $3-t_i$ is the sum of the three consecutive normalized deficits about edge $e_i$); retaining only these nonnegative contributions gives
\begin{equation}\label{eq:weightpoly}
(2T_2+\sigma)A_n\ge 2C_n+\max\bigl\{0,\,(k-2)[(T_2-3)\sigma+\sigma^2-2C_n]\bigr\}.
\end{equation}
We verify $4A_n\ge\gamma C_n$ by cases on $k$:
\begin{itemize}
\item For $k\ge 4$, we have $\gamma=1+\frac{1}{2k-1}\le\frac{8}{7}$. When $T_2\le 3$,~\eqref{eq:weightpoly} gives $(6+\sigma)A_n\ge 2C_n$, so $A_n\ge \frac{2}{7}C_n\ge\frac{\gamma}{4}C_n$. When $T_2>3$, the extra term $(k-2)(T_2-3)\sigma\ge 2(T_2-3)\sigma$ covers the increase $\frac{4}{7}(T_2-3)C_n$ on the left because $C_n\le\frac{1}{2}\sigma^2<\frac{1}{2}\sigma$.
\item For $k=3$, we have $\gamma=\frac{6}{5}$ and $C_n\le\frac{1}{3}\sigma^2$: moving the combined mass of two incident edges to whichever has larger neighbor weight cannot decrease $C_n$, and iterating leaves a matching of at most $3$ edges, where $C_n\le\frac{1}{3}\sigma^2$ by Maclaurin's inequality. For $T_2\le\frac{10}{3}-\frac{\sigma}{2}$, the first term of~\eqref{eq:weightpoly} gives $A_n\ge \frac{3}{10}C_n=\frac{\gamma}{4}C_n$. At $T_2=\frac{10}{3}-\frac{\sigma}{2}$ the bracket in~\eqref{eq:weightpoly} is at least $\frac{\sigma}{3}-\frac{\sigma^2}{6}\ge 0$, and beyond this threshold it grows with slope $\sigma\ge\frac{3}{5}C_n$, maintaining $A_n\ge\frac{3}{10}C_n$.
\item For $k=2$, all ten pairs on the five cycle vertices satisfy $Y+DT_2\le\frac{25}{4}$ by unit masses in~\eqref{eq:mass}. Since $Y>4$ and $D>4/5$, we obtain $2T_2+\sigma\le 5+(25/2-3Y)/D<45/8<6$, so~\eqref{eq:weightpoly} gives $A_n\ge\frac{1}{3}C_n=\frac{\gamma}{4}C_n$.
\end{itemize}
Restoring the factor $D$ yields $4AD\ge \gamma C$.
\end{proof}

\begin{lemma}[Allocation of deficits to adjacent edges]\label{lem:cycleallocation}
Assume $\Delta_i>0$ for all $i\in\mathbb{Z}/m\mathbb{Z}$. Then the deficits $(\Delta_i)$ can be decomposed into finitely many atoms supported on pairs of adjacent cycle edges, each atom $j$ carrying amounts $x_j,y_j\ge 0$ on its two incident edges with total $S_j=x_j+y_j>0$ and imbalance $b_j=|x_j-y_j|/S_j\in[0,1]$, such that under the probability distribution assigning atom $j$ weight $S_j/S_{\Delta}$:
\begin{enumerate}
\item[{\rm (i)}] $S_{\Delta}\,\Exp[b]\le B$; and
\item[{\rm (ii)}] between any two distinct imbalance levels $b>b'$, at least a fraction $(1+b)/2$ of the product of their total weights lies on nonincident cycle edges.
\end{enumerate}
\end{lemma}

\begin{proof}
Minimize the strictly convex quadratic $\sum_{i=0}^{m-1} \Delta_i h_i^2$ over $h_i\ge 0$ subject to $h_i+h_{i+1}\ge 1$ for all $i$ (truncating $h_i$ at $1$ cannot increase the objective). At the unique minimizer, no coordinate can equal $0$ (its neighbors would equal $1$, and simultaneously raising all zeros by $\varepsilon$ and lowering all ones by $\varepsilon$ preserves feasibility and strictly decreases the objective), nor can any coordinate equal $1$ once all $h_i>0$ (it could be lowered slightly). Hence $0<h_i<1$ for every $i$. By KKT optimality, there exist multipliers $\lambda_{i,i+1}\ge 0$ supported on tight pairs $h_i+h_{i+1}=1$ such that $\lambda_{i-1,i}+\lambda_{i,i+1}=2\Delta_i h_i$ for each $i$. For each pair $(i,i+1)$ with $\lambda_{i,i+1}>0$, form an atom allocating $x=\lambda_{i,i+1}/(2h_i)$ to edge $e_i$ and $y=\lambda_{i,i+1}/(2h_{i+1})$ to edge $e_{i+1}$. The allocations on edge $e_i$ sum to $(\lambda_{i-1,i}+\lambda_{i,i+1})/(2h_i)=\Delta_i$, and because $h_i+h_{i+1}=1$, the atom has imbalance $b=|h_i-h_{i+1}|$ with heavier endpoint at $h_{\min}=(1-b)/2$ carrying fraction $(1+b)/2$ of $S_j$.

All heavy endpoints of positive-imbalance atoms satisfy $h_i<\frac{1}{2}$ and therefore form a matching of cycle edges, since $h_i+h_{i+1}\ge 1$. Moreover, no edge with $h_i=\frac{1}{2}$ can be adjacent to a heavy endpoint. If any positive-imbalance atom exists, the remaining positions with $h_i=\frac{1}{2}$ induce disjoint paths, from each of which we may add a bipartition class carrying at least half its deficit weight to the matching; this yields $2T\ge S_{\Delta}(1+\Exp[b])$, proving~(i) (the case $\Exp[b]=0$ being immediate). Finally, if $b>b'$, the heavy endpoint of a class-$b$ atom has $h_i=(1-b)/2$, while both endpoints of a class-$b'$ atom have $h_r\le (1+b')/2<(1+b)/2$; thus their sum is strictly less than $1$, so they cannot be adjacent. Hence every heavy endpoint of class $b$ is nonincident to both edges of every class-$b'$ atom, establishing~(ii).
\end{proof}

\begin{proof}[Proof of Theorem~\ref{thm:cyclelaw}]
It remains to complete the case $m=2k+1\ge 5$ with $Y>m-1$. First, if $H\ge 2s$, then $z=(e-s)/(2s)\ge K$ maximizes the left-hand side of~\eqref{eq:weightrange} at $(e-s)^2/(4s)$. If instead $H<2s$, raise the comparison diameter by $\tau=s-H/2>0$ (replacing $D$ by $D'=D+\tau$ and each $\Delta_i$ by $\Delta'_i=\Delta_i+\tau$). Then $s'=H/2$, $e'=(m-1)s'+H$, and $e'-s'=2Ks'$. At $z=K$, the left-hand side of~\eqref{eq:weightrange} decreases by $K(m-2-K)\tau$, while $B'\le B$ because every matching has size at most $k$. Setting $A'=A+K(m-2-K)\tau$ preserves $4A'D'\ge \gamma C'$ since
\[
C'-C=(m-3)\tau(S_{\Delta}+m\tau/2)\le (m-3)\tau D' \quad\text{and}\quad 4K(m-2-K)=m(m-4)\ge \gamma(m-3),
\]
using $S_{\Delta}+m\tau/2=(S_{\Delta}+S'_{\Delta})/2\le D'$. Suppressing primes, in both cases we obtain a comparison diameter $D\ge\diam(d)$ with $D-\Delta_i=d_i$ and
\begin{equation}\label{eq:weightcenter}
e^2\le s\bigl(D+2e-B^2/D-4A\bigr), \qquad 4AD\ge \gamma C, \qquad c_0:=\frac{s}{e}\le \frac{1}{m+1}.
\end{equation}

Assume first that all $\Delta_i>0$, and apply Lemma~\ref{lem:cycleallocation}. For $0\le c\le c_0$, let $t_c(b)$ be the unique nonnegative root of
\[
(1-3c)t^2+(1-4c)t=c(1-b^2).
\]
Then $t_c(b)$ is continuous and nonincreasing in $b$, obeys $0\le t_c(b)\le c/(1-3c)\le 1/(m-2)$, and for $c>0$ satisfies $t/c=1+3t-b^2/(1+t)$. Evaluate at $c=c_0$ and put $E_0=S_{\Delta}\,\Exp[t_{c_0}]$ and $D_0=S_{\Delta}+E_0$. We claim that
\begin{equation}\label{eq:weightvariance}
S_{\Delta}\,\Exp\Bigl[\frac{b^2}{1+t_{c_0}}\Bigr]\le \frac{B^2+\gamma C}{D_0}.
\end{equation}
Indeed, expanding the identity
\[
(1+\Exp[t])\Exp\Bigl[\frac{b^2}{1+t}\Bigr]-(\Exp[b])^2
=\sum_{i<j} p_i p_j\biggl(b_i\sqrt{\frac{1+t_j}{1+t_i}}-b_j\sqrt{\frac{1+t_i}{1+t_j}}\biggr)^2
\]
over unordered pairs of atoms (with $p_j=S_j/S_{\Delta}$), terms with $b_i=b_j$ vanish, whereas for $b_i>b_j$ we have $t_i\le t_j$ and the squared parenthesis is at most $(1+\max t)b_i^2\le \gamma(1+b_i)/2$. Lemma~\ref{lem:cycleallocation} therefore bounds the sum by $\gamma C/S_{\Delta}^2$ and $(\Exp[b])^2$ by $B^2/S_{\Delta}^2$, proving~\eqref{eq:weightvariance}.

Define $f(y)=y/c_0-S_{\Delta}-3y+(B^2+\gamma C)/(S_{\Delta}+y)$ for $y\ge 0$. Taking expectations in $t/c_0=1+3t-b^2/(1+t)$ and using~\eqref{eq:weightvariance} gives $f(E_0)\ge 0$, whereas~\eqref{eq:weightcenter} gives $f(e)\le 0$. Since $B\le S_{\Delta}$ and $C\le S_{\Delta}^2/2$, we have $f'(y)\ge m+1-3-1-\gamma/2>0$ on $y\ge 0$, forcing $E_0\ge e$. By continuity from $c=0$ (where $S_{\Delta}\Exp[t_0]=0$), there exists $c\in (0,c_0]$ such that $S_{\Delta}\Exp[t_c]=e$.

For each atom $j$, put $e_j=S_j t_c(b_j)$ and $D_j=S_j+e_j$. Form the three-block partition that places the shared vertex of atom $j$'s adjacent pair into block $3$ and alternates blocks $1$ and $2$ along the complementary path. Assign the three block-pair distances $(D_j,\,D_j-x_j,\,D_j-y_j)$ to pairs $(12,13,23)$, which form a tripod with arms $(D_j+S_j b_j)/2$, $(D_j-S_j b_j)/2$, and $e_j/2$. Maximizing Zhou's quadratic mass over this three-point table (equivalently, applying the support formula~\eqref{eq:rectangle}) shows that its exact scaling factor to a normalized realizable three-point spectral table over $F$ is
\[
\Lambda_j=D_j+\frac{e_j^2}{D_j+2e_j-S_j^2 b_j^2/D_j}=D_j+ce_j,
\]
where the equality holds directly for $b_j=1$ (a cut with $e_j=0$) and follows from the defining quadratic of $t_c(b_j)$ for $b_j<1$. Because the scaled vector lies in the downward-closed convex hull of realizable three-point spectral tables over $F$, and because $\sum_j e_j=e$, $\sum_j D_j=D$, and
\[
\sum_j \Lambda_j=D+ce\le D+c_0 e=D+s=1,
\]
placing any remaining probability $1-\sum_j\Lambda_j$ on a constant partition yields a valid distribution of realizable three-block spectral tables. Each cycle edge $e_i$ receives expected distance at least $\sum_j D_j-\Delta_i=D-\Delta_i=d_i$, and the expected diameter is at least $\sum_j D_j=D$.

If some $\Delta_i=0$, perturb the deficits in Lemma~\ref{lem:cycleallocation} by adding $\eta>0$ to each $\Delta_i$ and pass to a convergent subsequence of the finitely many allocation weights as $\eta\downarrow 0$: vanishing atoms contribute zero to the bounded moments, and~\eqref{eq:weightcenter}--\eqref{eq:weightvariance} pass to the limit at fixed $c_0$ (note $S_{\Delta}=0$ is impossible, as~\eqref{eq:weightcenter} would give $e^2\le 3se$ while $e\ge (m+1)s>3s$). Finally, weighting the edge domination inequalities by $w\ge 0$ and combining with~\eqref{eq:intro-sandwich} and Proposition~\ref{prop:cyclethree} yields $\BP(C_m,w)=\Gamma_F(C_m,w)=L_3(C_m,w)$.
\end{proof}

\subsection{Uniform cycles, extremal rigidity, and cut chambers}
\label{sec:uniform}

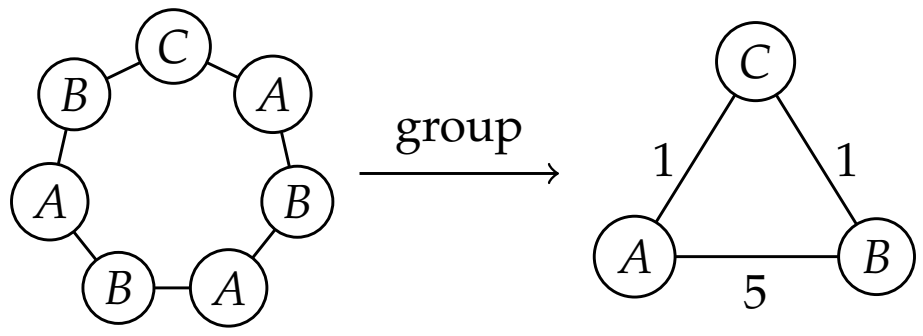
\begin{figure}[htbp]
\centering
\begin{tikzpicture}[x=1cm,y=1cm,line width=.55pt,every node/.style={font=\small}]
 \foreach \i in {0,...,6}{
   \coordinate (p\i) at ({90-360*\i/7}:0.84);
 }
 \draw (p0)--(p1)--(p2)--(p3)--(p4)--(p5)--(p6)--cycle;
 \foreach \i/\letter in {0/C,1/A,2/B,3/A,4/B,5/A,6/B}
   \node[circle,inner sep=1.7pt,draw,fill=white] at (p\i) {$\letter$};
 \draw[->] (1.23,0)--(2.55,0) node[midway,above] {group};
 \coordinate (A) at (3.05,-.55); \coordinate (B) at (4.65,-.55);
 \coordinate (C) at (3.85,.74);
 \draw (A)--node[below]{$5$}(B)--node[right]{$1$}(C)--node[left]{$1$}cycle;
 \foreach \v in {A,B,C}\node[circle,inner sep=2pt,draw,fill=white] at (\v) {$\v$};
\end{tikzpicture}
\caption{Three-block grouping of the seven-cycle $C_7$. Isolating one vertex in block $C$ and alternating $A,B$ along the complementary path produces crossing counts $(5,1,1)$, yielding the rectangular matrix multiplication tensor of dimensions $(n^5,n,n)$.}
\label{fig:group}
\end{figure}

We now specialize Theorem~\ref{thm:cyclelaw} and Theorem~\ref{thm:polygon} to unit-weight cycles and explicit weight regimes. For the uniform cycle $C_m$, the tensor $T_F(C_m,n)$ encodes the matrix-product trace $(A_0,\dots,A_{m-1})\mapsto\tr(A_0\cdots A_{m-1})$ on $n\times n$ matrices; write $r_F(m,n)=\rk_F(T_F(C_m,n))$. Isolating vertex $0$ in one block and alternating the remaining $m-1$ vertices between the other two blocks produces crossing counts $(m-2,1,1)$ (Figure~\ref{fig:group}).

\FloatBarrier

\begin{corollary}[Uniform odd cycles and finite-rank bounds]\label{cor:main}
Let $m\ge 3$ be odd and put $A=(2m-3)^2$ and $D_m=4(m-2)$. Over every field $F$:
\begin{enumerate}
\item[{\rm (i)}] $\Gamma_F(C_m)=\BP(C_m)=\dfrac{(2m-3)^2}{4(m-2)}=m-1+\dfrac{1}{4(m-2)}$;
\item[{\rm (ii)}] for every integer $n\ge 2$, $\ark_F(T_F(C_m,n))=n^{A/D_m}$ and $r_F(m,n)^{D_m}\ge n^{A}$;
\item[{\rm (iii)}] for every integer $q\ge 1$, there exist $t\ge 1$ and an exact $m$-factor decomposition of $T_F(C_m,2^t)$ over $F$ of length $r$ satisfying $r^{D_m q}\le 2^{(Aq+1)t}$.
\end{enumerate}
\end{corollary}

\begin{proof}
Part~(i) follows directly from Theorem~\ref{thm:cyclelaw} upon evaluating $U(m-2,1,1)=m-1+\frac{1}{4(m-2)}$. (A short direct proof of the upper bound also follows from Theorem~\ref{thm:polygon} and Lemma~\ref{lem:rows}: choosing diameter endpoints $a,b$ at distance $D$, placing mass $1$ at every cycle vertex and extra mass $k-1$ at each of $a,b$ gives total mass $4k-1$ and energy $M_d(u)=S+(k-1)(\rho(a)+\rho(b))+(k-1)^2 D\ge (2k-1)Y$, whence~\eqref{eq:mass} yields $4(2k-1)Y\le (4k-1)^2$; Appendix~\ref{app:local} supplies a third, completely local certificate using only $m$ fixed integer weight vectors and consecutive-quadruple Ptolemy inequalities.)

For~(ii), the finite lower bound is Lemma~\ref{lem:group} applied to the $(m-2,1,1)$ grouping. For general $n\ge 2$ and $j\ge 1$, put $p=\lfloor j\log_2 n\rfloor$; local truncation of edge alphabets yields $r_F(m,2^p)\le r_F(m,n^j)\le r_F(m,2^{p+1})$, and dividing logarithms by $j$ as $j\to\infty$ establishes $\ark_F(T_F(C_m,n))=n^{A/D_m}$. Part~(iii) follows immediately from~\eqref{eq:limit} by picking $t\ge 1$ with $t^{-1}\log_2 r_F(m,2^t)<A/D_m+1/(D_m q)$.
\end{proof}

For example, $\Gamma_F(C_5)=49/12$ sits strictly between its maximum cut $4$ and Zhou's conic upper bound $25/6$, while $\Gamma_F(C_7)=121/20$ and $\Gamma_F(C_9)=225/28$. The master identity~\eqref{eq:recur} in Theorem~\ref{thm:polygon} also completely classifies all extremal spectral pseudometrics on uniform odd cycles.

\begin{proposition}[Rigidity of extremal cycle pseudometrics]\label{prop:rigid}
Let $m=2k+1\ge 5$, set $x=m-2$, and let $d\in\mathcal{D}_{V(C_m)}$ attain the maximum perimeter $Y(d)=m-1+\frac{1}{4x}$ (in particular, any maximizing spectral pseudometric $d_\phi$ over a field $F$). Then up to cyclic relabeling and reversal, the zero-distance equivalence classes of $d$ have sizes $k,k,1$; removing the singleton leaves a path alternating between the two size-$k$ classes, whose mutual distance is
\[
p=1-\frac{1}{4x^2},
\]
while the singleton lies at distance
\[
q=\frac{1}{2}+\frac{1}{4x}
\]
from both classes. Consequently, at the all-ones weight vector $\onevec$, the right directional derivative of $\Gamma_F(C_m,\cdot)$ in any direction $v\in\mathbb{R}^m$ equals
\[
p\sum_{i=0}^{m-1} v_i+(q-p)\min_{0\le i<m}(v_{i-1}+v_i),
\]
and the left directional derivative replaces the minimum by the maximum.
\end{proposition}

\begin{proof}
Since $Y>m-1$, we have $D=\diam(d)>0$. For every diametral pair $\{a,b\}$, the mass test $u=\onevec+(k-1)(\mathbf{1}_a+\mathbf{1}_b)$ in the direct proof of Corollary~\ref{cor:main}(i) must saturate~\eqref{eq:mass}, forcing equality in Theorem~\ref{thm:polygon} and Lemma~\ref{lem:rows}. Rescaling by $D$ in~\eqref{eq:recur} gives $\Phi=0$ and hence $\mathcal{P}=0$; thus $(D-d_e)(D-d_f)=0$ for every pair of nonincident cycle edges $e,f$. Consequently, the cycle edges strictly shorter than $D$ pairwise meet, so for $m\ge 5$ there are at most two such edges, and if there are two they must be adjacent.

Returning to the original scale, since every entry of $u$ is strictly positive and $u$ saturates~\eqref{eq:mass}, first-order stationarity yields
\begin{equation}\label{eq:stationary}
\rho(i)+(k-1)\bigl(d(i,a)+d(i,b)\bigr)=\frac{4k-1}{2} \qquad (i\in\mathbb{Z}/m\mathbb{Z})
\end{equation}
for every diametral pair $\{a,b\}$. If two adjacent cycle edges $ab,bc$ both have length $D$, subtracting~\eqref{eq:stationary} for $ab$ and $bc$ gives $d(i,a)=d(i,c)$ for all $i$, whence $d(a,c)=0$. Along any path of length-$D$ edges, vertices of equal parity therefore collapse to zero distance. Not all $m$ edges can have length $D$ (oddness would collapse the entire cycle to a point), and if only one edge were shorter than $D$, its endpoints would coincide along the complementary even path, giving $Y=(m-1)D\le m-1$, a contradiction. Thus exactly two adjacent edges are shorter than $D$, and the complementary path of $2k-1$ length-$D$ edges collapses into two alternating classes of size $k$ at mutual distance $p=D$, leaving one vertex at distances $q_1,q_2$ from the two classes. Evaluating~\eqref{eq:stationary} on these three classes (with aggregate masses $x,x,1$) yields $xp+q_1=xp+q_2=x(q_1+q_2)=(2x+1)/2$, whose unique solution is the stated $p$ and $q_1=q_2=q>0$.

Because every cyclic rotation of this three-point table is realizable over $F$ and $\Gamma_F(C_m,\cdot)$ is the support function of a compact set, its one-sided directional derivative at $\onevec$ in direction $v$ is the maximum (for the right derivative) or minimum (for the left derivative) of $v\cdot d$ over all maximizing tables $d$, giving the displayed expression since $q-p<0$.
\end{proof}

We next record two weight-domain consequences: a general upper bound for any graph containing a positive odd cycle, and an elementary cut-equality criterion on weighted cycles.

\begin{theorem}[Two light edges in a general graph]\label{thm:light}
Let $(G,w)$ be a weighted graph of total weight $W$ containing a positive-weight odd cycle $C_m$ whose two smallest edge weights (counted with multiplicity) are $a\le b$. Then over every field $F$,
\[
\Gamma_F(G,w)\le \BP(G,w)\le U(W-a-b,\,b,\,a).
\]
If $G=C_m$ and the two lightest edges are adjacent, equality holds; for an arbitrary ordering on $C_m$, one has $\Gamma_F(C_m,w)=L_3(C_m,w)=W-a$ whenever $a(W-a)\le b(W-a-b)$.
\end{theorem}

\begin{proof}
Fix $d\in\mathcal{D}_V$ with global diameter $D$, cycle perimeter $Y$, and length $z$ on a marked cycle edge $e^*$ of weight $a$. Put $m=2k+1$, $n=m-1$, $H=W-a$, and $h=H/b-n\ge 0$. For any $s\in[0,1]$, assign mass $1$ to each endpoint of $e^*$, mass $1+s$ to the remaining $m-2$ vertices of $C_m$, and additional mass $(k-1)(1-s)+h\ge 0$ to each endpoint of a global diameter pair $\{P,Q\}$. By Theorem~\ref{thm:polygon}, the pairwise sum $S$ within $C_m$ satisfies $S\ge kY-k(k-1)D$. Moreover, for each cycle vertex $i$ outside $e^*$, shortcutting the alternating edges through $i$ bounds $\rho_{C_m}(i)$ from below; summing over all $i\notin e^*$ yields $\sum_{i\notin e^*}\rho_{C_m}(i)\ge kY-z$. Expanding the cycle mass energy $u_{\mathrm{cyc}}$ via $M_d(u_{\mathrm{cyc}})=(1-s^2)S+s(1+s)\sum_{i\notin e^*}\rho_{C_m}(i)+s^2 z$ therefore gives at least $k(1+s)Y-sz-k(k-1)(1-s^2)D$. Combining this with Lemma~\ref{lem:exterior} ($r_P+r_Q\ge Y+D$) and the triangle inequality on the extra mass $s$ off $e^*$, Zhou's inequality~\eqref{eq:mass} with total mass $2n-1+s+2h$ yields
\begin{equation}\label{eq:marked}
(n-1+h+s)(Y+hD)-sz\le \frac{(2n-1+s+2h)^2}{4} \qquad (0\le s\le 1).
\end{equation}
The total weighted score of $(G,w)$ is at most $b(Y+hD)-(b-a)z$ because the remaining weight $W-a-nb=bh$ outside the marked cycle costs at most $bhD$. Setting $s=(b-a)(H-b)/(ab)$ when $aH>b(H-b)$ (so $s\in[0,1)$) and $s=1$ otherwise substitutes into~\eqref{eq:marked} to give $U(H-b,b,a)$ and $H=W-a$, respectively. The remaining assertions follow immediately from Proposition~\ref{prop:cyclethree}.
\end{proof}

\begin{corollary}[Cut-equality chamber on weighted odd cycles]\label{cor:chamber}
Let $m\ge 7$ be odd, and let $w\in\mathbb{R}_{>0}^m$ have minimum weight $a=\min_i w_i>0$. If $w_{i-1}+w_i\ge \frac{9}{4}a$ for every $i\in\mathbb{Z}/m\mathbb{Z}$, then over every field $F$,
\[
\Gamma_F(C_m,w)=L_3(C_m,w)=\MC(C_m,w)=W-a.
\]
In particular, on any odd cycle $C_m$ ($m\ge 7$) with positive integer weights whose minimum satisfies $a\in\{1,2,3,4\}$, $\Gamma_F(C_m,w)$ equals $W-a$ whenever no two minimum-weight edges are adjacent.
\end{corollary}

\begin{proof}
By positive homogeneity, rescale so that $a=1$. By Theorem~\ref{thm:cyclelaw}, it suffices to verify that $U(W-x-y,x,y)\le W-1$ for every adjacent pair $(x,y)=(w_{i-1},w_i)$. Assume without loss of generality that $1\le x\le y$, with $x+y\ge \frac{9}{4}$, and put $Z=W-x-y$. Because the remaining $m-2\ge 5$ edges form a path of length at least $5$, pairing the first two and last two edges gives $Z\ge \frac{9}{4}+\frac{9}{4}+1=\frac{11}{2}$. Let $p\ge q\ge r$ be the sorted triple of $\{Z,x,y\}$, so $r\ge 1$ and $W-1=p+q+r-1$.

If $(p+q)r\le pq$, then $U(p,q,r)=p+q\le p+q+r-1$. Suppose instead that $(p+q)r>pq$. Using the identity $(p+q)r-pq=r^2-(p-r)(q-r)\le r^2$, we have
\[
U(p,q,r)-(p+q)=\frac{\bigl[(p+q)r-pq\bigr]^2}{4pqr}\le \frac{r^3}{4pq}.
\]
We check three ranges for $r$:
\begin{itemize}
\item If $r\ge \frac{4}{3}$, then $pq\ge r^2$ yields $\frac{r^3}{4pq}\le\frac{r}{4}\le r-1$.
\item If $\frac{9}{8}\le r<\frac{4}{3}$, then $Z\ge\frac{11}{2}$ forces $p\ge\frac{11}{2}$ and $q\ge r$, so $\frac{r^3}{4pq}\le\frac{r^2}{22}<r-1$ (as $22(r-1)-r^2$ increases on $[\frac{9}{8},\frac{4}{3}]$ and equals $\frac{95}{64}>0$ at $r=\frac{9}{8}$).
\item If $1\le r<\frac{9}{8}$, then $Z\ge\frac{11}{2}$ forces $r=x$, $p\ge\frac{11}{2}$, and $q\ge\min(Z,y)\ge\frac{9}{4}-r>\frac{9}{8}$, so $pq\ge \frac{99}{16}$. Moreover,
\[
0<(p+q)r-pq\le r^2-\Bigl(\frac{11}{2}-r\Bigr)\Bigl(\frac{9}{4}-2r\Bigr)
=(r-1)\Bigl(\frac{49}{4}-r\Bigr)-\frac{1}{8}<12(r-1),
\]
whence $U(p,q,r)-(p+q)<\frac{144(r-1)^2}{4(99/16)\cdot 1}=\frac{64}{11}(r-1)^2<\frac{8}{11}(r-1)\le r-1$.
\end{itemize}
Thus $U(p,q,r)\le W-1$ in every case. For positive integer weights with $a\in\{1,2,3,4\}$ and no two adjacent edges of weight $a$, every adjacent pair satisfies $w_{i-1}+w_i\ge 2a+1\ge \frac{9}{4}a$.
\end{proof}

Table~\ref{tab:cycles} illustrates the effect of edge ordering and the sharpness of the $\frac{9}{4}a$ threshold on weighted $5$- and $7$-cycles. The permutations $(1,1,2,2,2)$ and $(1,2,1,2,2)$ share the same multiset of weights yet yield distinct rank exponents ($169/24$ versus $7$), while the $7$-cycle $(5,6,5,6,5,6,6)$ has $a=5$ and adjacent sum $11<\frac{45}{4}$, attaining $34+\frac{1}{840}>34$ and witnessing the tightness of Corollary~\ref{cor:chamber} at integer weights.

\begin{table}[htbp]
\centering
\caption{Exact asymptotic rank exponents for representative weighted $5$- and $7$-cycles, illustrating sensitivity to cyclic ordering and sharpness of the cut-chamber threshold $9a/4$.}
\label{tab:cycles}
\begin{tabular}{@{}lll@{}}
\toprule
Cyclic edge weights & Maximum cut $M$ & Asymptotic rank exponent $\Gamma_F=L_3$ \\
\midrule
$(1,1,2,2,2)$ & $7$ & $169/24$ \\
$(1,2,1,2,2)$ & $7$ & $7$ \\
$(3,4,3,4,4)$ & $15$ & $7921/528$ \\
$(4,5,5,5,5)$ & $20$ & $961/48$ \\
$(5,6,5,6,5,6,6)$ & $34$ & $34+1/840$ \\
$(8,11,9,12,10,11,10)$ & $63$ & $63$ \\
\bottomrule
\end{tabular}
\end{table}

\subsection{Bipartite attachments and layered graphs}
\label{sec:extensions}

Because the dominating distribution in Theorem~\ref{thm:cyclelaw} controls the global diameter $D$ simultaneously with all cycle edges, it extends directly to bipartite graphs attached to an odd cycle and to graphs layered cyclically over an odd cycle.

\begin{corollary}[Odd cycle with bipartite attachments]\label{cor:attachments}
Attach arbitrary bipartite graphs to a simple odd cycle $C_m$ by identifying at most one vertex of each bipartite piece with a vertex of $C_m$, keeping their off-cycle vertex sets pairwise disjoint. Let $W$ be the total weight on $C_m$, let $H$ be the total weight of all off-cycle edges, and let $A_i$ be the total weight of off-cycle edges incident to cycle vertex $v_i$. Put
\[
L_{\mathrm{free}}=\max_{0\le i<m} U\bigl(W+H-w_{i-1}-w_i,\,w_{i-1},\,w_i\bigr).
\]
Then over every field $F$, $L_3(G,w)\le\Gamma_F(G,w)\le\BP(G,w)\le L_{\mathrm{free}}$, where
\[
L_3(G,w)=\max_{\substack{0\le i<m\\ \epsilon\in\{0,1\}}}
U\bigl(W+H-w_{i-1}-w_i-A_i,\,w_{i-1}+\epsilon A_i,\,w_i+(1-\epsilon)A_i\bigr).
\]
In particular, $\Gamma_F(G,w)=L_3(G,w)=L_{\mathrm{free}}$ whenever the bipartite components are disjoint from $C_m$, or more generally whenever some maximizer $i$ of $L_{\mathrm{free}}$ satisfies $A_i=0$.
\end{corollary}

\begin{proof}
Assume all cycle weights are positive (continuity covers zero weights). A binary cut misses only a lightest cycle edge. Suppose a three-block partition $c$ and a three-point spectral table $q$ (with sorted pair distances $q_1\ge q_2\ge q_3$) strictly beat this cut; then the unweighted losses $\sum_{e\in C_m}(1-q_e)$ sum to strictly less than $1$. Relabeling the two blocks attaining $q_1$ as blocks $1$ and $2$, there can be no monochromatic cycle edge, and parity of block degrees implies that each of the three color pairs appears an odd number of times on $C_m$. If either of the two shorter pairs $13$ or $23$ appeared at least three times, the cycle score would be at most $m-5+q_1+3q_2+q_3\le m-5+2q_1+2q_2+q_3\le m-5+U(2,2,1)=m-1$, contradicting loss less than $1$. Hence exactly one cycle vertex $v_i$ receives block $3$, while the remaining cycle vertices alternate between blocks $1$ and $2$.

Conditioned on this cycle coloring, any bipartite piece attached at a vertex of color $c_0$ can give all edges incident to its root the length $\max_{j\ne c_0} q_{c_0,j}$ and all remaining edges the maximum length $q_1$ (by coloring the opposite bipartition shore with the best neighbor of $c_0$ and the remaining vertices on the root shore with the other color in $\{1,2\}$). At the unique block-$3$ vertex $v_i$, the entire incident weight $A_i$ is therefore assigned to one of the two pairs $13$ or $23$, giving the stated formula for $L_3(G,w)$ (which also covers the cut bound via a lightest edge). For the upper bound, every off-cycle edge has length at most $D=\diam(d)$, so the total off-cycle contribution is at most $HD$. Applying Theorem~\ref{thm:cyclelaw} dominates the cycle edges and the ambient diameter $D$ simultaneously by three-block spectral tables, proving $\BP(G,w)\le L_{\mathrm{free}}$.
\end{proof}

For instance, taking cycle weights $(1,4,\frac{101}{100},\frac{101}{100},4)$ on $C_5$ and attaching a unit pendant edge away from the vertex between the two $\frac{101}{100}$-edges gives $\Gamma_F=L_3=\frac{4414201}{400000}>\frac{551}{50}=\MC$, and both the maximizing pair and the equality persist throughout an open neighborhood of independent weight perturbations.

\begin{corollary}[Layered graphs over an odd cycle]\label{cor:cyclelayers}
Let $V_0,V_1,\dots,V_{m-1}$ be pairwise disjoint vertex layers indexed cyclically for odd $m\ge 3$, with positive edges only between consecutive layers $V_i$ and $V_{i+1}$ and total inter-layer weight $a_i>0$. If the normalized inter-layer edge weights form the consecutive pairwise marginals of a single random tuple $(X_0,X_1,\dots,X_{m-1})\in\prod_{i=0}^{m-1} V_i$ (including the closing pair $(X_{m-1},X_0)$), then over every field $F$,
\[
\Gamma_F(G,w)=\BP(G,w)=L_3(G,w)=L_3(C_m,a).
\]
\end{corollary}

\begin{proof}
For any $d\in\mathcal{D}_V$, the weighted edge sum equals $\Exp\bigl[\sum_{i=0}^{m-1} a_i\, d(X_i,X_{i+1})\bigr]$, and for every realization $(X_0,\dots,X_{m-1})$, Theorem~\ref{thm:cyclelaw} bounds the cycle sum inside the expectation by $L_3(C_m,a)$. Conversely, lifting an optimal three-block partition of $(C_m,a)$ uniformly to each layer attains $L_3(C_m,a)$.
\end{proof}

To construct unbounded-size examples genuinely requiring two edge deletions to become bipartite, let each layer $V_i$ consist of two vertices $\{0,1\}$, include all four edges between $V_i$ and $V_{i+1}$ for $0\le i\le m-2$, and include only the two matching edges $\{(0,0),(1,1)\}$ between $V_{m-1}$ and $V_0$. Taking any full-support law on $(X_0,\dots,X_{m-2})$ and setting $X_{m-1}=X_0$ realizes arbitrary inter-layer weights $a_i>0$. Deleting the two closing edges makes the graph bipartite, while the two constant-coordinate odd cycles are edge-disjoint, so one deletion never suffices.

\section{Odd-Cycle Packings and Near-Bipartite Graphs}
\label{sec:deletions}

We now move from single cycles to graphs that become bipartite after deleting at most two edges---a class containing all odd cycles (one deletion), books, theta graphs, and $K_4$ (two deletions). We first develop general odd-cycle packing bounds and classify unit-weight single-deletion graphs (Section~\ref{sec:packings}), establish the uniform $100/93$ approximation ratio on the entire two-deletion class (Section~\ref{sec:ratio}), prove exactness on the cone of nonnegative triangle-indicator combinations (Section~\ref{sec:trianglecone}), record explicit exponents for weighted books, theta graphs, and subdivisions of $K_4$ (Section~\ref{sec:examples}), and finally prove exactness for all Ptolemy--Zhou pseudometrics taking at most two positive values (Section~\ref{sec:twoprices}).

\subsection{Fractional odd-cycle packings and single-deletion classification}
\label{sec:packings}

A \emph{fractional odd-cycle packing} of a weighted graph $(G,w)$ assigns nonnegative coefficients $\lambda_j>0$ to simple odd cycles $C_j$ of $G$ such that $\sum_{j:\,e\in C_j}\lambda_j\le w_e$ for every edge $e\in E$. Put $b=\sum_j \lambda_j$ for its total weight and $m_j=|C_j|$ for the length of $C_j$. Because the exterior-diameter inequality (Lemma~\ref{lem:exterior}) applies to every cycle simultaneously using the same global diameter endpoints $P,Q$, overlapping odd cycles can be bounded jointly.

\begin{theorem}[Common-diameter packing bound]\label{thm:pack}
Let $(G,w)$ be a weighted graph of total edge weight $W$ admitting a fractional odd-cycle packing of total weight $b>0$. If every packed cycle $C_j$ satisfies $m_j\le 2W/b-3$, then over every field $F$,
\[
\Gamma_F(G,w)\le \BP(G,w)\le W-b+\frac{b^2}{4(W-2b)}.
\]
Equality holds whenever $G$ admits a three-block partition with crossing weights $(W-2b,\,b,\,b)$.
\end{theorem}

\begin{proof}
Fix a pseudometric $d\in\mathcal{D}_V$ with global diameter $D=\diam(d)$ attained at $\{P,Q\}$. On any odd cycle $C$ of length $m=2k+1$ and perimeter $Y$, place mass $1$ at each cycle vertex and additional mass $t\ge 0$ at each of $P$ and $Q$ (accumulating masses if $P$ or $Q$ lies on $C$). Applying Theorem~\ref{thm:polygon} ($S\ge kY-k(k-1)D$) and Lemma~\ref{lem:exterior} ($r_P+r_Q\ge Y+D$), Zhou's inequality~\eqref{eq:mass} with total mass $m+2t$ gives
\[
(k+t)\bigl(Y+(t-k+1)D\bigr)\le \frac{(m+2t)^2}{4}.
\]
Set $H=W/b$ and $t_j=H-(m_j+3)/2\ge 0$ for each packed cycle $C_j$ of length $m_j=2k_j+1$. Since $k_j+t_j=H-2$ and $t_j-k_j+1=H-m_j$, dividing by $H-2>0$ yields
\[
Y_j+(H-m_j)D\le \frac{(2H-3)^2}{4(H-2)}=H-1+\frac{1}{4(H-2)}.
\]
The unallocated edge weight outside the packing equals $W-\sum_j\lambda_j m_j=\sum_j\lambda_j(H-m_j)$, and every unallocated edge has length at most $D$. Multiplying each cycle inequality by $\lambda_j$ and summing over $j$ gives
\[
\sum_{e\in E} w_e d_e\le \sum_j \lambda_j\bigl(Y_j+(H-m_j)D\bigr)
\le b\Bigl(H-1+\frac{1}{4(H-2)}\Bigr)=W-b+\frac{b^2}{4(W-2b)}.
\]
More generally, any allocation $H_j\ge (m_j+3)/2$ with $\sum_j\lambda_j H_j=W$ gives the upper bound $W-b+\sum_j \lambda_j/[4(H_j-2)]$, which by convexity is minimized at $H_j=\max\{(m_j+3)/2,h\}$ for a common threshold $h$. The equality condition follows directly from Lemma~\ref{lem:group} and~\eqref{eq:rectangle}.
\end{proof}

On graphs whose positive support becomes bipartite after deleting at most two edges, the maximum fractional odd-cycle packing matches the cut deficit $W-M$ exactly.

\begin{proposition}[Two-commodity packing duality]\label{prop:hu}
Let $(G,w)$ be a weighted graph with $w\ge 0$ whose positive-weight support becomes bipartite after deleting at most two edges, and let $M=\MC(G,w)$. Then the maximum total weight of a fractional odd-cycle packing of $(G,w)$ equals $W-M$.
\end{proposition}

\begin{proof}
Every odd cycle misses at least one edge of any maximum cut, so every fractional odd-cycle packing has total weight at most $W-M$. For the reverse inequality, fix a bipartition of $G$ whose monochromatic positive edges number at most two; the zero-edge case is trivial and the one-edge case reduces directly to the max-flow/min-cut theorem. Suppose there are two monochromatic edges $e_1,e_2$ of weights $p,q>0$. In the remaining bipartite graph $H$, let $h_1$, $h_2$, and $h_{12}$ denote the minimum capacities of cuts separating the endpoints of $e_1$, of $e_2$, and of both $e_1$ and $e_2$, respectively (taking $\infty$ if no such cut exists). Flipping an arbitrary subset of vertices relative to the reference bipartition shows that
\[
W-M=\min\{p+q,\,h_1+q,\,h_2+p,\,h_{12}\}.
\]
By Hu's undirected two-commodity flow theorem~\cite{hu} (which here follows directly by routing single-commodity flows for the sum and difference of the two terminal demands and taking half their sum and half their difference, using $|(x+y)/2|+|(x-y)/2|=\max(|x|,|y|)$ edgewise), there exist simultaneous flows in $H$ of values $u,v\ge 0$ between the respective endpoint pairs of $e_1,e_2$ whenever $u\le\min(p,h_1)$, $v\le\min(q,h_2)$, and $u+v\le h_{12}$. Since symmetric differences of cuts obey $h_{12}\le h_1+h_2$ whenever both are finite, the maximum feasible $u+v$ equals $\min\{p+q,h_1+q,h_2+p,h_{12}\}=W-M$. Decomposing each flow into simple paths in $H$ and closing each path with $e_1$ or $e_2$ yields a fractional odd-cycle packing of total weight $W-M$.
\end{proof}

Combining Theorem~\ref{thm:pack} with two cycle-covering loss estimates gives a complete classification for positive integer weights when deleting a single unit-weight edge leaves a bipartite graph.

\begin{theorem}[Single unit-weight deletion classification]\label{thm:unit}
Let $(G,w)$ be a weighted graph whose positive-weight support is connected and nonbipartite, with all positive weights integers, such that deleting a single edge of weight $1$ renders the support bipartite. Write $W=\sum_{e\in E} w_e$.
\begin{enumerate}
\item[{\rm (i)}] If there exists a vertex $x$ of weighted degree $2$ whose deletion renders the positive support bipartite, then over every field $F$,
\[
\Gamma_F(G,w)=\BP(G,w)=L_3(G,w)=W-1+\frac{1}{4(W-2)}.
\]
\item[{\rm (ii)}] In every other case, $\Gamma_F(G,w)=\BP(G,w)=L_3(G,w)=\MC(G,w)=W-1$.
\end{enumerate}
\end{theorem}

\begin{proof}
Replace each integer weight $w_e$ by $w_e$ parallel unit-weight edges (which preserves the graph tensor up to basis reindexing). Because the support is nonbipartite and deleting one unit edge $uv$ makes it bipartite, we have $M=W-1$. Selecting any single odd cycle with weight $1$ in Theorem~\ref{thm:pack} yields the upper bound $\BP\le W-1+\frac{1}{4(W-2)}$. In case~(i), a weighted-degree-$2$ odd-cycle transversal $x$ lies on an odd cycle and therefore has its two distinct neighbors on opposite shores of its connected component in $G-x$; placing $x$ alone in block $3$ and coloring $G-x$ properly in blocks $1,2$ gives crossing weights $(W-2,1,1)$ and matching lower bound $U(W-2,1,1)=W-1+\frac{1}{4(W-2)}$.

For case~(ii), assume no vertex of weighted degree $2$ meets every odd cycle. We use two loss estimates for a fixed $d\in\mathcal{D}_V$. First, consider a triangle on vertices $x_0,x_1,x_2$ together with three (possibly repeated) attached vertices $y_0,y_1,y_2$, and set $s_i=d(x_i,y_i)$ with $s_p=\max_i s_i$. We claim that the triangle perimeter $Y_3$ satisfies
\begin{equation}\label{eq:attached-tri}
Y_3\le 2+\sum_{i\ne p}(1-s_i).
\end{equation}
Sorting the three triangle edges and considering the three choices of $p$, exact rational polynomial identities over the $15$ pair distances express the nonnegative slack $2+\sum_{i\ne p}(1-s_i)-Y_3$ (under the assumption that it is negative, yielding one additional nonnegative affine factor) as a positive rational linear combination of Ptolemy slacks and products of affine triangle, unit-cap, and mass slacks, with $56$, $64$, and $64$ positive terms respectively (verified identically across all $408$ constant, linear, and quadratic coefficients). Applying~\eqref{eq:attached-tri} inductively along a fan rooted at a cycle vertex $v_p$ with maximal attachment $s_p$ on any odd cycle $C_m$ ($m=2k+1$) and shortcutting the intermediate fan edges yields
\begin{equation}\label{eq:attached-cycle}
Y_m\le m-1+\sum_{i\ne p}(1-s_i).
\end{equation}
Thus attaching a distinct unit edge at every vertex of $C_m$ forces total edge deficit $\sum_e (1-d_e)\ge 1$.

Second, suppose $h\le m-1$ additional unit edges cover all $m$ vertices of $C_m$ and have total length $T$. Assigning each vertex mass equal to its incident count among these $h$ edges (total mass $2h$) and applying the triangle inequality together with the endpoint row bound in~\eqref{eq:mass} yields $2h^2\ge hY_m+(2h-m+1)T$. Since $T\le h$, this implies $Y_m+T\le m+h-1$, again giving total deficit at least $1$.

Now return to our unit-edge multigraph $G$. If $G$ contains a unique odd cycle $C$, either some edge of $C$ has a parallel spare copy (which covers two cycle vertices and can be completed using at most $m-1$ total exterior edges to cover $C$, unless an uncovered vertex has degree $2$, which is impossible since it would be a degree-$2$ transversal), or else every cycle edge is simple and every cycle vertex has a distinct incident exterior edge, so~\eqref{eq:attached-cycle} forces a total deficit of at least $1$. If $G$ contains multiple odd cycles, let $uv$ be the deleted unit edge and consider the bipartite graph $H=G-uv$. Call an edge of $H$ \emph{mandatory} if it belongs to every $u\text{--}v$ path in $H$. Assigning capacity $2$ to mandatory edges and capacity $1$ to all other edges of $H$, the integral max-flow/min-cut theorem routes two $u\text{--}v$ paths $P_1,P_2$ in $H$ sharing only mandatory edges. Closing $P_1$ and $P_2$ with $uv$ forms two odd cycles $C_1,C_2$. Build two auxiliary multigraphs $G_1,G_2$ by doubling the edges exclusive to $P_r$ in $G_r$, retaining common edges once, and keeping all unused edges; the average of the edge weight vectors of $G_1$ and $G_2$ equals $w$. In each $G_r$, every exclusive edge of $C_r$ has a parallel spare copy covering its two endpoints, while any cycle vertex incident only to mandatory edges lies on every odd cycle of $G$ and therefore---by our assumption against a degree-$2$ transversal---possesses an additional incident exterior edge. Hence each $G_r$ admits an edge cover of $C_r$ with at most $|C_r|-1$ additional unit edges (or disjoint attachments at all vertices), forcing a total deficit of at least $1$. Averaging the two bounds yields $\sum_e w_e d_e\le W-1$.
\end{proof}

An analogous packing argument treats integer cut deficit $2$ and provides a sufficient weighted-degree condition for exact cut equality.

\begin{theorem}[Cut deficit two]\label{thm:two}
Let $(G,w)$ be a weighted graph with positive integer edge weights satisfying $\MC(G,w)=W-2$. Then over every field $F$,
\[
\Gamma_F(G,w)\le \BP(G,w)\le W-2+\frac{1}{W-4}.
\]
Equality holds whenever $G$ admits a three-block partition with crossing weights $(W-4,\,2,\,2)$.
\end{theorem}

\begin{proof}
Splitting integer weights into parallel unit edges, Proposition~\ref{prop:hu} gives a fractional odd-cycle packing of total weight $b=2$. To apply Theorem~\ref{thm:pack}, we need every packed cycle to have length at most $2W/2-3=W-3$. If some packed odd cycle $C$ has length $m>W-3$, then at most two unit edges lie outside $C$. Because adding a tree attachment or a single ear to an odd cycle leaves a graph made bipartite by one deletion (contradicting $W-M=2$), the two exterior edges must both be chords of $C$ (allowing parallel copies) whose induced even arcs along $C$ are edge-disjoint (otherwise deleting a shared edge of the two arcs would leave both chords joining opposite shores). Closing each chord with its even arc yields two odd cycles of length at most $W-3$ forming an alternative packing of weight $2$. Applying Theorem~\ref{thm:pack} completes the proof.
\end{proof}

\begin{remark}[Degree criterion for cut equality]\label{rem:degcut}
Suppose a weighted graph $(G,w)$ admits a fractional odd-cycle packing $\{\lambda_j,C_j\}$ of total weight $b=W-M$, and let $t_v=\sum_{j:\,v\in C_j}\lambda_j$ be the packed mass incident to vertex $v$. If $\deg_w(v)-2t_v\ge b$ holds for every vertex $v$ belonging to at least one packed cycle, then $\Gamma_F(G,w)=\BP(G,w)=L_3(G,w)=M$: allocating fraction $\lambda_j/b$ of all unpacked edges to cycle $C_j$ supplies each vertex of $C_j$ with at least $\lambda_j$ extra incident weight, so the fractional version of~\eqref{eq:attached-cycle} forces a deficit of at least $\lambda_j$ on $C_j$, summing to $b$. In particular, on any two-deletion graph, minimum weighted degree $3b$ ensures $\Gamma_F=M$ (for unit weights and cut deficit $2$, minimum degree $6$ suffices).
\end{remark}

\subsection{A uniform approximation bound on the two-deletion class}
\label{sec:ratio}

We now show that across the entire class of graphs made bipartite by deleting at most two edges---with arbitrary nonnegative real weights and arbitrary vertex count---the three-block lower bound $L_3(G,w)$ approximates $\Gamma_F(G,w)$ within a universal constant factor $100/93$. We first construct a single randomized three-block partition that jointly controls both individual edges and triangles.

\begin{lemma}[Joint control on edges and triangles]\label{lem:colbudget}
Let $G=(V,E)$ be a finite simple graph made bipartite by deleting at most two edges, and let $d\in\mathcal{D}_V$ (in particular, any realizable spectral pseudometric over a field $F$). There exists a random three-block partition of $V$, missing at most one edge of $G$ in every outcome, such that simultaneously for every edge $e\in E$ and every triangle $T\subseteq G$,
\begin{equation}\label{eq:jointbudget}
\Prb(e\text{ is monochromatic})\le 1-d_e,\qquad
\Prb(T\text{ is not rainbow})\le 2\Bigl(\frac{9}{4}-\sum_{e\in T} d_e\Bigr).
\end{equation}
\end{lemma}

\begin{proof}
If $G$ admits a proper $3$-coloring, choosing it deterministically satisfies~\eqref{eq:jointbudget} trivially (since $d_e\le 1$ and $\sum_{e\in T} d_e\le 9/4$ by uniform mass on $T$). Otherwise, fix a bipartition of $G$ with at most two monochromatic edges. Those two edges cannot lie on the same shore (otherwise their endpoints would have an independent cover of size at most $2$ on that shore, which could be recolored to color $3$), so they lie on opposite shores, say $01$ and $23$. Moreover, all four cross edges between $\{0,1\}$ and $\{2,3\}$ must be present (otherwise recoloring the endpoints of a missing cross pair to color $3$ would give a proper $3$-coloring). Hence $K=\{0,1,2,3\}$ induces a $K_4$, every exterior triangle of $G$ contains either $01$ or $23$, and for each of the six edges $e\subset K$ there is a global three-block partition whose \emph{sole} monochromatic edge in $G$ is $e$ (obtained by recoloring one endpoint of an exceptional edge, or one endpoint of each).

For each triangle $T$, set $a_T=\frac{9}{2}-2\sum_{e\in T} d_e\ge 0$, and write $a_i=a_{K\setminus\{i\}}$ ($i\in K$). For each edge $e\subset K$, define the cap $b_e=1-d_e$, except that on $e\in\{01,23\}$ we replace $b_e$ by $\min\{1-d_e,\,\min_{T\supset e,\,T\not\subset K} a_T\}$. A distribution over the six single-missed-edge partitions satisfying~\eqref{eq:jointbudget} corresponds to a fractional matching of total weight $1$ on the complementary pairs of $K$ with vertex caps $a_i$ and edge caps $b_e$. By linear programming duality, such a distribution exists if and only if every $y\in [0,1]^4$ satisfies
\begin{equation}\label{eq:lpdual}
\sum_{i\in K} y_i a_i+\sum_{\substack{e\subset K\\ |e|=2}} b_e\pos{1-\sum_{i\in K\setminus e} y_i}\ge 1.
\end{equation}
The extreme points of the subdivision of $[0,1]^4$ by the pair-sum hyperplanes have coordinates in $\{0,\frac{1}{2},1\}$ with either zero or at least three coordinates equal to $\frac{1}{2}$ (a fractional component requires an odd cycle of tight pair-sum equations).

For $y=0$, note that on any exterior triangle $T$ sharing an exceptional edge $e_0$, we have $2\sum_{e\in T} d_e-d_{e_0}\le U(2,2,1)=4$ (by Corollary~\ref{cor:four}(i)), so $a_T+d_{e_0}\ge\frac{1}{2}$ and hence $b_{e_0}+d_{e_0}\ge\frac{1}{2}$ for both $e_0\in\{01,23\}$. Uniform mass on $K$ in~\eqref{eq:mass} gives $\sum_{e\subset K} d_e\le 4$, whence $\sum_{e\subset K} b_e\ge b_{01}+d_{01}+b_{23}+d_{23}\ge 1$.

For every remaining extreme point $y\ne 0$, at most one of $01$ and $23$ carries a positive multiplier $\pos{1-\sum_{i\in K\setminus e} y_i}$ in~\eqref{eq:lpdual}. Selecting either the individual cap $1-d_e$ or the cap $a_T$ of a single exterior triangle on that exceptional edge involves at most five vertices of $G$ and puts the left-hand side of~\eqref{eq:lpdual} in the form $B-\sum k_e d_e$ with $k_e\ge 0$. By Theorem~\ref{thm:five}, $\sum k_e d_e\le L_3(k)$, so it suffices to check $B-L_3(k)\ge 1$ over the eight symmetry orbits of $2y$ (up to permutations within and between shores) and the three choices of exterior page (none, on $01$, or on $23$). Exact rational evaluation of~\eqref{eq:rectangle} over all $3^4$ partitions (fixing the color of vertex $0$) gives the exact minima
\[
\begin{array}{c|rrrrrrrr}
2y & 0002 & 0022 & 0111 & 0202 & 0222 & 1111 & 1112 & 2222 \\ \hline
\min\bigl(B-L_3(k)\bigr) & 1 & 1 & 1 & 1 & 1 & 1 & 7/6 & 2,
\end{array}
\]
verifying~\eqref{eq:lpdual} in every case. (Note that both constants $1$ and $2$ in~\eqref{eq:jointbudget} are individually sharp on $K_4$: a balanced binary cut of $K_4$ forces the triangle coefficient $2$, while an equilateral three-block table with one repeated pair on $K_4$ forces the edge coefficient $1$.)
\end{proof}

\begin{theorem}[Uniform comparison on the two-deletion class]\label{thm:globalratio}
Let $(G,w)$ be a finite simple graph with $w\ge 0$ whose positive-weight support becomes bipartite after deleting at most two edges, let $M=\MC(G,w)$, and let $b=W-M$. Over every fixed field $F$,
\[
L_3(G,w)\le \Gamma_F(G,w)\le \BP(G,w)\le \frac{100}{93}\,L_3(G,w),
\qquad \Gamma_F(G,w)\le M+\frac{b^2}{2M}\quad (M>0).
\]
If in addition the positive support contains no $K_4$, then $\Gamma_F(G,w)\le \BP(G,w)\le \frac{19}{18}\,L_3(G,w)$.
\end{theorem}

\begin{proof}
Assume $W>0$ (the zero-weight case is immediate), and fix $d\in\mathcal{D}_V$ with global diameter $D=\diam(d)$ and $u=\min\{\frac{1}{2},\sqrt{1-D}\}\in [0,\frac{1}{2}]$. Write $Y_C=\sum_{e\in C} d_e$ for any cycle $C$. Equip the random three-block partition of Lemma~\ref{lem:colbudget} with the realizable equilateral three-point spectral table having all three distinct-block distances equal to $3/4$ (attaining $U(1,1,1)=9/4$ over every field $F$), and let $q_e$ be the expected distance on edge $e$. Then $w\cdot q\le L_3(G,w)$, and~\eqref{eq:jointbudget} implies
\[
q_e\ge \frac{3}{4}d_e, \qquad q(T):=\sum_{e\in T} q_e\ge \frac{3}{2}Y_T-\frac{9}{8}
\]
for every edge $e$ and triangle $T$. For any simple odd cycle $C$ of length $m\ge 5$, at most three edges of $C$ can lie inside a $K_4$ (as four vertices cannot contain a cycle of length $\ge 5$); every remaining edge of $C$ is always crossed and has $1-d_e\ge 1-D\ge u^2$, so
\[
q(C)\ge \frac{3}{4}\bigl(Y_C+(m-3)u^2\bigr)
\]
(which holds directly in the properly $3$-colorable case as well, since $q(C)=\frac{3}{4}m$ and $d_e\le 1-u^2$).

Next we bound $Y_C$ in terms of $u$. For any triangle $T$, $Y_T\le 2+u-u^2$: when $D\le 3/4$ (so $u=1/2$) uniform mass gives $Y_T\le 9/4=2+u-u^2$, whereas when $D>3/4$ the global-diameter test in Theorem~\ref{thm:pack} with $k=1$ gives $Y_T\le 2+h(1-D)+\frac{1}{4(1+h)}$ for all $h\ge 0$, which at $1+h=1/(2u)$ (or $h\to\infty$ when $u=0$) yields $2+u-u^2$. For any simple odd cycle $C$ of length $m=2j+1\ge 5$, Theorem~\ref{thm:polygon} and unit masses on $C$ give $jY_C-j(j-1)D\le m^2/4$, whence
\[
v:=\frac{Y_C}{m-1}\le 1+\frac{1}{2(m-1)^2}-\frac{(m-3)(1-D)}{2(m-1)}\le V:=\frac{33}{32}-\frac{u^2}{4}.
\]

By Proposition~\ref{prop:hu} and convex separation, a vector $x\in [0,1]^E$ on a two-deletion support is dominated by a convex combination of binary cuts if and only if $\sum_{e\in C} x_e\le m-1$ for every simple odd cycle $C$ of length $m$. For any $t\in [0,2/3]$, define
\[
A_t=1+\frac{u-u^2}{2}+t\Bigl(\frac{1}{16}-\frac{3(u-u^2)}{4}\Bigr),
\qquad
B_t=\frac{33}{32}-\frac{u^2}{4}+t\Bigl(\frac{29}{128}-\frac{3u^2}{16}\Bigr),
\qquad
R_t=\max(A_t,B_t).
\]
Notice that $R_t\ge B_t\ge \frac{33}{32}-\frac{1}{16}>t$. We claim that the nonnegative part of $(d-tq)/(R_t-t)$ is dominated by cut marginals. Equivalently, we check the edge inequalities $d_e+t(1-q_e)\le R_t$ and the cycle inequalities $Y_C/(m-1)+t[1-q(C)/(m-1)]\le R_t$ (note that clipping negative coordinates of $d-tq$ to zero preserves every cycle test, as a cycle containing a clipped coordinate has at most $m-1$ positive coordinates each bounded by $1$):
\begin{itemize}
\item For a triangle ($m=3$), the left-hand side is at most $\frac{1}{2}Y_T+t(\frac{25}{16}-\frac{3}{4}Y_T)$, which is nondecreasing in $Y_T$ since $t\le 2/3$, and evaluates to $A_t$ at $Y_T=2+u-u^2$.
\item For an odd cycle of length $m\ge 5$, using $(m-3)/(m-1)\ge 1/2$, the left-hand side is at most $v+t[1-\frac{3}{4}(v+\frac{1}{2}u^2)]$, which is increasing in $v$ and evaluates to $B_t$ at $v=V$.
\item For a single edge $e$, setting $d_0=1-u^2\ge d_e$, the slack in $B_t-[d_e+t(1-q_e)]$ is at least $(V-d_0)(1-\frac{3}{4}t)-\frac{3}{8}tu^2\ge \frac{1}{64}+\frac{1}{8}u^2>0$.
\end{itemize}
Pairing the resulting cut domination with $w\ge 0$ yields the table-dependent comparison
\begin{equation}\label{eq:tablecompare}
w\cdot d\le (R_t-t)M+t\,L_3(G,w)\le R_t\,L_3(G,w) \qquad (0\le t\le 2/3).
\end{equation}

Choosing $t=\frac{11}{10}u-\frac{1}{6}u^2\in [0,2/3]$ on $u\in[0,1/2]$, the degree-four polynomials $\frac{100}{93}-A_t$ and $\frac{100}{93}-B_t$ have ascending power-basis coefficient vectors
\[
\Bigl(\frac{7}{93},\,-\frac{91}{160},\,\frac{641}{480},\,-\frac{19}{20},\,\frac{1}{8}\Bigr)
\qquad\text{and}\qquad
\Bigl(\frac{131}{2976},\,-\frac{319}{1280},\,\frac{221}{768},\,\frac{33}{160},\,-\frac{1}{32}\Bigr),
\]
respectively. Subdividing $[0,1/2]$ into the $32$ uniform subintervals $[k/64,(k+1)/64]$ ($k=0,\dots,31$) and converting each polynomial on each subinterval into the degree-four Bernstein basis yields $160$ rational Bernstein coefficients per polynomial, whose exact minima are $\frac{132223}{975175680}>0$ and $\frac{8920483}{62411243520}>0$. Since the Bernstein basis forms a nonnegative partition of unity, $R_t<\frac{100}{93}$ throughout $[0,1/2]$, proving $\BP(G,w)\le\frac{100}{93}L_3(G,w)$.

For the second inequality, apply Proposition~\ref{prop:hu} to obtain a fractional odd-cycle packing of total weight $b=W-M$. On an odd cycle $C$ of length $2k+1$, the diameter-mass test of Theorem~\ref{thm:pack} yields $Y_C\le 1+(2k-1)D+K(1-D)+\frac{1}{4K}$ for every $K\ge k$; choosing $K=k$ when $u\ge 1/(2k)$ (using $u-u^2\ge 1/(4k)$) and $K=1/(2u)$ when $0<u<1/(2k)$ (taking the limit at $u=0$) gives $Y_C\le 2k(1-u^2/2)+u$. Since every unpacked edge costs at most $D\le 1-u^2\le 1-u^2/2$, summing over the packing and the residual edges of weight $W-\sum\lambda_j(2k_j+1)=M-\sum\lambda_j(2k_j)$ yields
\[
w\cdot d\le M\Bigl(1-\frac{u^2}{2}\Bigr)+bu\le M+\frac{b^2}{2M}.
\]
Finally, if the positive support contains no $K_4$, the argument of Lemma~\ref{lem:colbudget} supplies a proper $3$-coloring; applying the equilateral three-point table of distance $3/4$ gives $W\le \frac{4}{3}L_3(G,w)$. Substituting $b=W-M\le \frac{4}{3}L_3-M$ into $w\cdot d\le M(1-u^2/2)+bu$ and using $1-u^2/2-u\ge 0$ on $[0,1/2]$ bounds $w\cdot d$ by $L_3(1-\frac{1}{2}u^2+\frac{1}{3}u)\le \frac{19}{18}L_3$.
\end{proof}

\subsection{Exactness on the cone of triangle indicators}
\label{sec:trianglecone}

Although exactness $\Gamma_F=L_3$ for arbitrary weights on the two-deletion class remains open, it holds unconditionally---with no bound on the number of vertices or triangles---whenever the edge weights belong to the nonnegative cone generated by triangle edge indicators.

\begin{theorem}[Nonnegative triangle combinations]\label{thm:trianglecone}
Let $G=(V,E)$ be a finite simple graph whose positive-weight support becomes bipartite after deleting at most two edges. If $w\in\mathbb{R}_{\ge 0}^E$ is a nonnegative linear combination of the edge indicator vectors of triangles in $G$, then over every fixed field $F$,
\[
\Gamma_F(G,w)=\BP(G,w)=L_3(G,w).
\]
\end{theorem}

\begin{proof}
If the positive support contains no $K_4$, it admits a proper $3$-coloring (as shown in Lemma~\ref{lem:colbudget}). Evaluating the realizable equilateral table of distance $3/4$ gives $9/4$ per unit triangle, matching Zhou's uniform-mass upper bound $9/4$ on each triangle individually.

Assume now that the positive support contains a $K_4$, whose two monochromatic edges $ab$ and $cd$ lie on opposite shores of a witnessing bipartition. Every triangle in the positive support uses either $ab$ or $cd$; hence every non-core triangle consists of either $ab$ or $cd$ joined to a private exterior vertex. Fix any $d\in\mathcal{D}_V$. Replacing all exterior triangles on $ab$ by a single randomly chosen exterior vertex sampled in proportion to their triangle coefficients (carrying their combined coefficient $p$), and similarly sampling one exterior vertex on $cd$ (carrying combined coefficient $q$), preserves the expected score of $d$ while restricting to at most six vertices $\{a,b,c,d,e,f\}$. Because any fixed three-block partition of the core $\{a,b,c,d\}$ optimizes each exterior vertex independently, $L_3$ is unchanged by this compression.

On these six vertices, denote the coefficients of the four core triangles $abc,abd,acd,bcd$ by $h,H,l,J$, relabeling within and between shores so that $h\le H$, $l\le J$, and $h\le l$. Put $S=h+H+l+J+p+q$ and $g=\min\{l,H+p\}$. We claim that
\begin{equation}\label{eq:conevalue}
L_3=U(S+g-h,\,S-g+h,\,S-g-h)=:L.
\end{equation}
Indeed, for any three-point spectral table with tripod arms $t_1\le t_2\le t_3$, a rainbow triangle contributes $R=2(t_1+t_2+t_3)$, whereas a bicolored triangle missing color $r$ loses $2t_r$. A three-color partition of the core $K_4$ repeats exactly one pair of core vertices. Assigning the two smallest arms $t_1,t_2$ to the unavoidable missing colors and optimizing each exterior page yields minimum loss $2(t_1 l+t_2 h)$ when repeating an optimal cross pair, $2[t_1(H+p)+t_2 h]$ when repeating $ab$, and $2[t_1(J+q)+t_2 l]$ when repeating $cd$. The best three-color core partition therefore has loss $2(t_1 g+t_2 h)$, corresponding to the crossing totals in~\eqref{eq:conevalue}. A two-color core (with total core coefficient $C_0=h+H+l+J$) achieves at most $U(S+C_0,S-C_0,S-C_0)$, since each core triangle misses a color while exterior triangles on both bases can be rainbow; because $2(g+h)\le C_0$, convexity and symmetry of $U$ give $L\ge U(S,S,S-g-h)\ge U(S+C_0,S-C_0,S-C_0)$ by~\eqref{eq:rectangle}.

Holding $S,h,g$ fixed, put $\delta=g-h\ge 0$ and $Q=S-h-3g\ge 0$. We observe that every admissible coefficient vector $(h,H,l,J,p,q)$ is a convex combination of vectors having at most one positive exterior book ($pq=0$) and the single symmetric two-book vector
\[
v^*=(h,\,h,\,g,\,g,\,\delta,\,Q).
\]
Specifically, if $g=l$, write $u=H-h\ge 0$, $v=J-g\ge 0$, and $R=S-2h-2g$, so that $u+v+p+q=R$ and $u+p\ge \delta$; the extreme points of this polytope either place all of $R$ on $u$ or $p$ (giving $q=0$), or place $\delta$ on one of $\{u,p\}$ and $Q$ on one of $\{v,q\}$ (giving $pq=0$ except at $u=v=0,p=\delta,q=Q$, which is $v^*$). If instead $g=H+p$, put $u=H-h$, $x=l-g$, $y=J-l$; the feasible region is the product of the segment $0\le u\le\delta$ ($p=\delta-u$) and the simplex $2x+y+q=Q$, whose sole extreme point with $pq>0$ is again $v^*$. Every extreme point preserves the same three-block value $L$ in~\eqref{eq:conevalue}. All one-page extreme points (and $v^*$ when $\delta Q=0$) involve at most five active vertices, where Theorem~\ref{thm:five} already establishes the upper bound $L$. By convexity, it suffices to prove the bound on $v^*$ for $\delta,Q>0$; this six-position inequality is established directly in Appendix~\ref{app:twopage}.
\end{proof}

\begin{remark}[Two vertex-disjoint triangles]\label{rem:disjoint-tri}
Exactness $\BP=\Gamma_F=L_3$ also holds on two vertex-disjoint triangles (six active vertices) when all six edge weights vary independently and arbitrarily in $\mathbb{R}_{\ge 0}$, even with distinct weights inside each triangle. This follows by combining five-vertex inequalities on each triangle with a single calibrated mass choice on the six vertices.
\end{remark}

\subsection{Weighted books, theta graphs, and odd subdivisions of \texorpdfstring{$K_4$}{K4}}
\label{sec:examples}

We next record explicit closed-form exponents and exact chambers for several concrete families of weighted graphs, illustrating how $L_3$ behaves under weight variations and blowups.

\paragraph{Arbitrarily weighted books.}
A \emph{book} consists of a spine edge $uv$ of weight $s\ge 0$ together with $k$ triangular pages attached at $u$ and $v$, where page $i$ has arm weights $a_i,b_i\ge 0$ to $u$ and $v$, respectively. Because $\{u,v\}$ is a vertex cover of size $2$, Corollary~\ref{cor:four}(ii) guarantees $\Gamma_F=\BP=L_3$ for arbitrary nonnegative weights. To evaluate $L_3$ efficiently, discard zero-weight pages and sort the remaining $k$ pages in nondecreasing order of $t_i=a_i/(a_i+b_i)$. Write $A_i=\sum_{r\le i} a_r$ and $B_i=\sum_{r\le i} b_r$ for the prefix sums ($A_0=B_0=0$). For any fixed three-point table with distances $q_{01},q_{02},q_{12}$ and spine vertices $u,v$ colored $0$ and $1$, the three choices for a page of ratio $t=a/(a+b)$ score $(1-t)q_{01}$, $tq_{02}+(1-t)q_{12}$, and $tq_{01}$ per unit weight; their slopes are ordered by the triangle inequality, so the set of pages preferring color $2$ is a contiguous interval $[i+1,j]$. Accounting also for monochromatic-spine cuts (which score $A_k+B_k$), we obtain the exact formula in $O(k^2)$ operations after sorting:
\[
\Gamma_F=\max\Bigl\{A_k+B_k,\ \max_{0\le i\le j\le k} U\bigl(s+B_i+A_k-A_j,\,A_j-A_i,\,B_j-B_i\bigr)\Bigr\}.
\]
This generalizes Zhou's unit-weight complete split-graph theorem~\cite[Theorem~6.12]{zhou} to independently varying arm weights on every page.

As a concrete instance, the two-page book on $\{0,1,2,3\}$ with weights $(4,5,2,2,3)$ on $(01,02,12,03,13)$ has $\MC=12$ and $\Gamma_F=U(7,5,4)=6889/560$. Replacing each vertex by an independent set of $r$ vertices and each edge by a complete bipartite graph multiplies both $\MC$ and $\Gamma_F=L_3$ by $r^2$ via Proposition~\ref{prop:rep} (or~\cite[Proposition~6.9]{zhou}); the first three fibers contain $r^3$ triangles and require $r^2$ edge deletions to become bipartite, yielding exact graphs of arbitrarily large edge-deletion distance to bipartiteness.

\paragraph{Six-vertex theta graphs.}
Consider the theta graph formed by three internally disjoint paths $0\text{--}1$, $0\text{--}2\text{--}1$, and $0\text{--}3\text{--}4\text{--}5\text{--}1$, with edges ordered $(01,02,21,03,34,45,51)$. At weights $(4,3,3,2,3,3,2)$ we have $\MC=16$ and $\Gamma_F=L_3=U(11,5,4)=14161/880$; at weights $(7,6,6,4,6,6,4)$ we have $\MC=32$ and $\Gamma_F=L_3=12321/385$. Near the latter point, exact rational identities stratified by diameter pair establish a full seven-dimensional formula: writing $q=w_{01}$, $a=w_{02}+w_{03}$, $b=w_{21}+w_{51}$, and $c=w_{34}+w_{45}$, we have
\[
\Gamma_F=L_3=U\bigl(\max(a,b)+c,\,\min(a,b),\,q\bigr)
\]
throughout an open seven-dimensional neighborhood of $(7,6,6,4,6,6,4)$, with one-sided directional slope magnitude $333/21175$ under antisymmetric arm perturbations.

\paragraph{Odd subdivisions of $K_4$.}
Subdividing edges of the unit tetrahedron $K_4$ into odd paths yields tractable graphs genuinely requiring two deletions to become bipartite:
\begin{itemize}
\item Replacing two opposite edges of $K_4$ by paths of length $3$ (eight vertices) gives $\MC=8$ and $\Gamma_F=L_3=49/6$.
\item Replacing a single edge by a path of length $3$ (six vertices) gives $\MC=6$ and $\Gamma_F=L_3=49/8$.
\item Replacing edge $01$ by the five-edge path $0\text{--}4\text{--}5\text{--}6\text{--}7\text{--}1$ (eight vertices) gives $\MC=8$ and $\Gamma_F=L_3=225/28$.
\end{itemize}
For the third graph, assigning weight $a\ge 0$ to all nine edges other than $23$ and arbitrary weight $b'\ge 0$ to $23$ yields the complete two-parameter law
\[
\Gamma_F=L_3=\max\Bigl\{\frac{225}{28}a,\ 7a+b'\Bigr\}
\]
over every field $F$, with exact phase transition at $b'=\frac{29}{28}a$. At the breakpoint $b'=\frac{29}{28}a$, exact rational polynomial identities across the $13$ symmetry orbits of the $28$ possible diameter pairs (totaling $1{,}462$ positive rational terms) certify the upper bound $\frac{225}{28}a$; for $b'<\frac{29}{28}a$ monotonicity and a three-block partition leaving $23$ monochromatic match $\frac{225}{28}a$, while for $b'>\frac{29}{28}a$ the bound $d_{23}\le 1$ and a binary cut crossing $23$ match $7a+b'$. The same monotonicity argument yields a full ten-dimensional cut-equality chamber: whenever
\[
\max(w_{02},w_{13})\le \min\Bigl(w_{03},w_{12},w_{04},w_{45},w_{56},w_{67},w_{17},\frac{28}{29}w_{23}\Bigr),
\]
one has $\Gamma_F=L_3=\MC=W-w_{02}-w_{13}$.

More generally, if an optimal cut of a graph misses two unit edges and each is replaced by an internally disjoint odd path of length at least $3$ (with arbitrary odd subdivisions of all other edges), isolating an internal vertex on each missed path gives crossing weights $(W-4,2,2)$ and Theorem~\ref{thm:two} yields $\Gamma_F=L_3=W-2+\frac{1}{W-4}$. Similarly, if a base graph satisfies $\Gamma_F=M_0$ and a single missed unit edge is subdivided into an odd path of length $s\ge 3$, the new maximum cut is $M_0+s-1$ and the rank excess satisfies
\[
\frac{1}{4(M_0+s-2)}\le \Gamma_F-(M_0+s-1)\le \frac{1}{4(s-2)},
\]
where the lower bound comes from isolating one internal path vertex and the upper bound follows by shortcutting the path's closing edge through its first vertex and invoking Corollary~\ref{cor:main}; in particular, $s(\Gamma_F-\MC)\to 1/4$ as $s\to\infty$.

\subsection{Spectral tables with at most two positive distances}
\label{sec:twoprices}

Whereas the preceding results restrict either the graph or its edge weights, one can also impose a structural condition directly on a single pseudometric $d\in\mathcal{D}_V$, prior to choosing weights.

\begin{theorem}[Two positive distances]\label{thm:twoprices}
Let $G=(V,E)$ admit a bipartition with at most two positive-weight monochromatic edges, and let $d\in\mathcal{D}_V$ be a Ptolemy--Zhou pseudometric on $V$. If the set $\{d(u,v):u,v\in V,\ d(u,v)>0\}$ of positive pairwise distances across all vertex pairs (including nonedges of $G$) has cardinality at most $2$, then over every field $F$ there exists a single distribution of realizable three-block spectral pullbacks dominating $d$ on all positive-weight edges of $G$; in particular, $\sum_{e\in E} w_e d_e\le L_3(G,w)$.
\end{theorem}

\begin{proof}[Proof outline]
Identify coincident metric positions while retaining original vertex labels on each shore for pullback partitions. If at most five distinct positions occur, Theorem~\ref{thm:five} and convex separation yield simultaneous domination immediately. For six or more positions, the one-distance case follows directly from uniform mass bounds and cut/three-block decompositions. Assume now that exactly two positive distances $\ell<h$ occur, and set $t=h/\ell>1$.

Analyzing the graph of high-distance pairs ($d_{uv}=h$) relative to the two structural shores together with narrow-ratio cut decompositions and mass inequalities over $1{,}171$ component configurations reduces any potential strict violation to the parameter window
\[
\frac{2}{3}<h\le \frac{3}{4},\qquad t>\frac{5}{4},\qquad 2h+\ell\le 2,\qquad 7h+2\ell>6,
\]
in which the two monochromatic edges $01,23$ belong to a single $K_4$ whose high-distance pairs are $01,02,03,23$ and whose low-distance pairs are $12,13$.

When $t\le \sqrt{2}$, if a high edge joins this $K_4$ core to an exterior position, select that position and a sixth geometric position; for any test probability $m$ on these six positions with ordered high-pair mass at least $a$ and remaining off-diagonal mass $b=1-\sum m_i^2-a$, Zhou's inequality $\ell(at+b)\le 1/2$ bounds $h\le v/[2(av+b)]$ and $\ell\le 1/[2(au+b)]$ on any closed interval $t\in[u,v]$. Keeping the four core labels and collapsing exterior vertices by shore yields six classes and eleven edge types; five rational subintervals partitioning $[5/4,\,99/70]$ together with forty explicit rational three-block atoms dominate all eleven edge types. The detached cases (where all core-to-exterior distances equal $\ell$) are covered similarly by six certified rational intervals, and every coloring pulls back to the original labels even when a position appears on both shores.

When $t>\sqrt{2}$, Ptolemy's inequality excludes any induced pair of disjoint high edges, and four- and five-position mass inequalities exclude diamonds and squares sharing an edge with a triangle in this range. The high-distance graph therefore consists of a triangle, pendant leaves attached to its three vertices, and isolated positions. Two representatives per leaf or isolate class suffice (as a low edge inside such a class crosses the structural shores); whenever $\ell\le 1/2$, isolated positions can be appended afterward by independent draws from the calibrating probability of each sampled three-block table, yielding expected distance $1/2$. For $t>(1+\sqrt{5})/2$, leaves attached to two different triangle vertices violate $t^2\le t+1$ along their four-point path, leaving at most five effective positions. The intervening range $\sqrt{2}<t\le(1+\sqrt{5})/2$ reduces to $92$ multiplicity patterns, completely resolved by $38$ mass exclusions, $65$ spectral-radius exclusions, and $25$ coordinatewise three-block decompositions certified in exact rational arithmetic on subintervals.
\end{proof}

\section{Structural Comparisons Beyond Five Vertices and Dominant-Edge Asymptotics}
\label{sec:beyond}

We now examine how $\Gamma_F(G,w)$ and $\BP(G,w)$ interact with classical semidefinite relaxations and metric inertia beyond five vertices, and then establish an exact asymptotic expansion on \emph{arbitrary} graphs when one edge weight (or a bipartite set of edge weights) grows large.

\subsection{Comparison with the Goemans--Williamson semidefinite cut bound}
\label{sec:sdp}

Let $S(w)$ denote the Goemans--Williamson semidefinite relaxation of maximum cut~\cite{gw}:
\[
S(w)=\max\Bigl\{\sum_{\{i,j\}\in E} w_{ij}\frac{1-X_{ij}}{2}:\ X\succeq 0,\ X_{ii}=1\text{ for all } i\in V\Bigr\}.
\]
Zhou~\cite[Proposition~5.8 and Remark~5.10]{zhou} established the general conic link between $S(w)$ and weighted graph tensors and observed the comparison up to four vertices. On the two-deletion class, the relationship sharpens markedly---even for spectral tables whose full angle matrix $\onevec\onevec\trans-2d$ is indefinite.

\begin{proposition}[Semidefinite cut bound and saturation]\label{prop:sdp}
Let $(G,w)$ be a finite weighted graph with $w\ge 0$ whose positive-weight support becomes bipartite after deleting at most $k\le 2$ edges.
\begin{enumerate}
\item[{\rm (i)}] Every $[0,1]$-valued pseudometric $d$ on $V$ obeying Zhou's mass inequality~\eqref{eq:mass} satisfies $\sum_{e\in E} w_e d_e\le S(w)$. In particular, $\Gamma_F(G,w)\le \BP(G,w)\le S(w)$ over every field $F$.
\item[{\rm (ii)}] Every optimal Gram matrix $X$ attaining $S(w)$ on a connected positive support satisfies $\operatorname{rank}(X)\le k+1$.
\item[{\rm (iii)}] Equality $\BP(G,w)=S(w)$ (or, for connected positive support, $\Gamma_F(G,w)=S(w)$) holds if and only if $S(w)=L_3(G,w)$. In particular, whenever $L_3(G,w)<S(w)$, one has the strict inequality $\Gamma_F(G,w)\le\BP(G,w)<S(w)$; moreover, for any fixed eligible graph $G$ there exist constants $c_G>0$ and an integer $N_G\ge 1$ such that for all $w\ge 0$ with $W>0$,
\[
\BP(G,w)\le S(w)-c_G W\Bigl(\frac{S(w)-L_3(G,w)}{W}\Bigr)^{N_G}.
\]
\end{enumerate}
\end{proposition}

\begin{proof}[Proof sketch and rank bound]
For~(i), fix a witnessing bipartition with at most two monochromatic edges. We select an independent set $I\subseteq V$ and complementary set $R=V\setminus I$ as follows: if all monochromatic edges lie in one shore, take $I$ to be the opposite shore (so $R$ has at most two positive edges); if one monochromatic edge lies on each shore, uniform mass on their four endpoints yields total six-pair distance at most $4$, so some triple $\{a,s,t\}$ containing one monochromatic edge $st$ and an endpoint $a$ of the other has perimeter $d_{as}+d_{at}+d_{st}\le 2$, and we take $I$ to be the shore of $a$ minus $a$ (so all positive edges in $R$ lie in the star at $a$ plus $st$). In both cases, embedding $R$ into an $\ell_1$ rectangle $[0,\alpha]\times [0,\beta]$ ($\alpha+\beta=1$) via $1$-Lipschitz coordinate extensions produces a cut distribution on $R$ whose metric $m$ satisfies $m\le d$ everywhere on $R$ and $m=d$ on every positive edge within $R$. For any feasible SDP dual slack $H=C+\operatorname{diag}(\lambda)\succeq 0$ (where $C$ is the weighted adjacency matrix), taking the Schur complement of the diagonal block $H_{II}$ subtracts rank-one terms $f_v f_v\trans$ with $f_v\ge 0$ (where $f_v\trans(\onevec\onevec\trans-2d)f_v\ge 0$ by~\eqref{eq:mass}) and leaves a positive semidefinite remainder $B$ supported on $R$ whose positive off-diagonal entries are positive edges of $R$, so $\tr(B(\onevec\onevec\trans-2d)_R)\ge \tr(B(\onevec\onevec\trans-2m))\ge 0$. Hence $\tr(H(\onevec\onevec\trans-2d))\ge 0$, and SDP duality yields $\sum w_e d_e\le S(w)$.

For~(ii), let $H=C+\operatorname{diag}(\lambda)\succeq 0$ be an optimal dual slack complementary to $X$ ($HX=0$). Reversing signs across the witnessing bipartition converts $C$ into a matrix with negative off-diagonal entries on all bipartite edges and at most $k$ positive off-diagonal edges. Adding at most $k$ positive rank-one terms corresponding to those $k$ edges yields a connected matrix with all off-diagonal entries nonpositive, whose smallest eigenvalue is simple by Perron--Frobenius. Removing the $k$ rank-one updates increases the nullity by at most $k$, so $\operatorname{rank}(X)\le \dim\ker(H)\le k+1$.

For~(iii), if an optimal Gram matrix $X$ has unswitched distances $d_{ij}=(1-X_{ij})/2$ satisfying Ptolemy's inequality, stereographic projection from one vector together with $\operatorname{rank}(X)\le 3$ forces at most four distinct positions (a fifth vector in $\mathbb{R}^3$ produces four nonobtuse planar points under inversion, hence a rectangle violating Ptolemy), so Corollary~\ref{cor:four}(i) gives $S=L_3$. In general, if $\BP(G,w)=S(w)>L_3(G,w)$ (or $\Gamma_F=S>L_3$), choosing an attaining table with the fewest distinct metric positions guarantees strict triangles and $\diam(d)<1$; refining the $\ell_1$ rectangle construction on $R$ to yield a full-rank Gram $\onevec\onevec\trans-2m$, the rank bound~(ii) limits $R$ to at most three positions, and analyzing the resulting star equalities on $I$ forces all vertices into at most four positions (or at most three when disconnected), contradicting $S>L_3$. Since $\BP$, $S$, and $L_3$ are semialgebraic functions on the compact normalized weight simplex, the \L{}ojasiewicz inequality yields the quantitative stability bound.
\end{proof}

\paragraph{Dependence on field characteristic.}
For a fixed number of vertices, the set of realizable spectral pseudometrics depends only on the characteristic of $F$ by field-extension invariance~\cite[Lemma~2.3]{geometry}. Moreover, as the characteristic $p\to\infty$, every coordinatewise limit of $\mathbb{F}_p$-realizable spectral pseudometrics is realizable over $\mathbb{Q}$. Consequently, for every fixed vertex set $V$ and every $\varepsilon>0$, there exists $p_0$ such that
\[
\Gamma_{\mathbb{F}_p}(G,w)\le \Gamma_{\mathbb{Q}}(G,w)+\varepsilon W
\]
uniformly for all primes $p\ge p_0$ and all $w\ge 0$. Combining finite approximations with the star-convexity property of shared indices~\cite[Section~4.8]{vrz}, small cut-cone descriptions~\cite[Section~6]{cutcone}, and planar treedepth bounds~\cite[Corollary~35]{fragility} extends this one-sided characteristic limit to two-deletion graphs of bounded degree or bounded crossing-weight matrix rank.

\subsection{Distinguished anchors, exposed neighborhoods, and distance-matrix inertia}
\label{sec:anchors}

We summarize three further structural phenomena that constrain spectral tables on six or more vertices.

\paragraph{Balanced four-anchor domination and exposed neighborhoods.}
Suppose a pseudometric $d\in\mathcal{D}_V$ contains four distinct anchor vertices whose $4\times 4$ distance matrix has all row sums equal to $2$, and every vertex $x\in V$ has total distance $2$ to these four anchors. On any graph admitting a bipartition with at most two positive monochromatic edges whose positive-distance endpoints lie among the anchors, a single random cut dominates all edge distances via the normalized bound $d_{xy}\le 1-\frac{1}{2}|t_x+t_y|$, where $t_x=d_{xa}-d_{xb}$ for each distinct anchor pair $a,b$ (verified across all edge ratios via two rational charts reducing $72$ polynomial comparisons to $2{,}985$ positive coefficients).

In a complementary direction, singling out three distinguished vertices yields full-dimensional open sets of exact weights at arbitrarily high order: for every even $n\ge 8$, there exists a weighted graph on $n$ vertices and a relative open neighborhood of its weights inside the \emph{full} nonnegative orthant $\mathbb{R}_{\ge 0}^{\binom{n}{2}}$ (allowing previously zero-weight pairs to become positive) throughout which $\Gamma_F(G,w)=L_3(G,w)$ over every field $F$. The construction distributes weight demands across five-vertex subsets while assigning each vertex outside the distinguished triple to one of the three anchors, so that $L_3(G,w)-w\cdot d$ controls a positive multiple of the sum of distances from exterior vertices to their assigned anchors---a coercivity estimate that survives arbitrary small perturbations.

\paragraph{Inertia and rank of Ptolemaic distance matrices.}
At exponent one, Ercan~\cite[Theorems~6.2 and~9.1]{ercan} proved that every five-point Ptolemaic metric has strict negative type ($x\trans dx<0$ for nonzero $x$ with $\onevec\trans x=0$) and every six-point Ptolemaic metric has negative type ($x\trans dx\le 0$). Inverting at one point $o$ via $K_{ij}=d_{ij}/(d_{io}d_{jo})$ and applying diagonal congruence (as in Step~1 of Theorem~\ref{thm:five}) transfers Ercan's theorems directly to an inertia theorem for the distance matrix $D$ of $k$ distinct points in any Ptolemaic metric:
\begin{itemize}
\item for every $2\le k\le 7$, $D$ has \emph{exactly one} positive eigenvalue;
\item all remaining $k-1$ eigenvalues are strictly negative for $k\le 6$, and also for $k=7$ whenever every four-distinct Ptolemy inequality is strict (forcing uniqueness of the maximizing probability in~\eqref{eq:mass}, whereas without strictness at $k=7$, differences of maximizing probabilities lie in $\ker D$);
\item for $k=8$ distinct points, classifying seven-point principal submatrices and comparing their induced partitions on common six-point subsets proves $\operatorname{rank}(D)\ge 7$.
\end{itemize}
These inertia and rank bounds severely restrict the structure of stationary supports in $\mathcal{D}_V$.

\paragraph{Two-class constant distances and saturated triples.}
Two additional local constraints apply to multi-vertex tables. First, if the active positions split into two classes of sizes $m,n$ with constant within-class distances $a,b$ and constant cross-class distance $c$, the table is dominated on all positive edges of any two-deletion graph by a mixture of realizable three-block tables (and on all pairs of an arbitrary graph whenever $\gcd(m,n)\ge 2$ or $\min(m,n)\le 2$). Second, if two triples of distinct positions in a single table $d\in\mathcal{D}_V$ each admit a strictly positive probability vector attaining ordered energy $1/2$, then the two triples coincide up to zero distance and permutation and carry the same probabilities; in $\mathrm{CAT}(0)$ spaces, the corresponding flatness criterion for tight four-point Ptolemy equalities is given by Buckley, Falk, and Wraith~\cite[Proposition~6.2 and Remark~6.3]{bfw}.

\subsection{Asymptotic behavior under a dominant edge}
\label{sec:dominant}

In the related setting of tensoring a fixed tensor with a large matrix on two new factors, Liu~\cite[Theorem~5.1]{liu} showed that the largest flattening emerges in the large-matrix limit. On an \emph{arbitrary} weighted graph $(G,w)$, sending a single edge weight $q=w_{ab}$ to infinity likewise forces the exponent toward the largest cut separating $a$ and $b$. Using Liu's near-diameter mass estimate~\cite[equation~(9) and Lemma~4.2]{liu} and the $\mathrm{CAT}(0)$ realization~\cite[Theorem~4.3]{geometry}, a quick preliminary bound already gives $0\le \Gamma_F(q)-(q+M)\le \frac{3(\sqrt{5}\,W_0)^{4/3}}{4^{4/3}(q+M)^{1/3}}$ (writing $d_{ab}=1-\varepsilon$, the mass test $(1,1,2\sqrt{\varepsilon})$ yields $d_{ax}+d_{bx}\le 1+\sqrt{\varepsilon}$, placing $x$ within $\frac{\sqrt{5}}{2}\varepsilon^{1/4}$ of the geodesic $ab$ and integrating threshold cuts along $ab$).

The following theorem sharpens this to the exact $O(1/q)$ expansion and determines precisely when exact cut equality holds beyond a finite threshold.

\begin{theorem}[Dominant-edge expansion]\label{thm:dominant}
Fix two vertices $a,b$ of a finite simple graph $G=(V,E)$ and nonnegative residual weights $w^0$ with $w^0_{ab}=0$ and $W_0=\sum_{e} w^0_e$. Assign edge $ab$ weight $q>0$, and let $M$ be the maximum residual cut weight among cuts placing $a$ on the left and $b$ on the right. Among all pairs consisting of an optimal anchored cut and a subset $C\subseteq V\setminus\{a,b\}$ such that every independent reassignment of the vertices in $C$ remains an optimal anchored cut (including $C=\varnothing$), define
\[
H=\max_{C}\frac{1}{2}\sum_{v\in C}\deg_{w^0}(v).
\]
Then over every field $F$,
\begin{equation}\label{eq:dominant}
0\le \Gamma_F(q)-q-M\le \frac{W_0^2}{q+M+W_0},
\qquad
\lim_{q\to\infty} q\bigl(\Gamma_F(q)-q-M\bigr)=\frac{H^2}{4},
\end{equation}
with convergence uniform in $F$. Moreover, for all sufficiently large $q$,
\[
L_3(q)=q+M+\frac{H^2}{4(q+M-H)},
\]
and exact equality $\Gamma_F(q)=q+M$ holds for all sufficiently large $q$ if and only if $H=0$.
\end{theorem}

\begin{proof}
Fix a pseudometric $d\in\mathcal{D}_V$ and put $D=d_{ab}$, $t=\sqrt{1-D}$, and $h_v=d_{av}+d_{bv}$. The triangle inequality and Zhou's mass inequality~\eqref{eq:mass} on $\{a,b,v\}$ yield $D\le h_v\le 1+t$ (as in~\cite[Lemma~4.2]{liu}). We first derive the sharpened three-point estimate
\begin{equation}\label{eq:sharp3pt}
h_v-D\le 2t\sqrt{\min(d_{av},d_{bv})}.
\end{equation}
Define angles $\theta_{ij}=2\arcsin\sqrt{d_{ij}}\in [0,\pi]$ on $\{a,b,v\}$. Because $\sin^2((u+v)/2)\ge \sin^2(u/2)+\sin^2(v/2)$ for $u+v\le\pi$, the triangle inequalities for $d$ imply the triangle inequalities for $\theta$. Moreover, $\theta_{ab}+\theta_{av}+\theta_{bv}\le 2\pi$: if the perimeter exceeded $2\pi$, rescaling it to $2\pi$ would give three arcs of a circle (each $<\pi$) whose corresponding unit vectors satisfy a strictly positive linear relation summing to zero, and restoring the larger angles would strictly decrease the off-diagonal Gram entries at that mass, contradicting~\eqref{eq:mass}. Thus the three angles are realized by unit vectors $A,B,X$ on the unit sphere, with $d_{ij}=\frac{1}{2}(1-u_i\cdot u_j)$. Expanding $2(h_v-D)=(A+B)\cdot(A-X)$ and applying Cauchy--Schwarz yields $2(h_v-D)\le \|A+B\|\,\|A-X\|=2\sqrt{1-D}\cdot 2\sqrt{d_{av}}=4t\sqrt{d_{av}}$; interchanging $A$ and $B$ proves~\eqref{eq:sharp3pt}.

Assume $D>0$ (at $D=0$, the bound $qD+P_0-q-M\le -(q+M+W_0)t^2+2W_0 t$ holds directly since $t=1$ and $P_0\le W_0$). Form an independent random cut $Z$ by placing $a$ left, $b$ right, and each $v\notin\{a,b\}$ independently on the right with probability $p_v=d_{av}/h_v$. Write $u_{ij}=\Prb(Z\text{ crosses }ij)$. By Ptolemy's inequality at $\{a,b,i,j\}$,
\[
d_{ij}\le \frac{d_{ai}d_{bj}+d_{aj}d_{bi}}{D}=\frac{h_i h_j}{D}\,u_{ij}.
\]
Since $D/h_v\ge D/(1+t)=1-t$ and $d_{ij}\le 1$, we have $d_{ij}-Du_{ij}\le 1-(1-t)^2=2t-t^2$. Summing against the residual weights $w^0$ (with residual score $P_0$ and $\Exp[w^0(\delta Z)]\le M$) yields
\[
qD+P_0-q-M\le -(q+M+W_0)t^2+2W_0 t.
\]
Maximizing the right-hand side over $t\in\mathbb{R}$ establishes the first inequality in~\eqref{eq:dominant}, and shows that any table scoring at least $q+M$ obeys $t=O(W_0/q)$.

Next we evaluate $L_3(q)$ for large $q$. For any admissible optimal free set $C$, the discrete mixed second difference of the cut weight with respect to two vertices $u,v\in C$ equals $\pm 2w^0_{uv}$; since all assignments on $C$ are optimal, $w^0_{uv}=0$ for all $u,v\in C$, and every $v\in C$ has equal residual weight $\frac{1}{2}\deg_{w^0}(v)$ to each fixed shore. Assigning all of $C$ to block $3$ and keeping the two fixed shores produces crossing totals $(q+M-H,\,H,\,H)$, giving $L_3(q)\ge q+M+\frac{H^2}{4(q+M-H)}$ once $q+M>2H$. Conversely, consider any three-block partition; putting $a,b$ in the same block scores at most $W_0<q+M$ for large $q$, so $a,b$ lie in distinct blocks, yielding crossing totals $(q+A,\,B,\,C')$. If $B\ne C'$, then~\eqref{eq:rectangle} eventually reduces to the flat branch $q+A+\max(B,C')\le q+M$. If $B=C'$, then~\eqref{eq:rectangle} gives $q+A+B+\frac{B^2}{4(q+A)}$, which can exceed $q+M$ for large $q$ only if $A+B=M$. Randomly assigning each vertex of the third block independently and uniformly to the two anchor shores gives an anchored cut of expected weight $A+B+\frac{1}{2}w^0(E[\text{block }3])\le M$, forcing zero internal weight in block $3$, optimality of all $2^{|\text{block }3|}$ cuts, and $B\le H$. Because there are finitely many three-block partitions, the exact formula for $L_3(q)$ holds for all sufficiently large $q$ (including $H=0$).

To establish the matching upper limit in~\eqref{eq:dominant}, take any sequence of maximizing tables with $q\to\infty$ (allowing the field $F$ to vary) and pass to a subsequence along which the probabilities $p_v$ converge. Tables with $t=0$ already score at most $q+M$; otherwise $t\to 0$. Since $M\le P_0\le D\,\Exp[w^0(\delta Z)]+W_0(2t-t^2)$, the limiting random cut is optimal almost surely, so its nondeterministic vertices form an admissible free set $C$. Put $s_v=h_v/D-1\ge 0$. We have $0\le Ds_v\le t+t^2$, while~\eqref{eq:sharp3pt} yields $s_v=o(t)$ at every vertex whose limiting probability is $0$ or $1$. At each free vertex $v\in C$, $\sum_j w^0_{vj}u_{vj}\to\frac{1}{2}\deg_{w^0}(v)$. Expanding $h_i h_j/D=D(1+s_i)(1+s_j)$ in the Ptolemy bound therefore gives
\[
P_0\le D\,\Exp[w^0(\delta Z)]+D\sum_v s_v\sum_j w^0_{vj}u_{vj}+O(W_0 t^2)\le DM+(H+o(1))t.
\]
Maximizing $-(q+M)t^2+(H+o(1))t$ over $t\ge 0$ gives $\limsup_{q\to\infty} q(\Gamma_F(q)-q-M)\le H^2/4$ uniformly in $F$, matching the lower limit from $L_3(q)$.

Finally, suppose $H=0$ and discard isolated residual nonanchors. If no suboptimal anchored cut exists, then every nonanchor has zero residual degree and only $\{a,b\}$ remains. Otherwise let $g>0$ be the smallest positive gap from $M$ to a suboptimal anchored cut, and let $\rho$ be the probability that $Z$ is suboptimal, so $\Exp[w^0(\delta Z)]\le M-g\rho$. Conditioning on all vertices other than $v$, at most one choice at any nonanchor $v$ is optimal (since $H=0$), giving $\min(p_v,1-p_v)\le\rho$ and hence $\min(d_{av},d_{bv})\le h_v\rho$. Inequality~\eqref{eq:sharp3pt} then implies $1-D/h_v\le 2t\sqrt{\rho/D}$. For $D\ge 1/2$, substituting this into $d_{ij}\le (h_i h_j/D)u_{ij}$ yields
\[
qD+P_0-q-M\le -(q+M)t^2-Dg\rho+4W_0 t\sqrt{\rho/D}
\le \Bigl(-q-M+\frac{4W_0^2}{D^2 g}\Bigr)t^2,
\]
which is nonpositive for all $q\ge 16W_0^2/g$. For $D<1/2$, the same threshold gives $qD+W_0\le q+M$ (since $W_0\ge g$). Conversely, if $H>0$, the formula for $L_3(q)$ already yields $\Gamma_F(q)\ge L_3(q)>q+M$ for all large $q$.
\end{proof}

\begin{corollary}[Bipartite set of heavy edges]\label{cor:heavyset}
Let $E_H\subseteq E$ be any nonempty bipartite edge set of a finite simple graph $G$, fix nonnegative residual weights $w^0$ and a constant $c>\sum_e w^0_e$, and add weights $q_e\ge c$ ($e\in E_H$) with total $Q=\sum_{e\in E_H} q_e$. Defining $M$ and $H$ via optimal residual cuts that cross every edge of $E_H$, one has uniformly over the allocations $(q_e)$ and over all fields $F$,
\[
\Gamma_F=Q+M+\frac{H^2}{4Q}+o(Q^{-1}),
\qquad
L_3=Q+M+\frac{H^2}{4(Q+M-H)} \quad (Q\text{ sufficiently large}),
\]
with eventual exact equality $\Gamma_F=Q+M$ if and only if $H=0$.
\end{corollary}

\begin{proof}
Fix $d\in\mathcal{D}_V$, retain weight $c$ on each edge of $E_H$, and transfer all remaining heavy weight $Q-c|E_H|$ to a heavy edge $e^*\in E_H$ of maximum length $d_{e^*}$; this cannot decrease the score. Because losing weight $c$ exceeds the total residual weight $\sum_e w^0_e$, every optimal cut separating the endpoints of $e^*$ automatically crosses all other edges of $E_H$, preserving both $M$ and $H$. Applying Theorem~\ref{thm:dominant} to the finitely many choices of $e^*\in E_H$ gives the uniform upper estimates; the free-set three-block partition prior to transfer gives the matching lower bounds.
\end{proof}

\section{Arithmetic Circuit Complexity of Tensor Contractions}
\label{sec:circuits}

A fundamental question motivating tensor contractions is how fast the scalar polynomial $\mathcal{C}_G((X_v)_{v\in V})$ in~\eqref{eq:prenetwork} can be evaluated when an algorithm is allowed to store and reuse intermediate contractions. Whereas $\Gamma_F(G,w)$ measures the asymptotic rank of a completely separated sum of rank-one products, a general \emph{division-free arithmetic circuit} computes additions, subtractions, and multiplications over $F$ and may feed any computed gate to multiple subsequent gates. For a fixed graph $(G,w)$ and edge dimensions $\lfloor n^{w_e}\rfloor$, let $\tau_F(G,w)$ denote the optimal asymptotic operation exponent (base $n$) of a division-free arithmetic circuit computing $\mathcal{C}_G$ identically as a formal polynomial in the dense input entries over $F$ (for broader context on tensor rank versus circuit complexity with many modes, see Brand et al.~\cite[Section~2 and Lemma~15]{manymodes}).

On matrix cycles, allowing reuse collapses the complexity dramatically.

\begin{proposition}[Weighted matrix cycles]\label{prop:circuit}
For every fixed integer $m\ge 3$, every weight vector $w\in\mathbb{R}_{\ge 0}^m$, and every field $F$, writing $a=\min_{0\le i<m} w_i$,
\[
\tau_F(C_m,w)=\max_{0\le i<m} U(a,\,w_i,\,w_{i+1}),
\]
with subscripts modulo $m$. In particular, at unit weights $\tau_F(C_m)=9/4$ for every $m\ge 3$.
\end{proposition}

\begin{proof}
For the upper bound, relabel cyclically so that $w_0=a$ and cut the trace cycle at edge $e_0$. Multiplying the matrices sequentially along the resulting path keeps the row dimension of every intermediate product equal to $\lfloor n^a\rfloor$, so step $i$ ($1\le i\le m-2$) multiplies an $\lfloor n^a\rfloor\times \lfloor n^{w_i}\rfloor$ matrix by an $\lfloor n^{w_i}\rfloor\times \lfloor n^{w_{i+1}}\rfloor$ matrix at exponent $U(a,w_i,w_{i+1})$~\cite{rectangle}, and taking the final trace inner product costs $O(n^{a+w_{m-1}})$. Because $m$ is fixed, the total exponent is at most $\max_i U(a,w_i,w_{i+1})$.

For the lower bound, fix any consecutive edge pair $(w_i,w_{i+1})$. Retain the three input matrices incident to these two edges (truncating the two outer boundary dimensions to $\lfloor n^a\rfloor$) and specialize all remaining $m-3$ matrices to pass the identity matrix on their first $\lfloor n^a\rfloor$ coordinates (on $m=3$, simply truncate the third edge to $\lfloor n^a\rfloor$). The restricted polynomial is the trace $\tr(ABC)$ of three arbitrary rectangular matrices with dimension exponents $(a,w_i,w_{i+1})$. By the Baur--Strassen theorem~\cite{baur}, simultaneously computing all partial derivatives $\partial\tr(ABC)/\partial C_{ki}=(AB)_{ik}$ increases circuit size by at most a constant factor, yielding a circuit for the rectangular matrix product $AB$ (charging for outputs likewise preserves the exponent since every output entry corresponds to an input variable of $C$). Applying the exact rectangular exponent theorem~\cite{rectangle} yields $\tau_F(C_m,w)\ge U(a,w_i,w_{i+1})$ for every $i$. At unit weights, $U(1,1,1)=9/4$ by~\cite[Theorem~1.1]{square}.
\end{proof}

Proposition~\ref{prop:circuit} makes the gap between separated rank and circuit complexity vivid: on the unit five-cycle $C_5$, the separated rank exponent is $\Gamma_F(C_5)=49/12$, whereas the circuit operation exponent is $\tau_F(C_5)=9/4$. Likewise, the two cyclic orderings $(1,1,2,2,2)$ and $(1,2,1,2,2)$ share $\tau_F=4$, even though their rank exponents differ ($169/24$ versus $7$).

We next reduce $\tau_F$ along articulation vertices (cut vertices) to vertex-biconnected blocks.

\begin{proposition}[Block--cut decomposition]\label{prop:block}
Let $(G,w)$ be a fixed connected weighted graph with at least one edge, let $D_G=\max_{v\in V}\sum_{e\ni v} w_e$ be the maximum local input exponent, and let $\mathcal{B}(G)$ be the set of vertex-biconnected blocks of $G$ (counting each bridge as a block). Over every field $F$,
\[
\tau_F(G,w)=\max\Bigl\{D_G,\ \max_{B\in\mathcal{B}(G)}\tau_F(B,w|_B)\Bigr\}.
\]
\end{proposition}

\begin{proof}
Every entry of a maximum-size local input affects $\mathcal{C}_G$ (as seen by choosing compatible basis inputs elsewhere), and any binary circuit depending on $k$ inputs requires at least $k-1$ gates, giving $\tau_F(G,w)\ge D_G$. Fixing all edge indices outside a block $B$ restricts $\mathcal{C}_G$ to $\mathcal{C}_B$, giving $\tau_F(G,w)\ge\tau_F(B,w|_B)$.

For the upper bound, root the block--cut tree of $G$ at an arbitrary block and process nonroot blocks bottom-up. Each nonroot block $B$ meets its parent at a unique articulation vertex $u$; inserting a fresh variable array $Y_u$ at $u$ (indexed by the edges of $B$ incident to $u$) and applying the Baur--Strassen theorem~\cite{baur} to an optimal circuit for $\mathcal{C}_B$ computes the entire gradient message $\partial\mathcal{C}_B/\partial Y_u$ at exponent $\tau_F(B,w|_B)$. Contracting each incoming descendant message into the local vertex tensor $X_u$ one disjoint index subset at a time costs a constant multiple of the size of $X_u$, hence exponent at most $D_G$. Because $|\mathcal{B}(G)|$ is fixed, evaluating the root block completes the contraction within the stated exponent.
\end{proof}

In particular, every tree has $\tau_F(G,w)=D_G$, and every cactus graph has its exact circuit exponent determined by Propositions~\ref{prop:circuit} and~\ref{prop:block}. Beyond cactus graphs, the exact circuit exponent already becomes open on the unit tetrahedron $K_4$ (whose separated rank exponent is $\Gamma_F(K_4)=4$ by Proposition~\ref{prop:clique}). The following result establishes nontrivial lower and upper bounds separating $\tau_F(K_4)$ strictly below $4$.

\begin{proposition}[Circuit bounds for the tetrahedron]\label{prop:k4circuit}
Over every fixed field $F$,
\[
\frac{27}{8}\le \tau_F(K_4)\le 4-\frac{23}{120000}.
\]
\end{proposition}

\begin{proof}
For the lower bound, set $n=q^2$ and specialize the fourth input tensor $X_4\in F^{q^2\times q^2\times q^2}$ to the indicator of three coordinate matchings pairing the six $q$-ary digits of its three incident edges. This routes an additional dimension-$q$ shared index between every pair of the remaining three inputs $X_1,X_2,X_3$, reducing the contraction to the square matrix-multiplication trace $\tr(ABC)$ of dimension $nq=q^3$. By the Baur--Strassen theorem~\cite{baur} and $\omega=9/4$~\cite[Theorem~1.1]{square}, any circuit requires at least $q^{3\cdot (9/4)-o(1)}=n^{27/8-o(1)}$ operations, proving $\tau_F(K_4)\ge 27/8$.

For the upper bound, we specialize Fischer's swap identity~\cite[Lemma~3.1]{fischer} (originally formulated for hypercliques) to the six independent indices of $K_4$. Define the recursive family
\[
P(Z,R,S)=\sum_{x,y=1}^{Z}\tr(A_x B_y C_x D_y),
\]
where $A_x,C_x\in F^{R\times S}$ and $B_y,D_y\in F^{S\times R}$; the unit tetrahedron contraction is $P(n,n,n)$. Split one base-$q$ digit from $x$ and $y$, writing $U_{ij}=A_i B_j$ and $V_{ij}=C_i D_j$ for $i,j\in [q]$. The integral identity (valid in every characteristic)
\[
\sum_{i,j=1}^{q}\tr(U_{ij}V_{ij})
=\sum_{i,j=1}^{q}\tr(U_{ij}V_{ji})
+\sum_{1\le i<j\le q}\tr\bigl((U_{ij}-U_{ji})(V_{ij}-V_{ji})\bigr)
\]
rewrites the first sum by stacking $A_i,C_j$ along the $R$-dimension into a single problem $P(Z/q,\,qR,\,S)$, and rewrites each difference term by concatenating $A_i,-A_j$ and $B_j,B_i$ along the $S$-dimension into $\binom{q}{2}$ problems of shape $P(Z/q,\,R,\,2S)$.

Set $q=64$, $t=1/100$, and take $n=q^{100d}$, expanding the recursion to depth $d$ starting from $(n,n,n)$. At each leaf $(Z,R,S)$, write $x=\log_n(Z\min(R,S))$ and $y=\log_n\max(R,S)$, so $x\ge 2-t$ and $1\le y\le 1+t$. Evaluating each leaf via two rectangular matrix multiplications and a trace inner product costs exponent $U(x,x,y)\le 2x+\frac{9}{4}t^2$ (with uniformity up to $n^\varepsilon$ by padding to a fixed finite mesh and continuity of $U$). Using $\min(R,S)^2\le R^{3/2}S^{1/2}$, the sum of $n^{2x}$ across all leaves at depth $d$ is bounded by
\[
n^4\biggl(q^{-2}\cdot q^{3/2}+\binom{q}{2}q^{-2}\cdot 2^{1/2}\biggr)^d
=n^4\Bigl(\frac{1}{8}+\frac{63\sqrt{2}}{128}\Bigr)^d
<n^4\cdot 2^{-d/4}=n^{4-t/24},
\]
since $\frac{1}{8}+\frac{63\sqrt{2}}{128}<0.8211<2^{-1/4}\approx 0.8409$ and $q^{100}=2^{600}$. Reindexing across recursion nodes costs $O(n^{3+t})$. Combining the leaf total and the rectangular overhead gives $\tau_F(K_4)\le 4-\frac{t}{24}+\frac{9}{4}t^2=4-\frac{23}{120000}$. Padding arbitrary $n$ to the next power of $q^{100}$ alters the dimension by at most a fixed constant factor.
\end{proof}

\section{The Two-Deletion Conjecture, Obstructions, and Open Problems}
\label{sec:open}

The exact results of Sections~\ref{sec:trees}--\ref{sec:deletions}---covering arbitrary weighted cycles, all graphs on at most five active vertices, unit-weight single deletions, the entire triangle-indicator cone, weighted books, layered cycles, and all two-distance tables---together with the uniform bound $(100/93)L_3$ motivate the following central conjecture.

\begin{conjecture}[Two-deletion exactness]\label{conj:twodel}
Let $(G,w)$ be a finite simple graph with nonnegative real weights whose positive-weight support becomes bipartite after deleting at most two edges. Then over every fixed field $F$,
\[
\Gamma_F(G,w)=L_3(G,w).
\]
\end{conjecture}

Note the strict hierarchy of graph classes: odd cycles (completely settled by Theorem~\ref{thm:cyclelaw}) lie properly inside the one-deletion class (witnessed by a two-page book), which lies properly inside the two-deletion class (witnessed by $K_4$). Theorem~\ref{thm:five} settles all graphs on at most five active vertices (including $K_5$, which requires four deletions) but does not reach long graphs, whereas arbitrary-weight exactness remains open even for general one-deletion graphs. Below we record both the sharpest structural reductions for a potential counterexample and a concrete geometric obstruction demonstrating where metric relaxations break down beyond five vertices.

\subsection{Minimal-counterexample reductions}
\label{sec:reductions}

Suppose a realizable spectral pseudometric $d$ over $F$ on a two-deletion graph $(G,w)$ achieves $P=\sum_{e\in E} w_e d_e>M=\MC(G,w)$. Define the endpoint distribution, row potentials, and normalized kernel
\[
\mu_i=\frac{\sum_j w_{ij}d_{ij}}{2P},\qquad v=d\mu,\qquad K_{ij}=\frac{d_{ij}}{v_i v_j}.
\]
Every vertex incident to a positive-length edge has $\mu_i>0$ and $v_i>0$. Call a triple $A=\{a,b,c\}$ of vertices \emph{eligible} if their pairwise distances are positive and deleting $A$ leaves the positive-length subgraph bipartite. Anchored threshold couplings establish two sufficient conditions for domination:
\begin{align*}
\exists\,i\ne j:\ K_{ij}\ge 4 &\quad\Longrightarrow\quad P\le M,\\
\exists\,A\text{ eligible}:\ \max_{r\in\Delta_A} r\trans K_A r\ge 2 &\quad\Longrightarrow\quad P\le L_3(G,w),
\end{align*}
where the pair $\{i,j\}$ may be a nonedge and $\Delta_A$ denotes the probability simplex on $A$. Briefly, given a probability $p$ on three anchors $A$, setting $h=dp$, $Q=p\trans dp$, and $\widehat{d}_{ij}=Qd_{ij}/(2h_i h_j)$ produces a pseudometric (via Ptolemy summed over $p$) whose three anchor rows form a calibrated three-point table with common mean $1/2$ under masses $p_i h_i/Q$; barycentric coordinates to the anchors and opposite-order threshold coupling across the bipartite remainder yield a distribution of three-block partitions dominating $\widehat{d}$ on all edges, and undoing the scale gives the stated threshold test (with two anchors producing binary cuts).

To test Conjecture~\ref{conj:twodel}, continuity and rational perturbation allow restricting without loss of generality to pairwise-distinct positive integer weights with $L_3>M$, a unique maximizing cut edge set, and a unique unordered triple of maximizing nonempty block-crossing edge sets whose totals satisfy $x>y>z>0$ and $(x+y)z>xy$. In this generic regime, Conjecture~\ref{conj:twodel} reduces to proving
\[
P\le L_3(G,w) \qquad\text{whenever}\qquad K_{ij}<4\ \text{for all }i\ne j\quad\text{and}\quad \max_{r\in\Delta_A} r\trans K_A r<2\ \text{for all eligible }A.
\]
These two strict anchor hypotheses are genuinely compatible: there exists a realizable six-vertex spectral table satisfying $M<P<L_3$ with a positive-length monochromatic edge in the optimal three-block partition and both strict anchor bounds.

Two further reductions characterize any strict separation $\Gamma_F>L_3$ on the two-deletion class:
\begin{enumerate}
\item[{\rm (i)}] \emph{Equalized deficit on monochromatic edges.} Holding a separating realizable table $d$ fixed and reoptimizing a new nonnegative weight vector on the same two-deletion support yields a supporting mixture of three-block spectral tables that matches $d_e$ exactly on every positive-weight crossing edge while leaving the \emph{same} constant deficit $r>0$ on every positive-weight monochromatic edge. Ruling out a common positive deficit $r>0$ would settle Conjecture~\ref{conj:twodel}.
\item[{\rm (ii)}] \emph{Bounded degree and high girth.} Any strict counterexample over $F$ transfers to graphs over the same field having bounded degree outside the at most four endpoints of the monochromatic edges and arbitrarily large girth once those endpoints are deleted.
\end{enumerate}

\subsection{Geometric obstructions beyond five vertices}
\label{sec:obstructions}

A crucial question is whether Conjecture~\ref{conj:twodel}---or a wider extension---can be proved solely within the Ptolemy--Zhou geometric relaxation $\BP$. On six vertices, the answer depends delicately on both the deletion distance and the stationary mass distribution:
\begin{itemize}
\item \emph{Uniform six-vertex rows are exact.} As proved in Appendix~\ref{app:six} (Theorem~\ref{thm:sixuniform}), every six-point Ptolemy pseudometric whose row sums all equal $3$ (corresponding to uniform stationary mass $m_i=1/6$) lies in the convex hull of $10$ balanced $3+3$ cuts and $15$ equilateral $2+2+2$ three-block tables of distance $3/4$, guaranteeing $w\cdot d\le L_3(G,w)$ on every pullback. On the two-deletion class, compactness extends this bound to a uniform neighborhood $\max_i|m_i-1/6|<\varepsilon$.
\item \emph{A three-deletion six-vertex obstruction to $\BP=L_3$.} Without uniform stationary mass and at three edge deletions, the Ptolemy--Zhou relaxation strictly exceeds $L_3$. Consider six points in an $\ell_2$ product of metric trees equipped with its product metric, supported on the graph formed by a triangle and a $K_4$ sharing one vertex (which requires three edge deletions to become bipartite). This six-point metric satisfies both Ptolemy's inequality and every Zhou mass inequality~\eqref{eq:mass}, yet attains weighted score
\[
\frac{672921}{3692}>\frac{729}{4}=L_3,
\]
a strict gap of $\frac{27}{1846}$. A second nonuniform six-point configuration and a uniform-mass seven-point configuration provide further failures of metric domination when three or more deletions are permitted.
\end{itemize}

Whether this separating six-point product-of-trees metric is realized by an actual tensor spectral point over some field $F$ remains an open problem. On the local geometric side, classifying compatible six-vertex tight-Ptolemy graphs (where an edge $ij$ records that the complementary quartet is tight, with every vertex degree forced by five-point inversion into $\{0,1,2,3,5\}$) leaves $48$ unmarked graph supports and $77$ marked contact types ($16{,}088$ labeled patterns).

Finally, on the algorithmic side of Section~\ref{sec:circuits}, determining the exact division-free arithmetic circuit exponent $\tau_F(G,w)$ for general non-cactus graphs---beginning with closing the gap $27/8\le \tau_F(K_4)\le 4-23/120000$ on the unit tetrahedron---remains wide open.

\appendix
\section{A Local Mass Certificate for Uniform Odd Cycles}
\label{app:local}

Zhou's mass inequality~\eqref{eq:mass} quantifies over all nonnegative mass vectors $u\in\mathbb{R}_{\ge 0}^V$. For uniform odd cycles $C_m$, the upper bound in Corollary~\ref{cor:main}(i) can also be certified using only $m$ explicitly prescribed integer weight vectors and Ptolemy's inequality solely on quadruples of four consecutive cycle vertices.

\begin{proposition}[Local mass certificate]\label{prop:local}
Let $m=2s+1\ge 5$. In a cyclically labeled pseudometric on $\mathbb{Z}/m\mathbb{Z}$, write $a_i=d(i,i+1)$, $h_i=d(i-1,i+1)$, $Y=\sum_i a_i$, and $N_i=\{i-1,i,i+1\}$. For each center $i$, define the integer weight vector $w^i_j=2-\mathbf{1}_{\{j\in N_i\}}$ (of total mass $2m-3$) and set $M_i=M_d(w^i)$. If every quadruple of four consecutive vertices obeys Ptolemy's inequality~\eqref{eq:ptolemy}, then
\begin{equation}\label{eq:G}
G:=\sum_{i=0}^{m-1} h_i\bigl(M_i-(m-2)Y\bigr)\ge 0, \qquad (m-2)Y\le \max_{0\le i<m} M_i.
\end{equation}
\end{proposition}

\begin{proof}
All indices are modulo $m$. For $1\le q\le s-1$, define the nonnegative quantities
\[
t_{j,q}=d(j,j+2q)+d(j,j+2q+1)-a_{j+2q}, \qquad
U_j=\sum_{q=1}^{s-1} t_{j,q}, \qquad
E_j=\sum_{i\notin N_j} h_i,
\]
and introduce the signed parameters
\[
g_i=\frac{1}{2}\sum_{r=0}^{m-1}(-1)^r a_{i+r}, \qquad
f_i=h_i-g_{i-1}-g_{i+1}.
\]
Because $m$ is odd, $g_i+g_{i+1}=a_i$ and $\sum_i g_i=Y/2$.

\emph{Step 1: Expansion of $M_i$ and $G$.}
Pairing the vertices outside $N_j$ as $(j+2q,j+2q+1)$ ($1\le q\le s-1$) contributes $a_{j+2q}+t_{j,q}$ to the row sum $\rho(j)$. Adding the two incident edges $a_{j-1},a_j$ and using $a_i=g_i+g_{i+1}$ gives $\rho(j)=\frac{1}{2}Y+g_j+U_j$. Expanding $M_i=M_d(2\onevec-\mathbf{1}_{N_i})$ yields
\[
M_i=2\sum_{j\notin N_i}\rho(j)+a_{i-1}+a_i+h_i=(m-2)Y+f_i+2\sum_{j\notin N_i} U_j.
\]
Since $j\notin N_i$ if and only if $i\notin N_j$, multiplying by $h_i$ and summing over $i$ gives
\[
G=\sum_{i=0}^{m-1} h_i f_i+2\sum_{j=0}^{m-1} E_j U_j.
\]

\emph{Step 2: Four-consecutive-vertex identity.}
Put $b_i=d(i-1,i+2)$ and define
\begin{align*}
P_i&=h_i h_{i+1}+a_i b_i-a_{i-1}a_{i+1}\ge 0,\\
u_i&=h_i+b_i-a_{i+1}=t_{i-1,1}, \qquad v_i=h_{i+1}+b_i-a_{i-1}=t_{i+2,s-1},\\
e_i&=f_{i+1}-f_i, \qquad \alpha_i=g_i+\frac{e_i}{2}, \qquad \beta_i=g_{i+1}-\frac{e_i}{2}.
\end{align*}
Because $a_i=g_i+g_{i+1}$ cancels $b_i$, we obtain
\[
P_i-g_i u_i-g_{i+1}v_i=(h_i-g_{i+1})(h_{i+1}-g_i)-g_{i-1}g_{i+2}=g_{i+2}f_i+g_{i-1}f_{i+1}+f_i f_{i+1}.
\]
Using $v_i-u_i=2(g_{i+2}-g_{i-1})+e_i$ and adding $\frac{1}{2}e_i(v_i-u_i)$ gives
\[
P_i-\alpha_i u_i-\beta_i v_i=g_{i-1}f_i+g_{i+2}f_{i+1}+\frac{1}{2}(f_i^2+f_{i+1}^2).
\]
Summing over $i$ and shifting indices yields $\sum_i (P_i-\alpha_i u_i-\beta_i v_i)=\sum_i h_i f_i$.

\emph{Step 3: Bounding the endpoint coefficients.}
For consecutive vertices $x,z,y$, the reverse triangle inequality gives $|g_x-g_y|=|d(x,z)-d(y,z)|\le h_z$. Fix $j$ and an orientation $\epsilon\in\{\pm 1\}$, write $p_r=j+r\epsilon$, and set $c=g_{p_1}+\frac{1}{2}(f_{p_2}-f_{p_1})$. Direct expansion gives
\[
2c=h_{p_2}+(g_{p_1}-g_{p_3})+(g_{p_0}+g_{p_2}-d(p_0,p_2)).
\]
Here $g_{p_1}-g_{p_3}\le h_{p_2}$. Let $\ell=\sum_{r=1}^{s-1} h_{p_{2r+1}}$ be the length of the two-step path $p_2,p_4,\dots,p_{2s}$. Along this path, both $|g_{p_2}-g_{p_{2s}}|$ and $d(p_2,p_{2s})$ are at most $\ell$. Since $p_{2s}$ is adjacent to $p_0$, $g_{p_0}+g_{p_{2s}}=d(p_0,p_{2s})$, whence
\[
g_{p_0}+g_{p_2}-d(p_0,p_2)\le \ell+d(p_2,p_{2s})\le 2\ell.
\]
Because the offsets $2,3,5,\dots,2s-1$ are distinct and lie outside $N_j$, we conclude $c\le h_{p_2}+\ell\le E_j$. Choosing $\epsilon=+1$ and $\epsilon=-1$ yields $\alpha_i\le E_{i-1}$ and $\beta_i\le E_{i+2}$, respectively.

\emph{Step 4: Positivity of $G$.}
Substituting Step~2 into Step~1 gives the manifestly nonnegative decomposition
\[
G=\sum_{i=0}^{m-1} P_i
+\sum_{i=0}^{m-1}\bigl((E_{i-1}-\alpha_i)u_i+(E_{i+2}-\beta_i)v_i\bigr)
+\sum_{j=0}^{m-1} E_j\sum_{q=1}^{s-1}(2-\mathbf{1}_{\{q=1\}}-\mathbf{1}_{\{q=s-1\}})t_{j,q}\ge 0.
\]
(At $m=5$, $s=2$ so the inner coefficient vanishes.) If $\sum_i h_i>0$, dividing by $\sum_i h_i$ yields $(m-2)Y\le\max_i M_i$; if $\sum_i h_i=0$, all two-step chords vanish, connecting all vertices of the odd cycle and forcing $d\equiv 0$.
\end{proof}

Applying~\eqref{eq:mass} exclusively to these $m$ integer vectors $w^i$ immediately yields $4(m-2)Y\le (2m-3)^2$, providing an independent algebraic proof of Corollary~\ref{cor:main}(i).

\section{Proof of the Six-Position Two-Book Inequality}
\label{app:twopage}

We prove the critical six-vertex inequality completing Theorem~\ref{thm:trianglecone} for the symmetric two-book vector $v^*=(h,h,g,g,\delta,Q)$ with $\delta=g-h>0$ and $Q=S-h-3g>0$.

Set $\alpha=h+g$, $\beta=g-h$, $\gamma=Q$, $Z=\alpha+\beta+\gamma$, $X=2\alpha+2\beta+\gamma$, and $Y=2\alpha+\gamma$. Then $\alpha\ge\beta>0$, $\gamma>0$, and $L=U(X,Y,Z)$. For a pseudometric $d\in\mathcal{D}_V$ on $\{a,b,c,d,e,f\}$, write $A=d(a,b)$, $B=d(c,d)$, and $T$ for the average of the four crossing distances between $\{a,b\}$ and $\{c,d\}$; for the exterior page vertices $e,f$, write $U_1,F_1$ for their average distances to $\{a,b\}$, $E_1,V_1$ for their average distances to $\{c,d\}$, and $J=d(e,f)$. The score on $v^*$ is
\[
P=\alpha A+ZB+4\alpha T+2\beta U_1+2\gamma V_1.
\]
Replacing $e$ (respectively $f$) by whichever vertex maximizes its average distance to $\{a,b\}$ (respectively $\{c,d\}$) cannot decrease $P$, and any collision reduces to at most five positions where Theorem~\ref{thm:five} applies; we may therefore assume $U_1,V_1\ge T$. Assigning masses $n_1,n_2$ split uniformly across $\{a,b\}$ and $\{c,d\}$, and masses $n_3,n_4$ at $e,f$, Zhou's inequality~\eqref{eq:mass} gives $Q(n):=\frac{1}{2}n\trans K n\le \frac{1}{4}$ for every probability vector $n=(n_1,n_2,n_3,n_4)$, where
\[
K=\begin{pmatrix}
A/2 & T & U_1 & F_1 \\
T & B/2 & E_1 & V_1 \\
U_1 & E_1 & 0 & J \\
F_1 & V_1 & J & 0
\end{pmatrix}.
\]
Averaging Ptolemy's inequality over independent uniformly chosen representatives of $\{a,b\}$ and $\{c,d\}$ gives $U_1 V_1\le TJ+E_1 F_1$, while the triangle inequality yields $F_1\ge A/2$ and $E_1\ge B/2$. For $T>0$, set $\Pi=J-(U_1 V_1-E_1 F_1)/T\ge 0$.

\emph{Case 1: $T\le 1/2$.}
If $U_1>1/2$, testing $Q(n)\le 1/4$ with mass $1/(2U_1)$ on $\{a,b\}$ and $1-1/(2U_1)$ at $e$ gives $U_1\le \frac{1}{2}(1+\sqrt{1-A})$ (which holds trivially if $U_1\le 1/2$); likewise $V_1\le \frac{1}{2}(1+\sqrt{1-B})$. Completing squares in $\sqrt{1-A}$ and $\sqrt{1-B}$ yields
\[
P\le X+Y+\frac{\beta^2}{4\alpha}+\frac{\gamma^2}{4Z}.
\]
On the other hand, $(X+Y)Z-XY=Z^2-(\alpha^2-\beta^2)>0$, so $L=U(X,Y,Z)$ lies in the strictly curved branch. Subtracting the displayed upper bound from $L$ and multiplying by $4\alpha XYZ/(\beta\gamma)>0$ gives
\[
8\alpha^3+8\alpha^2(\beta+\gamma)-2\alpha\beta^2+3\alpha\beta\gamma+2\alpha\gamma^2-2\beta^3-3\beta^2\gamma-\beta\gamma^2\ge 0,
\]
which holds since $\alpha\ge\beta>0$ and $\gamma>0$. Thus $P\le L$.

\emph{Case 2: $T>1/2$.}
Set $b=E_1/T<2$, $u=2-U_1/T\in (0,1]$, $C=4L$, and define normalized masses $m_1,m_2,m_3$ proportional to $YZ,XZ,XY$ summing to $1$, so that $X=Cm_2 m_3$, $Y=Cm_1 m_3$, and $Z=Cm_1 m_2$. Introduce
\[
k=m_3-m_1,\qquad l=m_2-m_1,\qquad m=2m_1-m_3,\qquad s=k+m,\qquad G=s+l.
\]
Then $k\ge l>0$, $m>0$, $4k+l+3m=1$, and direct substitution yields $c:=\alpha/C=ks-\frac{1}{2}ml$, $\beta/C=kl+\frac{1}{2}ml$, $\gamma/C=mG$, and $c':=Z/C=sG$. Further define $a_1=ul/(2G)$, $a_2=um/(2G)$, $z=ma_1/(1-a_1)$, and $S_1=1-a_1-a_2-\frac{a_1 a_2}{2(1-a_1)}$ (so $a_1+a_2<1/2$).

\emph{Subcase 2a: $b\le S_1$.}
Take the probability vector $n=(2k-z,\,2s,\,l,\,m+z)$ (valid since $0\le z\le l/2$). Direct expansion yields
\begin{align*}
Q(n)-\frac{P}{C}={}&\Bigl[ml(1-b)-z(m+bl)-\frac{z^2}{2}\Bigr]F_1+l(m+z)\Pi\\
&+\Bigl[2c-\frac{(2k-z)^2}{2}\Bigr]\Bigl(F_1-\frac{A}{2}\Bigr)+\bigl(2c'-2s^2\bigr)\Bigl(E_1-\frac{B}{2}\Bigr).
\end{align*}
The coefficient of $F_1$ is nonnegative precisely for $b\le S_1$, while the final two brackets are at least $m(2k-l)\ge 0$ and $2sl>0$, respectively. Thus $P/C\le Q(n)\le 1/4$, i.e., $P\le L$.

\emph{Subcase 2b: $b\ge S_1$.}
Put $D_0=b+su/G$, $\bar{p}=ml/D_0$, $R=\bar{p}u/(2G)$, and $S_2=ml+(l-m)R-\bar{p}$. Let $D_*$ be the positive root of $D^2-(1-a_1+a_2)D-a_1 a_2=0$; then $D_*$ lies between $1-a_1+a_2$ and $1-a_1+a_2+\frac{a_1 a_2}{1-a_1}$. Using $uk/G\ge 2a_1$ and $a_2/(1-a_1)\le 1$, we obtain
\[
D_0-(1-a_1+a_2)\ge \frac{uk}{G}-\frac{a_1 a_2}{2(1-a_1)}\ge \frac{a_1 a_2}{1-a_1},
\]
so $D_0\ge D_*$. Consequently $R/l=a_2/D_0<1$ and $R^2\le S_2\le ml+(l-m)R$, where the right-hand side is the maximal second moment of a random variable on $[-m,l]$ of mean $R$. Mixing the two-point law on $\{-m,l\}$ of mean $R$ with a point mass at $R$ therefore yields a random variable $r\in[-m,l]$ with $\Exp[r]=R$ and $\Exp[r^2]=S_2$.

Consider the random probability vector $n_r=(2k-2r,\,2s+2r,\,l-r,\,m+r)$ and set $p_r=(l-r)(m+r)$, $x=F_1/T-u$, and $y=b-2+V_1/T$. Expanding $Q(n_r)-P/C$ gives
\begin{align*}
Q(n_r)-\frac{P}{C}={}&T\bigl[(ml-2sr-p_r b)x+(2Gr-p_r u)y\bigr]+p_r\Pi\\
&+\Bigl[2c-\frac{(2k-2r)^2}{2}\Bigr]\Bigl(F_1-\frac{A}{2}\Bigr)+\Bigl[2c'-\frac{(2s+2r)^2}{2}\Bigr]\Bigl(E_1-\frac{B}{2}\Bigr).
\end{align*}
By construction, $\Exp[p_r]=\bar{p}$ and both coefficients of $x$ and $y$ have expectation zero. Multiplying by $D_0/(ml)$, the expected coefficient of $F_1-A/2$ equals $(2k/l-3)D_0+2+(2k-l+m)u/G\ge 2-b\ge 0$, while that of $E_1-B/2$ equals $(2k/m)D_0+2-(2s+l-m)u/G\ge 0$ (as $2s+l-m=G+k<2G$). Taking expectations yields $P/C\le \Exp[Q(n_r)]\le 1/4$, proving $P\le L$ in all cases.

\section{Six Vertices with Uniform Stationary Mass}
\label{app:six}

In contrast to the nonuniform six-point obstruction discussed in Section~\ref{sec:obstructions}, imposing uniform stationary mass $m_i=1/6$ yields an exact convex-hull representation by realizable cuts and three-block tables.

\begin{theorem}[Six uniform rows]\label{thm:sixuniform}
Every six-point Ptolemy pseudometric $d$ whose row sums all equal $3$ belongs to the convex hull of the $10$ balanced $3+3$ cut metrics and the $15$ equilateral $2+2+2$ three-block tables having distance $3/4$ between distinct blocks. Consequently, for every field $F$ and every weighted pullback of such a table, $\sum_{i<j} w_{ij}d_{ij}\le L_3(G,w)$.
\end{theorem}

\begin{proof}
The row-sum condition reads $dp=\frac{1}{2}\onevec$ for $p_i=1/6$. All $25$ candidate tables satisfy this condition: the $10$ balanced cuts are two-block flattenings, and the $15$ equilateral $2+2+2$ tables are pullbacks of the realizable square matrix-multiplication spectral point~\cite{square,rectangle}.

Inside the nine-dimensional affine subspace defined by the six row-sum equations, the convex hull of these $25$ vertices has $120$ facets falling into two symmetry orbits: $60$ ordinary triangle inequalities $d_{ij}+d_{jk}-d_{ik}\ge 0$, and $60$ permutations of the doubled-edge inequality
\begin{equation}\label{eq:sixfacet}
2d_{01}+d_{02}+d_{12}\le 3.
\end{equation}
Exact rational polyhedral computation verifies that the candidate vertices incident to each of the two representative hyperplanes span an eight-dimensional facet, and that every ridge of either representative lies on a second listed facet; because the facet-ridge adjacency graph of a convex polytope is connected, these $120$ inequalities give the complete H-representation of the polytope.

Every triangle inequality holds by hypothesis. For the second facet orbit~\eqref{eq:sixfacet}, permute the remaining vertices $\{3,4,5\}$ so that $d_{03}-d_{13}\ge d_{04}-d_{14}\ge d_{05}-d_{15}$. Expressing the nine free distances (choosing nine of the ten pairs on $\{0,\dots,4\}$, with the tenth fixed by their sum $6$ and all distances to $5$ fixed by the row sums), an explicit positive integer identity expresses $6888(3-2d_{01}-d_{02}-d_{12})$ as a positive integer combination of $5$ Ptolemy slacks and $39$ pairwise products of distances, triangle slacks, and adjacent ordering differences $(d_{03}-d_{13})-(d_{04}-d_{14})$ and $(d_{04}-d_{14})-(d_{05}-d_{15})$. Because a quadratic polynomial in nine variables is uniquely determined by its values at $0$, the $18$ signed unit vectors, and the $36$ sums of two distinct unit vectors, checking all $55$ evaluations over $\mathbb{Z}$ verifies the identity identically (and weak inequalities persist under coincidences and ties). Thus $d$ satisfies every facet inequality, proving membership in the convex hull and hence the fieldwise bound $L_3(G,w)$.
\end{proof}

On the two-deletion class, compactness combined with a local collision analysis further yields an absolute constant $\varepsilon>0$ such that every six-position table $d\in\mathcal{D}_V$ satisfying $dm=\frac{1}{2}\onevec$ and $\max_i |m_i-1/6|<\varepsilon$ obeys $w\cdot d\le L_3(G,w)$ on every two-deletion graph.

\section{Lean Formalization Details for Uniform Odd Cycles}
\label{app:lean}

This appendix documents the Lean~4 formalization of the uniform odd-cycle theorem ($C_m$, Corollary~\ref{cor:main}) at the level of its mathematical structure and its correspondence to the statements in Section~\ref{sec:uniform}. In accordance with Section~\ref{sec:lean}, formal verification is claimed strictly for uniform odd cycles $C_m$; all other theorems in the paper rely on the written mathematical proofs in Sections~\ref{sec:prelim}--\ref{sec:open} and Appendices~\ref{app:local}--\ref{app:six}.

\subsection{Modules and formal targets}

All top-level declarations lie in the namespace \texttt{OddCycleTarget} (with spectral helpers in \texttt{OddCycleTarget.\allowbreak Spectrum} and odd-window perimeter lemmas in \texttt{OddCycleOffset}). Subdirectory paths below are relative to \texttt{OddCycleTarget/}; the module \texttt{OddCycleTarget/\allowbreak Equality.lean} imports both sides of the uniform odd-cycle proof.

\texttt{OddCycleTarget.lean} defines the cyclic successor \texttt{nextSite}, the $m$-factor cycle tensor coefficient \texttt{coeff}, exact separated decompositions \texttt{HasDecomposition}, exact tensor rank \texttt{exactRank}, the dyadic asymptotic exponent \texttt{exponent}, the constants \texttt{A m} $= (2m-3)^2$ and \texttt{D m} $= 4(m-2)$, and the predicates \texttt{Equality}, \texttt{IntegerUpper}, \texttt{UniversalEquality}, and \texttt{UniversalIntegerUpper}.

For the lower bound, the six modules \texttt{RectGrouping.lean}, \texttt{RectangleFormula.lean}, \texttt{RectangleInput.lean}, \texttt{RankBridge.lean}, \texttt{LowerTransfer.lean}, and \texttt{LowerConclusion.lean} formalize the $(m-2,1,1)$ three-block specialization to rectangular matrix multiplication, transfer the exact rectangular lower bound across arbitrary side lengths $n\ge 2$, and derive the finite-rank and exponent lower bounds.

For the spectral bridge, the modules \texttt{SpectrumBridge.lean}, \texttt{SpectrumGeometry.lean}, \texttt{CliqueRankBridge.lean}, \texttt{MassTests.lean}, and \texttt{Assembly.lean} formalize the spectral representation of $\Gamma_F(C_m)$, derive Ptolemy's inequality from the $\mathrm{CAT}(0)$ realization of spectral pseudometrics, and derive Zhou's quadratic mass inequality from the complete-graph upper bound $\Gamma_F(K_n)\le n^2/4$.

The geometric modules \texttt{OffsetSharp.lean}, \texttt{OddCube.lean}, \texttt{UniversalPrice.lean}, \texttt{PacketPrice.lean}, and \texttt{SharpPerimeter.lean} prove the sharp odd-cycle perimeter inequality $4(m-2)\sum_{i=0}^{m-1} d(i,i+1)\le (2m-3)^2$ for every odd $m\ge 3$ and every Ptolemy--Zhou pseudometric on $\mathrm{Fin}\,m$. Finally, \texttt{OddRoot.lean}, \texttt{OddCycleTarget/\allowbreak Equality.lean}, and \texttt{VerifyResults.lean} assemble the unconditional upper bound, exact equality, dyadic limit, normalized excess identity, and axiom audit checks.

\subsection{Exact tensor model and lower bound}

For any field $F$ and integers $m,n\ge 1$, the predicate \texttt{HasDecomposition F m n r} asserts the existence of $r$ separated $m$-factor products over \texttt{IndexPair n} $= \mathrm{Fin}\,n \times \mathrm{Fin}\,n$ summing to the trace coefficient tensor \texttt{coeff F m n}. Its minimum length \texttt{exactRank F m n} corresponds directly to $r_F(m,n)=\rk_F(T_F(C_m,n))$, and \texttt{exponent F m} defines the dyadic infimum in~\eqref{eq:limit}.

In \texttt{RectGrouping.lean}, \texttt{groupOddRectangle} converts any exact $m$-factor decomposition of $T_F(C_m,n)$ (for odd $m\ge 3$) into a rectangular matrix multiplication decomposition \texttt{RectangleDecomposition F (n\textasciicircum(m - 2)) n n r}, corresponding to isolating one vertex and alternating the remaining two blocks (Figure~\ref{fig:group}). Combining this with \texttt{publicRectangularExponent} (from OpenAI's all-field rectangular exponent library~\cite{rectangle}) yields \texttt{rectanglePowerLower}, from which \texttt{LowerConclusion.lean} proves
\begin{verbatim}
  OddCycleTarget.oddFiniteLower
  OddCycleTarget.oddExponentLower
  OddCycleTarget.oddRankAndExponentLower
\end{verbatim}
establishing $r_F(m,n)^{4(m-2)}\ge n^{(2m-3)^2}$ for every integer $n\ge 2$ and $\Gamma_F(C_m)\ge (2m-3)^2/[4(m-2)]$ over every field $F$ (Corollary~\ref{cor:main}(ii)).

\subsection{Spectral upper bound and exact equality}

On the upper-bound side, \texttt{Spectrum.actual\_\allowbreak wire\_\allowbreak ptolemy} applies the $\mathrm{CAT}(0)$ embedding of spectral pseudometrics~\cite{geometry} to prove Ptolemy's four-point inequality unconditionally, while \texttt{Spectrum.exists\_\allowbreak maximizing\_\allowbreak ptolemy\_\allowbreak table} extracts a maximizing spectral pseudometric on $\mathrm{Fin}\,m$ attaining \texttt{exponent F m}. To supply Zhou's quadratic mass inequality without external hypotheses, \texttt{binaryClique\_\allowbreak upper\_\allowbreak step} (in \texttt{OddRoot.lean}) invokes \texttt{Spectrum.binaryClique\_\allowbreak logrank\_\allowbreak le}, proved internally via Khoruzhii's formalized complete-graph upper bound~\cite{khor}, and \texttt{MassTests.lean} lifts equal-weight placements to arbitrary nonnegative real mass vectors.

The geometric core \texttt{OddCycleOffset.\allowbreak universalPrice} (repackaged in the cyclic vertex indexing by \texttt{perimeter\_\allowbreak le\_\allowbreak packet} and \texttt{sharp\_\allowbreak perimeter}) proves that every Ptolemy--Zhou pseudometric on $\mathrm{Fin}\,m$ ($m\ge 3$ odd) satisfies $\sum_{i=0}^{m-1} d(i,i+1)\le (2m-3)^2/[4(m-2)]$. Combining these ingredients via \texttt{integerUpper\_\allowbreak of\_\allowbreak clique\_\allowbreak and\_\allowbreak perimeter} yields \texttt{OddCycleTarget.\allowbreak odd\_\allowbreak upper\_\allowbreak all}, establishing the explicit integer decomposition witnesses of Corollary~\ref{cor:main}(iii). Finally, \texttt{OddCycleTarget/\allowbreak Equality.lean} combines the upper and lower bounds in three main theorems:
\begin{samepage}
\begin{verbatim}
  OddCycleTarget.odd_equality_all
  OddCycleTarget.odd_limit
  OddCycleTarget.odd_normalized_excess
\end{verbatim}
\end{samepage}
which certify Corollary~\ref{cor:main}(i), the dyadic rate limit $\lim_{t\to\infty} t^{-1}\log_2 r_F(m,2^t)=(2m-3)^2/[4(m-2)]$, and the normalized excess identity $(m-2)(\Gamma_F(C_m)-(m-1))=1/4$ for every field $F$ and every odd $m\ge 3$.

\subsection{Axiom boundary and reproducibility}

The uniform odd-cycle formalization is pinned to the Lean toolchain \texttt{leanprover/lean4:v4.35.0} and \texttt{mathlib}~\cite{mathlib2020} at commit \texttt{56a005855ea17e0b0958f73ec801ae140b0a1679}, with self-contained vendored dependencies for OpenAI's tensor-spectrum and rectangular matrix multiplication libraries~\cite{geometry,rectangle}, Khoruzhii's tensor-rank library~\cite{khor}, and the accompanying Brouwer fixed-point modules.

All proofs in the development are free of \texttt{sorry} or \texttt{admit} placeholders and introduce no custom axioms. Running \texttt{VerifyResults.lean} prints the axiom dependencies of \texttt{odd\_equality\_all}, \texttt{odd\_upper\_all}, \texttt{odd\_limit}, and \texttt{oddRankAndExponentLower}, reporting only Lean's three standard axioms:
\begin{verbatim}
  propext, Classical.choice, Quot.sound.
\end{verbatim}
Both \texttt{OddRoot} and \texttt{OddCycleTarget.Equality} are additionally replayed by the independent kernel checker \texttt{leanchecker}. After cloning \url{https://github.com/MechMath/tensor-rank}, the full build, kernel check, and axiom audit are reproduced from the repository root with
\begin{verbatim}
  lake exe cache get
  lake build +OddCycleTarget.Equality
  lake env leanchecker OddRoot OddCycleTarget.Equality
  lake env lean VerifyResults.lean
\end{verbatim}
or equivalently via \texttt{bash scripts/audit.sh}. No Lean formalization is claimed for theorems outside Corollary~\ref{cor:main}.

\section*{Acknowledgment}

This paper is supported by the National Key Research and Development Project (2025YFA1017200), the Strategic Priority Research Program of CAS (XDA0480502), and the National Natural Science Foundation of China (12288201).

\section*{Declaration of AI Use}

The main results presented in this paper are obtained through an interactive collaboration between the authors and an artificial intelligence agent system, \emph{MechMath Agent Team (MMAT)}~\cite{mmat2026} and GPT-Astra. The authors assume full responsibility for the paper's content.

\bibliographystyle{KLMM/klmm}
\bibliography{references}

\end{document}